\documentclass[10pt,english,prd,twocolumn,superscriptaddress,nofootinbib,preprintnumbers,showpacs,floatfix]{revtex4-1}
\usepackage[utf8]{inputenc}
\usepackage{amsmath}
\usepackage{amsfonts}
\usepackage{amssymb}
\usepackage{graphicx}
\usepackage{array}
\usepackage{booktabs}
\usepackage{multirow}
\usepackage{mathtools}
\usepackage{subcaption}
\usepackage{makecell}
\DeclareMathOperator{\arccot}{arccot}
\DeclareMathOperator{\arcsinh}{arcsinh}
\DeclareMathOperator{\csch}{csch}
\DeclarePairedDelimiter\abs{\lvert}{\rvert}%
\usepackage[usenames,dvipsnames]{xcolor}
\usepackage{hyperref}   
\hypersetup{colorlinks=true, 
citecolor=MidnightBlue, linkcolor=MidnightBlue, urlcolor=Cyan}
\usepackage{tikz}
\usepackage{verbatim} % comment
\definecolor{purple}{rgb}{1,0,1}
\definecolor{lime}{HTML}{A6CE39} % needs xcolor

\newcommand{\orcidicon}{%
	\begin{tikzpicture}
	\draw[lime, fill=lime] (0,0) 
		circle [radius=0.16] 
		node[white] {{\fontfamily{qag}\selectfont \tiny ID}};
	\draw[white, fill=white] (-0.0625,0.095) 
		circle [radius=0.007];
	\end{tikzpicture}	\hspace{-2mm}
}
\newcommand\orcidFrancisco{{\href{https://orcid.org/0000-0002-9388-8373}{\orcidicon}}}
\newcommand\orcidManuel{{\href{https://orcid.org/0000-0001-8586-0285}{\orcidicon}}}
\newcommand\orcidTarciso{{\href{https://orcid.org/0009-0007-0450-2672}{\orcidicon}}}
\newcommand\orcidGabriel{{\href{https://orcid.org/0009-0007-3770-8976}{\orcidicon}}}
\begin{document}
%========================================================
\title{Novel electrically charged wormhole, black hole and black bounce exact solutions in hybrid metric-Palatini gravity}

%========================================================
\author{Gabriel I. R\'{o}is\orcidGabriel\!\!}
\email{fc54507@alunos.fc.ul.pt}
\affiliation{Instituto de Astrof\'{i}sica e Ci\^{e}ncias do Espa\c{c}o, Faculdade de Ci\^{e}ncias da Universidade de Lisboa, Edifício C8, Campo Grande, P-1749-016 Lisbon, Portugal}
\affiliation{Departamento de F\'{i}sica, Faculdade de Ci\^{e}ncias da Universidade de Lisboa, Edif\'{i}cio C8, Campo Grande, P-1749-016 Lisbon, Portugal}
%-----------------------------------------------------------------
\author{Jos\'{e} Tarciso S. S. Junior\orcidTarciso\!\!}
\email{tarcisojunior17@gmail.com}
\affiliation{Faculdade de F\'{\i}sica, Programa de P\'{o}s-Gradua\c{c}\~{a}o em 
F\'isica, Universidade Federal do 
 Par\'{a},  66075-110, Bel\'{e}m, Par\'{a}, Brazil}

%=================================================================
\author{Francisco S. N. Lobo\orcidFrancisco\!\!} 
\email{fslobo@ciencias.ulisboa.pt}
\affiliation{Instituto de Astrof\'{i}sica e Ci\^{e}ncias do Espa\c{c}o, Faculdade de Ci\^{e}ncias da Universidade de Lisboa, Edifício C8, Campo Grande, P-1749-016 Lisbon, Portugal}
\affiliation{Departamento de F\'{i}sica, Faculdade de Ci\^{e}ncias da Universidade de Lisboa, Edif\'{i}cio C8, Campo Grande, P-1749-016 Lisbon, Portugal}

%=================================================================
\author{Manuel E. Rodrigues\orcidManuel\!\!}
\email{esialg@gmail.com}
\affiliation{Faculdade de F\'{\i}sica, Programa de P\'{o}s-Gradua\c{c}\~{a}o em 
F\'isica, Universidade Federal do 
 Par\'{a},  66075-110, Bel\'{e}m, Par\'{a}, Brazil}
\affiliation{Faculdade de Ci\^{e}ncias Exatas e Tecnologia, 
Universidade Federal do Par\'{a}\\
Campus Universit\'{a}rio de Abaetetuba, 68440-000, Abaetetuba, Par\'{a}, 
Brazil}
%========================================================

\date{\LaTeX-ed \today}
%%%%%%%%%%%%%%
%========================================================
\begin{abstract}
%========================================================
This paper presents a systematic exploration of exact solutions for electrically charged wormholes, black holes, and black bounces within the hybrid metric-Palatini gravity (HMPG) framework. HMPG combines features of the metric and Palatini formulations of modified gravity, offering a powerful approach to address challenges in General Relativity, particularly those related to cosmic acceleration and dark matter. We examine configurations characterized by a zero scalar potential under spherical symmetry, and present solutions in both the Jordan and Einstein conformal frames. A diverse set of solutions emerges, including traversable wormholes, black holes with double horizons, and ``black universe'' models in which spacetime beyond the horizon leads to an expanding cosmological solution rather than a singularity.
Each configuration is categorized according to the properties of the scalar field, with an in-depth analysis of the horizon and throat structures, asymptotic behaviour, and singularities. These findings underscore the versatility of HMPG in capturing complex gravitational phenomena and broadens the scope of the theory, offering a robust framework for modelling gravitational phenomena across a range of astrophysical contexts.
Future work will benefit from extending these solutions to include scalar potentials, addressing both early-universe inflation and late-time acceleration, and applying observational data, such as gravitational lensing and gravitational wave measurements.
%========================================================
\end{abstract}
%========================================================
%\pacs{04.50.Kd,04.70.Bw}
%=================================================================
\maketitle
%=================================================================
%\def\HMS{{\scriptscriptstyle{\rm HMS}}}
%========================================================
%\bigskip
%\hrule
%\tableofcontents
%\bigskip
%\hrule
%========================================================
%\parindent0pt
%\parskip7pt
%========================================================

%%%%%%%%%%%%%%%%%%%%%%%%%%%%%%%%%%%%%%%%%%%%%%%%%%%%%%%%%%%%
\section{Introduction}
%%%%%%%%%%%%%%%%%%%%%%%%%%%%%%%%%%%%%%%%%%%%%%%%%%%%%%%%%%%%

The pursuit of a comprehensive theory of gravity has become one of the most significant challenges in modern theoretical physics. While General Relativity (GR) \cite{Einstein:1915ca,Einstein:1916vd} remains an extraordinarily successful theory, accurately describing a wide range of gravitational phenomena, it faces significant challenges when confronted with certain observations. For instance, the late-time cosmic acceleration, initially observed through distant supernovae \cite{SupernovaSearchTeam:1998fmf,SupernovaCosmologyProject:1998vns}, suggests that the universe is expanding at an accelerating rate, an effect that GR cannot fully explain without invoking a mysterious form of energy known as dark energy \cite{Copeland:2006wr}. Additionally, the presence of dark matter, inferred from the rotational curves of galaxies and large-scale structure formation \cite{Navarro:1995iw,Moore:1999nt,Bertone:2004pz,Arkani-Hamed:2008hhe}, presents yet another challenge that GR faces within its framework. The early universe, dominated by high-energy conditions and rapid expansion, further complicates the issue \cite{Kolb:1990vq}. The standard cosmological model, based on GR, requires an inflationary period to resolve several initial condition problems  \cite{Starobinsky:1980te,Guth:1980zm,Linde:1981mu,Albrecht:1982wi}. However, the nature of this inflationary period and its underlying mechanisms remain speculative, pointing to potential limitations in GR's applicability to extreme conditions.
In light of the limitations of GR, alternative theories of gravity have been widely investigated in the scientific literature. These theories seek to extend or modify the foundational principles of GR, offering a more comprehensive framework capable of addressing observations that GR alone has difficulty explaining.

One of the primary approaches involves modifying the Einstein-Hilbert action, the fundamental action underlying GR, such as $f(R)$ gravity, where the Ricci scalar is replaced by a more general function of the scalar curvature \cite{Capozziello:2002rd,Nojiri:2006ri,Sotiriou:2008rp,Olmo:2011uz,Capozziello:2011et,Clifton:2011jh,Harko:2018ayt,CANTATA:2021ktz}. This modification allows for additional degrees of freedom, which can potentially explain cosmic acceleration without the need for dark energy.
Another significant direction in alternative theories of gravity is exploring different geometrical formulations of spacetime. While GR is based on Riemannian geometry, other approaches consider different geometric structures, where the spacetime curvature is replaced by torsion \cite{Aldrovandi:2013wha,Cai:2015emx} and non-metricity \cite{BeltranJimenez:2017tkd,BeltranJimenez:2019esp}.
Furthermore, there are theories that introduce additional fields to the existing framework of GR. These include scalar-tensor theories \cite{Brans:1961sx,Brans:1962wsm,Brans:1962zz}, where a scalar field is coupled to the metric tensor, leading to varying gravitational `constants' over time or space, and Horndeski theories \cite{Horndeski:1974wa,Gleyzes:2014dya,Gleyzes:2014qga}, which represent the most general scalar-tensor theories with second-order field equations. These theories can accommodate a wide range of cosmological behaviours, including those that could explain the late-time cosmic acceleration. 
Overall, these alternative theories of gravity are motivated by the need to reconcile GR with observational data that suggest the presence of phenomena beyond the scope of the standard model of cosmology. By modifying the fundamental principles of GR or exploring entirely new geometrical structures, these theories aim to provide a more comprehensive understanding of the universe, addressing both the successes and challenges of GR while offering novel predictions that can be tested through future observations and experiments.

Indeed, research in extended theories of gravity reveal a key difference between the metric \cite{Sotiriou:2008rp,Capozziello:2011et,Clifton:2011jh} and Palatini (or metric-affine) \cite{Olmo:2011uz} formulations of gravity. In the metric approach, the metric tensor is treated as the sole dynamical variable in the action. Conversely, the Palatini approach considers the connection as an independent variable, separate from the metric tensor. Consequently, in the Palatini formulation, the field equations are derived by varying the action with respect to both the metric and the connection. 
While the metric approach often leads to higher-order derivative equations, the Palatini approach consistently results in second-order equations. However, this simplicity in the Palatini formulation comes with algebraic relations between the matter fields and the affine connection, coupling the connection with both matter and the metric.
In $f(R)$ theories, this distinction is particularly clear. In the metric formulation, $\phi \equiv \text{d}f/\text{d}R$ acts as a dynamical scalar field governed by a second-order equation, with its impact on large-scale phenomena depending on the field’s mass and interaction range. Light scalar fields are tightly constrained by laboratory and Solar System tests unless screening mechanisms are taken into account \cite{Brax:2012gr,Joyce:2014kja}.
Conversely, in the Palatini approach, the scalar field satisfies an algebraic equation, making $\phi$ a function of the trace of the stress-energy tensor, $\phi=\phi(T)$.

To address and potentially overcome specific limitations present in both Palatini and metric formulations of $f(R)$ gravity, the hybrid metric-Palatini gravity (HMPG) theories were introduced. This approach was first proposed in \cite{Harko:2011nh} and has since evolved significantly through extensive research across a variety of contexts. Key developments in HMPG have been made within cosmology, such as those presented in Refs. \cite{Capozziello:2012ny,Bohmer:2013ffw,Carloni:2015bua,Capozziello:2015lza,Santos:2016tds,Sa:2020qfd,Harko:2020ibn,Paliathanasis:2020fyp,Lobo:2023ddb}. These studies explore HMPG’s potential to model the universe's large-scale behaviour and its implications for the cosmological expansion history.
Beyond cosmology, HMPG has also been explored under observational constraints, and notable contributions in this domain include the analysis in Refs. \cite{Lima:2014aza,Lima:2015nma,Leanizbarrutia:2017xyd,Dyadina:2023exu}, where the compatibility of HMPG with observed phenomena has been analysed to establish feasible parameter spaces for the theory. Furthermore, HMPG has been applied to inflationary models, as demonstrated in studies like those presented in Refs. \cite{Sadeghi:2022tzd,He:2022xef,Shahid:2023dea,Asfour:2024mfr}, which investigate its ability to describe the universe’s accelerated expansion during the inflationary epoch.
The scope of HMPG extends into alternative fields of gravitational phenomena, such as in dark matter models \cite{Capozziello:2012qt,Capozziello:2013uya,Capozziello:2013yha} and compact objects \cite{Capozziello:2012hr,Danila:2016lqx,Rosa:2018jwp,Danila:2018xya,Avdeev:2020jqo,Chen:2020evr,KordZangeneh:2020ixt}. 
Within these frameworks, HMPG has been examined as a viable mechanism for describing the gravitational interactions of astrophysical bodies under extreme conditions. Additionally, stability considerations within HMPG, as discussed in Refs. \cite{Koivisto:2013kwa,Capozziello:2013gza,Bombacigno:2019did}, have been pivotal in evaluating the theoretical consistency and physical feasibility of solutions within this theory.

The theoretical framework of HMPG has been further expanded through generalized formulations and additional applications, as explored in \cite{Tamanini:2013ltp,Rosa:2017jld,Rosa:2019ejh,Rosa:2021ish,Borowiec:2020lfx,Rosa:2021lhc}, providing an enriched structural foundation for the theory. Explorations of Noether symmetries in HMPG \cite{Borowiec:2014wva} have provided insights into conserved quantities and symmetry properties of the model, adding depth to its mathematical formulation. 
Furthermore, the versatility of HMPG has been demonstrated across a range of contexts, including braneworld scenarios \cite{Fu:2016szo,Rosa:2020uli}, where it provides a framework for addressing extra-dimensional effects, and screening mechanisms \cite{VargasdosSantos:2017ggl}, which allow it to recover GR in local settings. Additionally, HMPG has shown promise in gravitational wave predictions \cite{Kausar:2018ipo}, offering insights that align with current observations, as well as in cosmic string models \cite{Harko:2020oxq,daSilva:2021dsq}, which connect to topological defects and early-universe phenomena. These applications significantly broaden HMPG’s relevance across a diverse spectrum of gravitational phenomena.
A notable strength of the HMPG framework lies in its scalar-tensor representation, which incorporates long-range scalar interactions while remaining consistent with Solar System observational tests \cite{Harko:2011nh}. This intrinsic alignment with established observational constraints enhances the viability of HMPG as a compelling alternative to traditional modified theories of gravity. Indeed, by offering a coherent description of gravitational interactions across multiple scales, HMPG emerges as a robust candidate for addressing a wide array of challenges in gravitational physics.

An intriguing application of HMPG theories lies in the study of spherically symmetric compact objects. Notably, the HMPG framework has enabled detailed explorations of vacuum static spherically symmetric solutions within its scalar-tensor representation, as carried out numerically in \cite{Danila:2018xya}. In this analysis, black hole formation is indicated by singularities emerging in the components of the metric tensor. This study also provides a comprehensive investigation of the thermodynamic properties of these black holes, examining key aspects such as the horizon temperature, specific heat, entropy, and the characteristic time scale for black hole evaporation.
Further examination of vacuum geometries in the scalar-tensor representation of HMPG was undertaken in \cite{Bronnikov:2019ugl,Bronnikov:2020vgg}. In \cite{Bronnikov:2019ugl}, solutions were analysed under the assumption of a zero scalar field potential, where the theory aligns closely with GR but with the addition of a phantom conformally coupled scalar field acting as the gravitational source. In this configuration, the solutions reveal a diversity of behaviours, such as, generic asymptotically flat solutions tend to manifest either as naked central singularities or as traversable wormholes. In fact, a particular two-parameter family of solutions results in globally regular black holes with double horizons. Additionally, a unique one-parameter family of solutions presents an intriguing structure featuring an infinite sequence of double horizons interspersed between static regions, where the spherical radius varies monotonically across these regions.

The study was extended further in \cite{Bronnikov:2020vgg}, where analytical solutions incorporating non-zero scalar potentials were considered. In these cases, black hole solutions featuring conventional horizons typically need negative potential values when canonical scalars are involved. For phantom scalars, however, a different class of solutions, termed ``black universe'' solutions, emerges. In these configurations, instead of leading to a central singularity, the geometry beyond the horizon transitions into an expanding universe. Stability analysis reveals that, while most solutions tend to be unstable under scalar monopole perturbations, certain specialized black hole configurations remain stable.
The study was further explored in the generalized HMPG theory \cite{Tamanini:2013ltp}, where specific solutions were first derived in the Einstein-frame and then interpreted in the Jordan-frame \cite{Bronnikov:2021tie}. Under the assumption of a zero potential, the general solutions frequently exhibit naked singularities or traversable wormholes, with special cases giving rise to black holes characterized by double horizons. When a non-zero potential is introduced, the analytical solutions typically lead to configurations with naked singularities. Even when the Einstein-frame metric $g^E_{\mu\nu}$ can be obtained analytically, solving the scalar field equations often requires numerical methods. Additionally, any horizon observed in $g^E_{\mu\nu}$ transforms into a singularity in the Jordan-frame as a result of the conformal factor, underscoring the complexity introduced by the scalar field dynamics in HMPG theories.
In this work, we build and extend on these earlier studies to explore novel configurations under certain conditions. More specifically, we focus on electrically charged scenarios with spherical symmetry where the scalar potential is identically zero, presenting these within both the Jordan and Einstein frames, and analysing them in the former. Such a setup allows for significant simplifications while still encompassing a rich variety of physical phenomena.

This paper is organized in the following manner:
In Sec. \ref{chap 1}, we present the framework of the HMPG theories, detailing the formulation of the action and the scalar-tensor representation used to derive the field equations. The transition between the Jordan and Einstein frames is outlined to simplify the analysis and solution derivation.
In Sec. \ref{SecIII}, we restrict the analysis to configurations with a zero scalar potential, focusing on solutions in the Jordan-frame. A transformation to the Einstein-frame is performed, and specific line elements are constructed to explore static, spherically symmetric solutions with electric charge.
In Secs. \ref{secIV} and \ref{secIVb}, we systematically examine different classes of solutions, 
%categorized based on the scalar field behaviour and asymptotic properties, 
and extensively discuss the influence of the parameters on the horizon structure and potential singularities, distinguishing between naked singularities, traversable wormholes, black holes, and unique ``black universe'' models.
Finally, in Sec. \ref{Conclusion}, we summarize our results, emphasizing the versatility of HMPG in yielding diverse gravitational configurations, and briefly explore potential directions for future research.

Throughout this paper, we consider the metric signature $(+--\,-)$.

%%%%%%%%%%%%%%%%%%%%%%%%%%%%%%%%%%%%%%%%%%%%%%%%%%%%%%%%%%%%
\section{The HMPG theory} \label{chap 1}
%%%%%%%%%%%%%%%%%%%%%%%%%%%%%%%%%%%%%%%%%%%%%%%%%%%%%%%%%%%%

The action of the HMPG theory is given by \cite{Harko:2011nh}:
\begin{equation}\label{2}
    S=\int d^4x\sqrt{-g}[R+f(\mathcal{R})]+S_m\,\,,
\end{equation}
where $g=$ det$(g_{\mu\nu})$, $g_{\mu\nu}$ is the Riemannian metric, $R$ is the metric Ricci scalar, associated with $g_{\mu\nu}$. $S_m$ is the matter action defined as $S_m=\int d^4x \sqrt{-g} \mathcal{L}_m$, where $\mathcal{L}_m$ is the matter Lagrangean.
In particular, the Palatini scalar curvature $\mathcal{R}$ is defined as $\mathcal{R}\equiv g^{\mu\nu}\mathcal{R_{\mu\nu}}$,where $\mathcal{R_{\mu\nu}}$ is the Palatini Ricci tensor defined through an independent connection $\hat{\Gamma}$
\begin{equation}
\mathcal{R}
\equiv  g^{\mu\nu}\mathcal{R}_{\mu\nu} \equiv g^{\mu\nu}\left(
\hat{\Gamma}^\alpha_{\mu\nu , \alpha}
       -\hat{\Gamma}^\alpha_{\mu\alpha , \nu} 
       +
\hat{\Gamma}^\alpha_{\alpha\lambda}\hat{\Gamma}^\lambda_{\mu\nu} -
\hat{\Gamma}^\alpha_{\mu\lambda}\hat{\Gamma}^\lambda_{\alpha\nu}\right).\label{r_def}
\end{equation}

It is extremely useful to represent the HMPG theory in the scalar-tensor representation, and rather than outline all the details here, we refer the reader to \cite{Danila:2018xya} for more details. 
More specifically, by introducing a new auxiliary field $E$,  the hybrid metric-Palatini action (\ref{2}) can be reformulated in the equivalent form of a scalar-tensor theory, given by
\begin{equation}\label{eq2sst}
S=\int\mathrm{d}^{4}x\sqrt{-g}[R+f(E)+f^{\prime }(E)(%
\mathcal{R}-E)]+S_m,
\end{equation}
where for $E=\mathcal{R}$, the action (\ref{eq2sst}) reduces to action (\ref{2}).
By introducing the following definitions:
\begin{equation}\label{3}
    \phi\equiv f^\prime(E)\,, \qquad  V(\phi)=Ef^\prime(E)-f(E)\,\,,
\end{equation}
it is possible to reformulate Eq. (\ref{2}) into a scalar-tensor theory with the following action \cite{Harko:2011nh}:
\begin{equation}\label{4}
    S=\int d^4x\sqrt{-g}[R+\phi\mathcal{R}-V(\phi)]+S_m\,\,.
\end{equation}

Thus, by varying the action (\ref{4}) with respect to the metric, $\phi$ and the independent connection, and using the respective field equations, it is possible to relate $\mathcal{R}_{\mu\nu}$ and $R_{\mu\nu}$ given by 
\begin{equation} \label{eq:conformal_Rmn}
\mathcal{R}_{\mu\nu}=R_{\mu\nu}+\frac{3}{2\phi^2}\partial_\mu \phi \partial_\nu \phi-\frac{1}{\phi}\left(\nabla_\mu
\nabla_\nu \phi+\frac{1}{2}g_{\mu\nu}\Box\phi\right) \,,
\end{equation}
and finally rewrite the action as (we refer the reader to \cite{Harko:2011nh} for more details):
\begin{equation}\label{5_}
    S=\int d^4x\sqrt{-g}\left[(1+\phi)R+\frac{3}{2\phi}(\partial\phi)^2 -V(\phi)\right]+S_m\,\,.
\end{equation}

It is interesting to note that this action is, in fact, a special case of the Bergmann-Wagoner-Nordtvedt scalar-tensor theories \cite{Bergmann:1968ve,Wagoner:1970vr,Nordtvedt:1970uv} , given by \cite{Bronnikov:2020vgg}:
\begin{equation}\label{6}
    S=\int d^4x\sqrt{-g}\left[f(\phi)R+g(\phi)(\partial\phi)^2 -V(\phi)\right]+S_m\,\,,
\end{equation}
where $f(\phi)$, $g(\phi)$ and $V(\phi)$ are arbitrary functions of $\phi$, which, in the case of the HMPG theory, by comparison with action (\ref{5_}), are given by
\begin{equation}\label{7}
     f(\phi)=1+\phi\,,\;\;\;\;\;\; g(\phi)=\frac{3}{2\phi}\,\,,
\end{equation}
and $V(\phi)$ is the scalar field potential, as defined before. Associating our theory with the Bergmann-Wagoner-Nordtvedt action is, indeed, of great importance, as in these theories it is possible to apply a transformation from the Jordan conformal frame, corresponding to Eq. (\ref{5_}), to the Einstein conformal frame, which greatly simplifies solving the field equations. 

In general, and in the particular case of our theory, given the expressions in Eq. (\ref{7}), this transformation is given, respectively, by \cite{Bronnikov:2020vgg,Wagoner:1970vr}:
\begin{eqnarray}
\bar{g}_{\mu\nu} &=& f(\phi) g_{\mu\nu}\,, 
	\nonumber \\
	\frac{d\phi}{d\bar{\phi}} &=& f(\phi)\left|f(\phi)g(\phi)-\frac{3}{2}\left(\frac{df}{d\phi}\right)^2\right|^{-1/2} \,, 
	\nonumber
\end{eqnarray}
which provides the following relations
\begin{eqnarray} \label{8}
	\bar{g}_{\mu\nu} &=& (1+\phi)g_{\mu\nu}\,\,, 
		\notag \\
	\phi &=& -\tanh^2\frac{\bar{\phi}}{\sqrt{6}}\,, \quad \text{if  } \quad -1<\phi<0\,\,, 
		\notag \\
	\phi &=&\tan^2\frac{\bar{\phi}}{\sqrt{6}}\,, \quad \text{if  } \quad  \phi>0\,\, \notag.
\end{eqnarray}

With this, we obtain the general form of the HMPG action in the Einstein-frame, in which bars are related to the transformed quantities \cite{Bronnikov:2020vgg,Bronnikov:2024uyb}:
\begin{eqnarray}\label{9_}
	S_E &=& \int d^4x\sqrt{-\bar{g}}\Bigg[\bar{R}-n\bar{g}^{\mu\nu}\bar{\phi}_{,\mu}\bar{\phi}_{,\nu}  
		\nonumber \\
	&& \hspace{2cm}  -\frac{V(\phi)}{(1+\phi)^2}+ \frac{\bar{\mathcal{L}}_m}{(1+\phi)^2} \Bigg]\,\,,
\end{eqnarray}
where $n$ is related to the sign of the scalar field kinetic term in action (\ref{5_}), which depends on the sign of the field itself. Specifically, $n=-\text{sign }(\frac{3}{2\phi})=-\text{sign }\phi$. Furthermore, if $n=+1$, $\phi$ is a canonical scalar field, whereas if $n=-1$, $\phi$ is a phantom scalar field. Apart from this, by transforming the matter action, $S_m$, that last term is obtained, in which $\bar{\mathcal{L}}_m$ represents the transformed matter Lagrangean, that depends on its yet unknown expression. Note that $V(\phi)$ remains without any bars, as its expression is still unknown and is irrelevant when obtaining the field equations from the action, as is its transformation, unlike the matter Lagrangean.
%, as we are going to see. 

%%%%%%%%%%%%%%%%%%%%%%%%%%%%%%%%%%%%%%%%%%%%%%%%%%%%%%%%%%%%
\section{Action and general solution for $V(\phi)\equiv 0$}\label{SecIII}
%%%%%%%%%%%%%%%%%%%%%%%%%%%%%%%%%%%%%%%%%%%%%%%%%%%%%%%%%%%%

Now, for our analysis, we are interested in studying electrically charged solutions to the HMPG theory, represented, in general, by action (\ref{5_}). Note that if we were to consider a magnetic charge, or both simultaneously, the results would be analogous, as explained in \cite{Bronnikov:2024uyb}. Furthermore, we focus on the case in which $V(\phi)\equiv 0$. For this study, in the Jordan conformal frame, we consider the following action:
\begin{equation}\label{10}
    S=\int d^4x\sqrt{-g}\left[(1+\phi)R+\frac{3}{2\phi}(\partial\phi)^2 +F_{\mu\nu}F^{\mu\nu}\right].
\end{equation}
Here, $\mathcal{L}_m= F_{\mu\nu}F^{\mu\nu}$, which is a pure Maxwell term, where $F_{\mu\nu}\equiv \nabla_\mu A_\nu -\nabla_\nu A_\mu$ is the electromagnetic field tensor or Maxwell tensor, and $A_\mu$ is the electromagnetic 4-potential.
Applying the transformation to the Einstein conformal frame, using Eq. (\ref{8}), we obtain the following action \cite{Bronnikov:2024uyb,Bronnikov:1999wh}:
\begin{equation}\label{11}
    S_E=\int d^4x\sqrt{-\bar{g}}\left[\bar{R}-n\bar{g}^{\mu\nu}\bar{\phi}_{,\mu}\bar{\phi}_{,\nu} +F_{\mu\nu}F^{\mu\nu}\right]\,\,.
\end{equation}
As we can see, this transformation yields no changes in the form of the matter action, since $\bar{\mathcal{L}}_m=(1+\phi)^2F_{\mu\nu}F^{\mu\nu}$. 

Now, in order to find static, spherically symmetric solutions to the field equations derived from this action, we define the line element in the following manner: 
\begin{equation}\label{12}
    ds^2=e^{2\gamma(u)}dt^2-e^{2\alpha(u)}du^2-e^{2\beta(u)}d\Omega^2\,,
\end{equation}
where  $\gamma$, $\alpha$ and $\beta$ are arbitrary functions of $u$, an arbitrarily chosen radial coordinate, and $d\Omega^2=d\theta^2+\sin^2\theta \,d\varphi^2$. Apart from this, using the harmonic coordinate condition, which will allow the solutions for the canonical and phantom scalar fields to be written in a unified form \cite{Bronnikov:2020vgg}, we also define the function $\alpha(u)$ as:
\begin{equation}\label{13}
    \alpha(u)=2\beta(u)+\gamma(u)\,.
\end{equation}

Taking into account these conditions, we obtain a general solution to the field equations that emerge from the action in Eq. (\ref{11}).
A specific solution, given in the Einstein-frame and for a source that is solely electrically charged, which is the focus of our study, was derived in \cite{Bronnikov:1973fh, Bronnikov:1999wh, Bronnikov:2024uyb} and is expressed as:
\begin{eqnarray}
	ds_E^2=\frac{dt^2}{q^2 s^2(h,u+u_1)}-\frac{q^2s^2(h,u+u_1)}{s^2(k,u)}\times
		\nonumber \\
	\times \left[\frac{du^2}{s^2(k,u)}+d\Omega^2\right]\,,\label{14}
\end{eqnarray}
\begin{equation}
	F_{\mu\nu}=(\delta_{\mu0}\delta_{\nu1}-\delta_{\nu0}\delta_{\mu1})\frac{1}{q^2 s^2(h,u+u_1)}\,,\label{15}
\end{equation}
\begin{equation}
	\bar{\phi}(u)=\bar{C}u+\bar{\phi}_0\,,\label{16}
\end{equation}
where the integration constants $q$ and $\bar{C}$ are the electric and scalar charges, respectively, $u_1$ will be defined below, and the functions $s^2(k,u)$ and $s^2(h,u+u_1)$ are defined as:
\begin{align}%\label{17}
s^2(k,u)=
  \begin{cases}
      k^{-2}\sinh^2ku\,\,,  & k>0\,,\\
      u^2\,\,,&k=0\,\,,\\
      k^{-2}\sin^2 ku\,\,,&k<0\,,
   \end{cases}
\end{align}
and
\begin{align}\label{17}
   s^2(h,u+u_1)=
   \begin{cases}
       h^{-2}\sinh^2[h(u+u_1)]\,\,,  & h>0\,,\\
      (u+u_1)^2\,\,,&h=0\,\,,\\
      h^{-2}\sin^2 [h(u+u_1)]\,\,,&h<0\,,
   \end{cases}
\end{align}
respectively, where the integration constants $\bar{C}$, $k$ and $h$ follow the relation:
\begin{equation}\label{18}
    2k^2\text{sign }k=n\bar{C}^2+2h^2\text{sign }h\,.
\end{equation}
These constants may be positive, negative or zero, except for $\bar{C}$, which we must consider to be different than 0, since in that case the scalar field would be constant, as can be seen from Eq. (\ref{16}), which bears no interest for our study.
Apart from these, $\bar{\phi}_0\coloneqq\bar{\phi}(0)$ is also an integration constant. Note that the solution for the scalar field, in Eq. (\ref{16}), was obtained in \cite{Bronnikov:1973fh, Bronnikov:2024uyb}, without loss of generality, considering that constant to be zero. However, in our case, we consider $\bar{\phi}_0\not\equiv0$, since this leads to different and more interesting results, as explained further below.

Concerning the coordinate $u$, we will, as in \cite{Bronnikov:1999wh, Bronnikov:2024uyb}, define it in the interval $0<u<u_{\text{max}}\leq \infty$, unless stated otherwise, where $u=0$ is spatial infinity, i.e., it is a regular point and $r(u)\equiv \sqrt{-g_{22}}$, which is the spherical radius, approaches infinity. $u_{\text{max}}$, as the name suggests, corresponds to the maximum value of $u$, allowed by the metric behaviour, and may correspond to a singular point, where spacetime ends, a second spatial infinity, or a regular centre, depending on the metric functions and on its regularity. Defining $u$ to be positive is just a choice, as this coordinate could be defined in the range $u<0$ instead. However, when $u=0$ is spatial infinity, that would lead to the same results, since we are dealing with hyperbolic and trigonometric functions, thus, by choosing adequate values for the constants, we obtain the same results for both ranges. In certain classes this choice is also discussed. When $u=0$ is not spatial infinity, $u$ is no longer defined in the interval $0<u<u_{\text{max}}\leq \infty$, and it is also no longer restricted to being either positive or negative. Regarding the integration constant $u_1$, its definition (see \cite{Bronnikov:1999wh, Bronnikov:2024uyb}) is imposed, without loss of generality, by $\bar{g}_{00}(0)=1$, which occurs if the Einstein-frame metric is asymptotically flat, in particular Minkowskian, at $u=0$, for which we obtain: 
\begin{equation}\label{u1}
    s^2(h,\,u_1)=1/q^2\,\,.
\end{equation}
From this definition, we see that this constant can never be null, but may have both signs.
Furthermore, taking into account the possible combinations of integration constants, and also the results obtained later in the analysis, we obtain the following classification for the solutions of the family (\ref{14})-(\ref{16}) \cite{Bronnikov:1999wh, Bronnikov:2024uyb}:
\begin{equation}\label{classes +}
    \begin{minipage}{0.4\textwidth}
        \begin{align}
    [1+] \quad&n=+1, \quad k>h>0;\nonumber\\
    [2+] \quad&n=+1, \quad k>h=0;\nonumber\\
    [3+] \quad&n=+1, \quad h<0\nonumber;
        \end{align}
    \end{minipage}
\end{equation}
and
\begin{equation}\label{classes -}
    \begin{minipage}{0.4\textwidth}
        \begin{align}
    [1-] \quad&n=-1, \quad h>k\geq 0;\nonumber\\
    [2-] \quad&n=-1, \quad h\geq0,\,k<0;\nonumber\\
    [3-] \quad&n=-1, \quad 0>h>k.\nonumber
        \end{align}
    \end{minipage}
\end{equation}

Before we start analysing these classes, we first need to obtain the metric back in the Jordan-frame, since those are the solutions we are interested in. Inverting the metric transformation equation and using the transformation of $\phi$, for both types of scalar field, from Eq. (\ref{8}), we obtain:
\begin{align}
      g_{\mu\nu}=\cosh^2(\bar{\phi}/\sqrt{6})\bar{g}_{\mu\nu}\,\,,\,\, &\text{if  }\,\, -1<\phi<0\,\,,\,n=+1\,\,,\label{g munu can}\\
      g_{\mu\nu}=\cos^2(\bar{\phi}/\sqrt{6})\bar{g}_{\mu\nu}\,\,,\,\, &\text{if  }\,\, \phi>0\,\,,\,n=-1\,\,.\label{g munu pha}
\end{align}
Defining $\psi(u)\coloneqq\bar{\phi}(u)/\sqrt{6}=Cu+\psi_0$, where $C=\bar{C}/\sqrt{6}$ and $\psi_0=\bar{\phi}_0/\sqrt{6}$, for notational simplicity, we obtain the following relations between line elements:
\begin{align}
        ds_J^2=\cosh^2\psi\, ds_E^2\,\,,\,\, &\text{if  }\,\, -1<\phi<0\,\,,\,n=+1\,\,, \label{sJ can}\\
        ds_J^2=\cos^2\psi\, ds_E^2\,\,,\,\, &\text{if  }\,\, \phi>0\,\,,\,n=-1\,\,,\label{sJ pha}
\end{align}
where the subscripts $J$ and $E$ are relative to the Jordan and Einstein frames, respectively. Using $C$, we have the following relation between constants, which will be of extreme importance, instead of Eq. (\ref{18}):
\begin{equation}\label{25}
    k^2\text{sign }k=n\,3C^2+h^2\text{sign }h\,.
\end{equation}

From Eqs. (\ref{sJ can}) and (\ref{sJ pha}), we see why the choice $\bar{\phi}_0\equiv0$ should not be considered in Eq. (\ref{16}). In both canonical and phantom cases, the choice of $\bar{\phi}_0\not =0$, hence $\psi_0\neq 0$, affects the behaviour, in particular as $u\to 0$, of the conformal factors $\cosh^2\psi$ and $\cos^2\psi$, affecting the asymptotic behaviour of the metrics. Apart from this, in the case of a phantom field, the conformal factor $\cos^2\psi$ may go to zero at different values of $u$ according to the value of $\psi_0$, which may lead to different possibilities for those spacetimes. Conversely, in the case of a canonical field, adding $\psi_0\not=0$ in the conformal factor has no effects in that sense, however, it may affect its relations with other functions in the metric. These behaviours will be explored in greater detail when analysing the solutions.

%%%%%%%%%%%%%%%%%%%%%%%%%%%%%%%%%%%%%%%%%%%%%%%%%%%%%%%%%%%%
\section{Systematic strategy for analysing solutions} \label{secIV}
%%%%%%%%%%%%%%%%%%%%%%%%%%%%%%%%%%%%%%%%%%%%%%%%%%%%%%%%%%%%

We now examine each of the aforementioned classes of solutions, following a systematic approach for each. To begin, we analyse the asymptotic behaviour of the metric as $u\to 0^+$, which generally corresponds to spatial infinity, as previously discussed.

In fact, if there are other points with these features, we may consider the spatial infinity to be located at any one of them. However, in our analysis, whenever $u=0$ presents those features, which happens in the vast majority of the cases, we consider spatial infinity to be located at that point, even if there are other possible points. Even though alternative points may exist, selecting any of them would yield the same results. 
On the other hand, if $u=0$ does not have those features, we must search for a point that does and define it as spatial infinity. In this particular case, both ranges $u>0$ and $u<0$ may be considered simultaneously, as the coordinate is no longer restricted to one of them, as mentioned before. Thus, for simplicity, in the following explanations we are always considering spatial infinity to be located at $u=0$.

In order to simplify the analysis of spatial infinity, we transform the coordinate $u$, using the ``quasiglobal" coordinate condition:
\begin{equation}
    \alpha(u)+\gamma(u)=0\,,
\end{equation}
where these functions are present in Eq. (\ref{12}). In general, by doing this, we transform $u$ into a ``quasiglobal" coordinate $x$, obtained by the integral $x=\int e^{\alpha+\gamma}du$, which leads to $g_{00}=-g^{11}$. 
The ideal approach would be to apply this condition directly to the metric in the Jordan-frame. However, integrating with the conformal factor included in the metric functions proves to be impossible, at least when considering only real solutions. 
Thus, for this transformation only (the Jordan-frame metric is still the one of interest to analyse), we are considering the Einstein-frame's metric functions. This way, unlike what usually happens when we define a ``quasiglobal" coordinate, in this case, in the Jordan-frame, we do not get $g_{00}=-g^{11}$ because the conformal factor is not included in the definition of the coordinate transformation. 
Thus, we are going to transform $u$ into what we call the Einstein-frame's ``quasiglobal" coordinate $x$, which is obtained by using the given condition and solving the corresponding integral for the Einstein-frame's metric. %, where we have $\bar{g}_{00}=-\bar{g}^{11}$.

Taking into account the line element (\ref{14}) and the function inside the integral $x=\int e^{\alpha+\gamma}du$, we find that the transformation to the coordinate $x$ only depends on the sign of $k$, which may vary for different classes of solutions. This way, we get three different transformations, which read \cite{Bronnikov:2024uyb}: 
\begin{equation}
e^{-2ku}=1-\frac{2k}{x} \; \Rightarrow \; u=-\frac{\log (1-\frac{2k}{x})}{2k}, \;\;\text{if  }k>0\;,\label{transformation +}
\end{equation}
\begin{equation}
u=\frac{1}{x}\,,\; \text{if} \;\; k=0\;,\label{transformation 0}
\end{equation}
\begin{eqnarray}
x &=& |k|\cot{(|k|u)} \; \Rightarrow \; u=\frac{\arccot(\frac{x}{|k|})+c_1\pi}{|k|}\,,
		\nonumber  \\    
    && \qquad \qquad c_1\in \mathbb{Z}\;,\;\text{if  }k<0\;\;\label{transformation -}.
\end{eqnarray}
In this last expression, we have to consider the term $c_1\pi$, because the $\cot$ function is periodic and $\arccot$ is only defined over a single interval of length $\pi$. As discussed before, when $u=0$ is spatial infinity, we define it to be restricted to the range $u>0$, in which case the variable $c_1$ must be a positive integer ($c_1\in \mathbb{Z}^+_0$). Nevertheless, when that point is not spatial infinity, $c_1$ may have any integer value, being different values of it associated with different intervals, with a length of $\pi$, of the coordinate $u$. In particular, for the case $c_1=0$, the $\arccot$ function must cover the range $0<x<\pi$ of the $\cot$ function, instead of $-\frac{\pi}{2}<x<\frac{\pi}{2}$, as defined in some softwares.

In all transformations, we find that the point $u=0$, in particular, $u\to 0^+$, corresponds to $x\to\infty$, with $c_1=0$ in the case $k<0$, which means that this limit of the coordinate $x$ corresponds to a spatial infinity. Accordingly, we find that this transformation reverses the direction in which the spherical radius decreases. Thus, in the same way that we have defined $u_\text{max}$ for the coordinate $u$, we can define $x_\text{min}$ for the coordinate $x$, so that this coordinate ranges from that point to infinity. Using the previous transformations, we are able to restrict it for each case. Taking into account that $u_\text{max}\leq \infty$, using Eqs. (\ref{transformation +})--(\ref{transformation -}), we have, respectively: $x_\text{min}\geq 2k$, $x_\text{min}\geq 0$ and $x_\text{min} \geq -\infty$ (in this case the value of $c_1$ must also be mentioned).

The choice between this new coordinate and $u$ is arbitrary, depending on which is more convenient for the analysis being performed. In particular, using the coordinate $x$ is very useful when studying the asymptotic behaviour, as mentioned before, because if we analyse the metric at a spatial infinity, in this case located at $x\to\infty$, and we get that both $g_{00}$ and $g_{11}$ are constant and finite non-zero, then we have an asymptotically flat spacetime. A particular case, that is also flat, happens when $g_{00}=-g_{11}=1$, and it corresponds to an asymptotically Minkowskian spacetime. On the other hand, if $g_{00}=0$ at spatial infinity, then there are several possibilities, for which we need to analyse curvature invariants, in particular the Kretschmann scalar. In general, at infinity, if it diverges, there is a singularity there; if it is non-zero finite, then we have an asymptotically non-flat spacetime; if it is null, as is the case when $g_{00}$ and $g_{11}$ are finite non-zero, we have an asymptotically flat spacetime (a discussion about this scalar will be addressed shortly). 

%We find that in these spacetimes, the curvature invariants, in particular the Kretschmann scalar, which is going to be explored in greater detail below, are null at infinity. A particular case, that is also flat, happens when $g_{00}=-g_{11}=1$, and it corresponds to an asymptotically Minkowskian spacetime. On the other hand, if $g_{00}=0$ at spatial infinity, then there are several possibilities, for which we need to analyse the Kretschmann scalar as well, for example. If that scalar diverges, there is a singularity at infinity, if it is finite, then we have an asymptotically non-flat spacetime and if it is null, we have an asymptotically flat spacetime (a discussion about this scalar will be addressed shortly). 

For asymptotically flat spacetimes, we are able to determine the mass of the configuration. In order to do this, we compare the asymptotic behaviour of the metric, at spatial infinity, with the Schwarzschild solution \cite{dInverno:1992gxs}, through a series expansion of $g_{00}$. By doing this, we are able to determine what is called the Schwarzschild mass. However, to obtain it, we are, in fact, going to use a general expression, used in \cite{Bronnikov:2020vgg}, given by
\begin{equation}\label{massa}
  m=\lim_{x\to\infty} {\rm e}^\beta \gamma '/\beta '\,\,,
\end{equation}
where the limit as $x\to\infty$ is going to be used in our analysis, but can be adapted to other coordinates and other possibilities of spatial infinity. For example, if we consider the coordinate $u$, this expression can also be used in the limit $u\to 0^+$ (or $u\to 0^-$, according to the range of $u$ being analysed). Note that, in the case of this coordinate, the direction of the limit is crucial, as we verify in our analysis that the mass expressions above and below a given point of spatial infinity are symmetric, as they correspond to the limits $x\to\infty$ and $x\to-\infty$, respectively. In our analysis, we impose a positive mass, $m>0$, whenever it is determined relatively to the first spatial infinity.

After performing this analysis, we test the regularity of the metric. To do this, we analyse the Kretschmann scalar, $K$, which can be written in the form:
\begin{equation}\label{K}
  K=4K_1^2+8K_2^2+8K_3^2+4K_4^2\,\,,
\end{equation}
where each of its components is given by
\begin{eqnarray} \label{Ki}
  K_1=g^{11}R^0_{101}\,,\quad K_2=g^{22}R^0_{202}\,, 
	\nonumber \\  
   K_3=g^{22}R^1_{212}\,,\quad K_4=g^{33}R^2_{323}\,\,.
   %\nonumber
\end{eqnarray}

This way, in order to analyse $K$, we consider each of these separately. A singularity corresponds to a point of $x$, or $u$, where $K$ diverges, which happens if any of the four components diverges. Accordingly, if at a given point of $x$, or $u$, none of the four components diverges, then spacetime is regular at that point. Let us represent a singular point by $x_s$, or $u_s$. We have that if $K$ diverges at several points, then the singularity is located at the first one from spatial infinity, since spacetime ends there, and at the considered point we define $x_{\text{min}}=x_s$, or $u_\text{max}=u_s$. In fact, this happens unless $r\to 0$, without being a minimum (or maximum), or $r\to\infty$ at a regular point before that. In these cases, there is a regular centre and a second spatial infinity, respectively, and no singularity, since spacetime does not even reach the point $x=x_s$, or $u=u_s$, being always regular. In this case, we define there $x_{\text{min}}$, or $u_\text{max}$. In general, their locations vary with the metric, as we will see in the analysis of each class of solutions, being possible to coincide with the end of the validity interval opposite to the first spatial infinity, or any other point before that. Thus, their actual locations are only determined after a thorough analysis of the metric.

In the case there is a singularity, we may even verify if it is attractive or repulsive. This classification is based on the fact that in an asymptotically flat spacetime, at $r(x)\to\infty$, we have that $g_{00}=1+U(x)$, where $U(x)$ is the Newtonian gravitational potential (weak-field limit). For $r(x)\geq 0$, we may instead consider a generalized Newtonian potential, according to which $g_{00}$ serves as an analogue of the gravitational potential. Accordingly, its derivative with respect to $x$ is symmetric to the gravitational force, and so, we may classify the singularity in the following way: if $g_{00}'>0$, as $x$ approaches $x_s$ from above, or if $g_{00}'<0$, as $x$ approaches $x_s$ from below, it is attractive, because the gravitational force in its vicinity is attractive. Conversely, if $g_{00}'<0$, as $x$ approaches $x_s$ from above, or if $g_{00}'>0$, as $x$ approaches $x_s$ from below, it is repulsive, due to the forces being repulsive. To better visualize these relations, think of a minimum as attractive and a maximum as repulsive. The same classification can, and will, be used for the coordinate $u$ (accordingly, in that case the potential and the spherical radius are functions of $u$). Note that in our case, in which $x\to\infty$ corresponds to the spatial infinity, $x$ always approaches $x_s$ from above. 
However, if spatial infinity is located at $x\to -\infty$, for example, or in the case of the coordinate $u$ ($u>0$ and infinity at $u=0$), $x$ and $u$ approach $x_s$ and $u_s$, respectively, from below.
Another way of analysing the singularity, based on the previous explanation, instead of analysing the $g_{00}$ derivative function directly, is: if $g_{00}\to0^+$($0^-$) at $x_s$ or $u_s$, it is attractive (repulsive); if $g_{00} \to \infty$ ($-\infty$) at that point, it is repulsive (attractive). 
Although this only applies to those specific values, it works better due to the absence of numerical errors.

Apart from this classification, there are also time-like and space-like singularities (apart from light-like ones, which will be discussed shortly), similar to the ones present in the Reissner-Nordstr\"{o}m and Schwarzschild solutions \cite{dInverno:1992gxs}, respectively. The former occurs if at its vicinity the metric signature is $(+--\,-)$, whereas the latter occurs if at its vicinity the time and radial coordinates have their roles interchanged, by which the metric signature is $(-+-\,-)$.

After this, we may then analyse the existence of horizons in the metric. These occur at regular points of $x$, or $u$, always preceding a singularity, at which $g_{00}=0$. If this equation has a double root, making it degenerate, which is always the case in our analysis whenever a horizon exists, it is referred to as a double horizon. We will represent them by $x_H$, or $u_H$. Note that the metric signature remains unchanged beyond this type of horizon. Relating the existence (or absence) of horizons with the existence of a singularity, we have that if there is at least one horizon, then it is a black hole solution, whereas if there is none, then it is a naked singularity solution. For instance, consider the Reissner-Nordstr\"{o}m solution as an example of one that may present 0, 1 (extremal) or even 2 horizons, according to the relation between the electric charge, $q$, and the mass, $m$ \cite{dInverno:1992gxs}. Besides all of this, if $g_{00}=0$ at a singular point, then it is a light-like, or null, singularity, also called a singular horizon \cite{Clement:2009ai}.

Now, regarding the spherical radius function, $r(x)\equiv\sqrt{-g_{22}}$, or $r(u)$, if it approaches infinity at a given point of $x$, or $u$, it corresponds to a surface located at infinity, with infinite area; if it is non-zero finite we have a sphere with non-zero finite area; if $r\to 0$ at a singular point or a regular one which is not a minimum (or a maximum) of $r$, as aforementioned, then it is a centre. 

Finally, we are also interested in identifying throats within the metric. To do this, we need to analyse the radius function, particularly its derivative, in greater detail.
A throat occurs at a regular minimum point of this function (a zero of its derivative), once again, always before a singular point, and under the assumption $g_{00}\geq0$. In the particular case of $g_{00}=0$ it is referred to as an extremal null throat \cite{Simpson:2018tsi}. However, if $g_{00}<0$ at a regular minimum of the function, a bounce occurs at that point instead \cite{Simpson:2018tsi}. 
Apart from these structures, when analysing the radius function, if at a given point it has a regular local maximum (also a zero of the derivative) and $g_{00}>0$, then it is denoted an anti-throat \cite{Rodrigues:2022mdm}. For the other signs of $g_{00}$ we use the same type of terminology (``anti-'') applied to those cases. In our analysis, the existence of these structures always require the existence of their ``normal'' counterpart.

The existence of a throat (with no anti-throat) is, in general, associated with a wormhole solution. In this type of geometry, the spacetime is entirely regular and presents a throat and a second spatial infinity past that structure. These geometries may be classified in two different ways: if $g_{00}\not=0$ at any point, then it corresponds to a two-way traversable wormhole; if $g_{00}=0$ at the throat, then the geometry corresponds to a one-way traversable wormhole \cite{Simpson:2018tsi}. Nonetheless, there are cases where a throat is not associated with a wormhole geometry, such as when an anti-throat is present, as this precludes the existence of a second spatial infinity.

Another interesting example, in which the geometry is not a wormhole, is if $g_{00}=0$ at a point (or points), corresponding to a horizon (or horizons), before a throat, an extremal null throat or a bounce, in which case this geometry is denoted as a ``black bounce". More specifically, it corresponds to a type of regular black hole in which the singularity has been replaced by one of these structures (a particular case of a ``black universe") \cite{Simpson:2018tsi}.

A throat (or a bounce), in the context of a wormhole or black bounce, is a region of spacetime that connects two distinct, separate regions. These regions could both belong to our own universe, or different universes. Note that the metric can be asymmetric or symmetric relative to the location of the throat (or bounce).
To verify this, we have to see if all metric functions are invariant to the transformation $x\to -x$, in the case that the structure is located at $x=0$. If it is located at any other point, we may apply a translation coordinate transformation, so that it is located at zero, to verify that. If all functions exhibit the considered symmetry, then the metric, and thus the spacetime, do as well. We can begin by analysing the spherical radius function (since we already use it to identify these structures) to determine if it is asymmetric with respect to the throat (or bounce). If this is the case, we might observe, for instance, horizons on one side of it but none on the other, or different evolutions of $r$ on each side. Regardless, in addition to the well-known scenarios discussed thus far, specific cases like those mentioned above may arise in the subsequent analyses, where they will be carefully examined and discussed.

Considering all of this analysis, it is also possible to construct the Penrose diagrams for each of the spacetime geometries under study. The key structures that require more attention in constructing these diagrams are spatial infinities, horizons, throats, bounces, and singularities.

%%%%%%%%%%%%%%%%%%%%%%%%%%%%%%%%%%%%%%%%%%%%%%%%%%%%%%%%%%%%
\section{Solutions analysis} \label{secIVb}
%%%%%%%%%%%%%%%%%%%%%%%%%%%%%%%%%%%%%%%%%%%%%%%%%%%%%%%%%%%%

Now, taking into account the previous section, we finally have the necessary conditions to analyse each of the classes of solutions shown in (\ref{classes +}) and (\ref{classes -}).
Let us start by analysing the classes of (\ref{classes +}), that correspond to a canonical scalar field. In these cases, the conformal factor, showed before in Eq. (\ref{sJ can}), is a $\cosh$, which diverges as its argument goes to infinity. Accordingly, at that point of $x$ or $u$, a singularity is expected. However, this is only the case unless spacetime ends (another point of divergence), $r$ goes to infinity, or a regular centre exists, at some other point before that, or even if the remaining metric ``cancels" that divergence and it corresponds to a regular point instead. This way, a detailed analysis is necessary in each and every case. 

%%%%%%%%%%%%%%%%%%%%%%%%%%%%%%%%%%%%%%%%%%%%%%%%%%%%%%%%%%%%
\subsection{Class [1+]} \label{chap 1+}
%%%%%%%%%%%%%%%%%%%%%%%%%%%%%%%%%%%%%%%%%%%%%%%%%%%%%%%%%%%%

That being said, let us start by analysing class $[1+]$. In this case, according to Eq. (\ref{25}), we have the relation $k=\sqrt{3C^2+h^2}$. From Eq. (\ref{u1}), in this class we have $u_1=\pm \arcsinh (\frac{h}{q})/h$.
Given the signs of $n$, $k$ and $h$ and according to Eqs. (\ref{14}), (\ref{17}) and (\ref{sJ can}), in this case, the line element is given by
\begin{eqnarray}
    ds_J^2 &=& \cosh^2(Cu+\psi_0)\Bigg\{\frac{h^2dt^2}{q^2 \sinh^2[h(u+u_1)]}
		\nonumber \\    
    && \hspace{-0.75cm} -\frac{k^2 q^2 \sinh^2[h(u+u_1)]} {h^2 \sinh^2(ku)}\left[\frac{k^2 du^2}{\sinh^2(ku)}+d\Omega^2\right]\Bigg\} .
\end{eqnarray}

Now, applying the transformation present in Eq. (\ref{transformation +}), we obtain the line element with the coordinate $x$:
\begin{eqnarray}\label{1+}
    ds_J^2 &=& \cosh^2\left(C \bar{u} +\psi_0\right)\Bigg\{\frac{h^2dt^2}{q^2 \sinh^2\left[h\left(\bar{u}+u_1\right)\right]}
		\nonumber \\    
   && -\frac{q^2 \sinh^2[h(\bar{u}+u_1)]} {h^2}\left[dx^2+x^2\bar{x}\,d\Omega^2\right]\Big\}\,,
\end{eqnarray}
where we have defined $\bar{u}=\frac{\log (\bar{x})}{-2k}$, being equivalent to $u$, with $\bar{x} = 1-\frac{2k}{x}$, for notational simplicity.

Taking into account the discussion carried out after Eq. (\ref{transformation +}), we have that $x$ ranges from $x_{\text{min}}\geq 2k$ to $\infty$.
Analysing the latter line element, as $x\to\infty$, which corresponds to spatial infinity, and using the expression aforementioned for $u_1$, we find $g_{00}=-g_{11}=\cosh^2 \psi_0$. Thus, this is an asymptotically flat spacetime, in particular, Minkowskian for $\psi_0=0$, according to our previous discussion. For the Schwarzschild mass of this configuration, we have obtained, using Eq. (\ref{massa}), $m=h \coth(h u_1)\cosh\psi_0- C\sinh\psi_0$. By imposing $m>0$, we obtain the constraint $C<h\coth(h\,u_1) \coth\psi_0$ (assuming that this constant may be positive or negative, but never zero, as explained before), if $\psi_0>0$, whereas if $\psi_0<0$, $C$ has to be strictly greater than this expression. However, as one can see, this expression is only valid for $\psi_0\not=0$. In this case, $m>0$ allows both positive and negative values of $u_1$. If $u_1<0$, from that constraint, we find that the signs of $C$ and $\psi_0$ have to be opposite. On the other hand, in the particular case of $\psi_0=0$, we have $m=h \coth(h u_1)$, which, by imposing $m>0$, imposes $u_1>0$. Nevertheless, as we are not imposing $\psi_0$ to be null, as already explained, we analyse both signs of $u_1$, which may, in fact, lead to different results. In the following analysis we always assume combinations of constants that ensure $m>0$. 

Now, we analyse the Kretschmann scalar, $K$, in order to test the regularity of the metric. To compute this scalar, we use the latter line element (more convenient in this case) and begin with the analysis of $K_1$, which is given by the expression:
\begin{eqnarray}\label{K 1+}
   K_1 &=& \frac{1}{q^2 x^2 (x-2 k)^2}h^2 \text{sech}^2\left(C\bar{u}+\psi_0\right) \times
		\nonumber \\   
  && \;\; \times  \text{csch}^2\left(h
   \left[\bar{u}+u_1\right]\right)  \left\{-2 h \coth \left(h
   \left[\bar{u}+u_1\right]\right)
    \right. \notag\\
   &&\left. 
   \;\; \times \left[C \tanh \left(C\bar{u}+\psi_0\right)-k+x \right]
   \right. \notag\\
   && \left. 
   +C\, \text{sech}^2\left(C\bar{u}+\psi_0\right)  \times
    \right. \notag\\
   && \left. 
   \;\; \times\left[(-k+x) \sinh \left(2C\bar{u}+2\psi_0\right)+C \right] 
   \right. \notag\\
   && \left. 
   +2 h^2 \coth^2\left(h \left[\bar{u}+u_1\right]\right)
   \right. \notag\\
   && \left. 
   +h^2 \text{csch}^2\left(h \left[\bar{u}+u_1\right]\right)\right\}\,.
\end{eqnarray}

First of all, as $x\to \infty$, or $\bar{u}\to 0^+$, by analysing Eq. (\ref{K 1+}), we find that this term of the Kretschmann scalar never diverges, being actually null. By analysing the remaining terms of $K$, we find the same results, supporting our previous description of the spatial infinity. Apart from that, in the above equation, there is the term $\frac{1}{(x-2 k)^2}$, which diverges as $x\to2k$. Additionally, it contains hyperbolic trigonometric functions, with $\sinh$ diverging as $\bar{u} \to \infty$, thus, as $x\to 2k$, and $\tanh$ being finite in all its domain. It also contains the reciprocals, which approach $0$ at that limit, except for $\coth$. Accordingly, if we exclude the first quotient, we actually find that all terms in $K_1$ involving these functions are finite at that point. This way, the behaviour of this term of the Kretschmann scalar, as $x$ approaches that point, will depend on which terms dominate: if it is the term $\frac{1}{(x-2 k)^2}$, then $K_1$ diverges, but if the terms involving the hyperbolic functions (and reciprocals) dominate, then $K_1$ is non-zero finite. 

We split the analysis according to the sign of $u_1$, because if it is positive, 
then $\sinh\left(h\left[\bar{u}+u_1\right]\right)$ has no zeros in the domain $\bar{u}>0$ (which is the domain we are considering), and so, in that case, apart from $x=2k$, there are no other possibilities of $K_1$ going to infinity. However, if it is negative then there is a zero at $\bar{u}=-u_1$, at which the reciprocal functions $\coth$ and $ \text{csch}$ diverge, which cause $K_1$ to also diverge. Note that, as mentioned before, due to the freedom of choice of the constants, if we were to consider $u<0$ --- here the mass is symmetric to $m$ --- and $u_1>0$, or $u_1<0$, we would obtain the same results as in the case $u>0$ and $u_1<0$, or $u_1>0$, and that is why we only consider this last range of $u$.
   
This way, considering first $u_1>0$, we actually find two different behaviours related to the constants $C$ and $h$: if $\abs{C}=h$, $K_1$ is non-zero finite in the entire domain, including the point $x=2k$; if $\abs{C}\neq h$, $K_1$ only diverges at that point. Analysing the remaining terms of the Kretschmann scalar, we actually find the same behaviours, so it is only interesting to show $K_1$. In both cases, $r$ is always non-zero finite before that point (there is neither a second spatial infinity, nor a regular centre). In the case $\abs{C}=h$, we find that $K$ is finite throughout the entire domain, thus, there is no singularity in this case and we define $x_\text{min}=2k$. Conversely, when $\abs{C}\neq h$ there is a singularity at $x_{\text{min}}=x_s=2k$. 
Proceeding with the analysis of the metric, we find that $g_{00}\neq0$ is always verified at any $x>x_\text{min}$, meaning that there are no horizons in any case, implying that the metric signature never changes and all singularities are naked. At $x=x_\text{min}$, we actually find $g_{00}=0$ if $\abs{C}<h$, meaning that, in this case, there is a light-like and attractive singularity, whereas if $\abs{C}>h$, $g_{00}$ approaches infinity, and so, there is a time-like and repulsive singularity. When $\abs{C}=h$, $g_{00}$ has a non-zero finite value.

Before proceeding further with the analysis, in the case $\abs{C}=h$, apart from $g_{00}$ being non-zero finite, we find that at $x=2k$ there is a regular sphere with a non-zero finite radius. This way, we need, if possible, to analyse the metric in values of $x$ smaller than that. The apparent problem is that this value of $x$ corresponds to $u\to\infty$, which means that values of the former that are smaller than $2k$ would essentially correspond to values of $u$ higher than $\infty$, which is not possible. However, by transforming the coordinate $u$ into a coordinate like $x$, in which $\infty$ is transformed into a finite value, we might be able to extend the solution beyond that point \cite{Bronnikov:2006bv}. This is known as analytical continuation. To do this, after transforming the coordinate $u$, we need to obtain a new line element, using the new coordinate, as already obtained for $x$ in Eq. (\ref{1+}). Now, using this, the coordinate $x$ is, in fact, no longer restricted to the range imposed by the definition in Eq. (\ref{transformation +}), being now restricted to the range of values that result in a real-valued metric. Analysing the line element is more straightforward if we use the exponential form of the hyperbolic trigonometric functions, which yields:
\begin{eqnarray}\label{1+ aberto}
        ds_J^2 &=&\frac{1}{4}e^{-2\psi_0}\left(e^{2\psi_0}+
        \bar{x}^{\frac{C}{k}}\right)^2 \bar{x}^{-\frac{C}{k}}\left\{\frac{4 h^2 e^{2 h u_1} \bar{x}^{\frac{h}{k}}}{q^2 \left(e^{2 h
   u_1}-\bar{x}^{\frac{h}{k}}\right)^2}
   \right.
   \nonumber \\
    &&\left.
    -\frac{q^2 \left(e^{2 h u_1}-\bar{x}^{\frac{h}{k}}\right)^2}{4 h^2 e^{2 h u_1} \bar{x}^{\frac{h}{k}}}\left(dx^2+x^2\bar{x} \, d\Omega^2\right)\right\}\,,
\end{eqnarray}
where we have defined $\bar{x} = 1-\frac{2k}{x}$ for notational simplicity, once again.
We find that $x$ still ranges from $2k$ to $\infty$, because the exponents $C/k$ and $h/k$ are always equal to $\frac{1}{2}$, since $\abs{C}=h$ in this case, and so, $k=2C=2h$, which leads to complex values if $x$ is below $2k$. This way, it is, in fact, not possible to extend the solution beyond the regular sphere mentioned above. Thus, this scenario is non-physical, as there is a non-analytical region where particles are forbidden. As a result, we are not going to analyse this case any further.

Now, in the cases in which $\abs{C}\neq h$, we must continue the analysis. Accordingly, if $\abs{C}<h$ we find that $r\to \infty$ as $x\to x_s$ and if $\abs{C}>h$, in that limit of $x$, $r\to 0$. Apart from this, in the former case, there is always a throat at some point $x_T>x_s$, which is closer (in terms of $x$) to $x_s$ than to spatial infinity, which is important when constructing the Penrose diagram. Conversely, in the latter case, the results depend on the relation between the constants $q$, $h$, $C$ and $\psi_0$. For a given combination of $h$, $C$, that satisfies $\abs{C}>h$, and $\psi_0$, with the same sign of $C$, there is a critical value of $q$, named $q_c$. In fact, there are two symmetric critical values, as the sign of $q$ has no effects in the metric, and so, we will represent them just as one, $\abs{q_c}$. We have that when $q<\abs{q_c}$, the derivative of $r$ has no zeros; when $q=\abs{q_c}$, it has one zero, corresponding to an inflection point in $r$, which has no significant physical meaning; when $q>\abs{q_c}$ it has two zeros, corresponding to a throat and an anti-throat, located at $x_T$ and $x_{aT}$, respectively, being that $x_T>x_{aT}>x_s$. We find that both of these structures are closer to $x_s$ than to spatial infinity (not relevant for the Penrose diagram in this case). When $C$ and $\psi_0$ have opposite signs, which always includes the particular case $\psi_0=0$, this critical behaviour does not apply, thus, the existence of both structures is forbidden. 

About the critical values, note that they are not unique, being always dependent on the values of the other constants. Moreover, these constants also present critical behaviour, which means we could analyse any of them instead. Regardless, to determine $\abs{q_c}$, we started by requiring that the derivative of $r$ and its second derivative are null simultaneously. However, we were not able to solve that analytically, not being able to obtain a general analytic expression for it, but we were able to obtain numerical solutions, by fixing the other constants in advance. This means that we always need to follow this method in order to obtain different critical values. For example, following this approach, by fixing the values $h=2$, $C=2.1$ and $\psi_0=2$, which were arbitrarily chosen, but within the requirements aforementioned, we obtain $\abs{q_c}=3.1871$, being the inflection point located at $x=8.3509$.

In short, for $u_1>0$, we have: if $|C|=h$ it is a non-physical scenario; if $|C|<h$ there is a light-like, naked, attractive singularity at infinity situated beyond a throat, also known as ``space pocket" \cite{Jordan:1955}; if $|C|>h$ there is a time-like, naked, repulsive central singularity situated beyond a throat and anti-throat in some cases, or there may be no throats at all.

Now, considering $u_1<0$, we verify that, apart from $x=2k$, $K_1$ also diverges at the point $\bar{u}=-u_1$, as explained before, which occurs before the former point. When analysing the remaining terms of $K$, we do not find any other divergence between $\bar{u}=0$ (spatial infinity) and $\bar{u}=-u_1$, and also $r$ remains non-zero finite in this open range, which means the latter point is a singularity. This way, in terms of $u$, instead of $x$, we have $u_\text{max}=u_s=-u_1$. Furthermore, by analysing $g_{00}$, we verify that there are no horizons, and so, the metric signature remains unchanged. At $u=u_s$, this function goes to infinity. By analysing the radius function, we find that $r\to 0$ as $u\to u_s$ and that it has no regular minima. In short, in this case, at $u=u_\text{max}$, there is a time-like, naked, repulsive central singularity.

At last, we are also interested, as mentioned before, in obtaining the Penrose diagrams of all of these spacetime geometries. Given the detailed analysis performed so far, we are able to construct them. Let us start with the cases relative to $u_1>0$. When $\abs{C}<h$, we obtain a diagram similar to one of an asymmetric two-way traversable wormhole, with the throat (long-dashed line) closer to the second spatial infinity, however, with a light-like singularity located at that infinity. This diagram can be seen in the left plot  of Fig. \ref{Fig1}. When $\abs{C}>h$, in the case there are no throats, we have the same diagram of the Reissner-Nordstr\"{o}m solution with $q^2>m^2$ \cite{dInverno:1992gxs}, as can be seen in the middle plot of Fig. \ref{Fig1}. In the case there is a throat and an anti-throat (short-dashed line), we obtain a similar one, but with those structures, as can be seen in the right plot of Fig. \ref{Fig1}. Note that, in that diagram, after the throat and anti-throat it is a Parallel Universe, as in the left plot of Fig. \ref{Fig1}. In the case $u_1<0$, we obtain the diagram of middle plot of Fig. \ref{Fig1} as well, however, according to our analysis, where there is ``$x=x_s$" should be ``$u=u_s$".

%%%%%%%%%%%%%%%%%%%%%%%%%%%%%%%%%%%%%%%%%%%%%%%%%%%%%%%%%%%%
 \subsection{Class [2+]}
 %%%%%%%%%%%%%%%%%%%%%%%%%%%%%%%%%%%%%%%%%%%%%%%%%%%%%%%%%%%%

In this class, according to Eq. (\ref{25}), we always have the relation $k=\sqrt{3}\abs{C}$. From Eq. (\ref{u1}), in this class we have $u_1=\pm 1/q$.
Given the signs of $n$, $k$ and $h$ and according to the equations (\ref{14}), (\ref{17}) and (\ref{sJ can}), in this case, the line element is given by
\begin{eqnarray}
    ds_J^2 &=& \cosh^2(Cu+\psi_0)\Bigg\{\frac{dt^2}{q^2 (u+u_1)^2}
		\nonumber \\    
    && -\frac{k^2 q^2 (u+u_1)^2} { \sinh^2(ku)}\left[\frac{k^2 du^2}{\sinh^2(ku)}+d\Omega^2\right]\Bigg\}\,.
\end{eqnarray}

Using the transformation from Eq. (\ref{transformation +}), we obtain the following line element with the coordinate $x$:
\begin{eqnarray}
    ds_J^2 &=& \cosh^2\left(C\bar{u}+\psi_0\right)\Bigg\{\frac{dt^2}{q^2 
    \left(\bar{u}+u_1\right)^2}
		\nonumber \\    
    && -q^2 \left(\bar{u}+u_1\right)^2\left[dx^2+x^2\bar{x}\,d\Omega^2\right]\Bigg\}\,,
\end{eqnarray}
where we have used the definition $\bar{u}=\frac{\log (\bar{x})}{-2k}$, with $\bar{x} = 1-\frac{2k}{x}$, for notational simplicity.
Analysing this metric as $x\to\infty$, which is spatial infinity, in the same way we did for the previous class, 
we also find $g_{00}=-g_{11}=\cosh^2\psi_0$. This way, this is an asymptotically flat spacetime, which is Minkowskian if $\psi_0=0$. 
The Schwarzschild mass is now given by $m=\cosh(\psi_0)/u_1-C\sinh
\psi_0$. By imposing $m>0$, we obtain the constraint $C<\coth(\psi_0)/u_1$, if $\psi_0>0$, and if $
\psi_0<0$, $C$ has to be strictly greater than that. Once again, this imposition on $C$ is only allowed if 
$\psi_0\neq 0$, in which case both signs of $u_1$ are allowed. If $u_1<0$ we know beforehand that, in 
order to have $m>0$, the signs of $C$ and $\psi_0$ have to be opposite. In the case of $\psi_0=0$ we 
obtain $m= 1/u_1$, which, by imposing $m>0$, imposes $u_1>0$. Using the expression for 
$u_1$ showed before, we have $m=\abs{q}$. In the following analysis both signs of $u_1$ are going to be 
analysed, being that we always assume combinations of constants that guarantee $m>0$.

Moving on to the analysis of the regularity of the metric, we may start by analysing the term $K_1$ of the Kretschmann scalar associated with the latter line element (more convenient in this case), which is given by
\begin{eqnarray}
     K_1&=&\frac{-4k^2(\text{sech}^2(C\bar{u}+\psi_0))}{q^2 x^2 (x-2 k)^2 \left[-2k(\bar{u}+ u_1)\right]^4} %\text{sech}^2(C\bar{u}+\psi_0)
		\nonumber \\     
    && \quad \times \{4k^2\,[(2u_1x-2ku_1-3)+2(x-k)\bar{u}] 
    	\nonumber \\
    &&+2Ck(\bar{u}+u_1)[-2Ck(\bar{u}+u_1)
    \nonumber \\
    &&
    \quad \times\text{sech}^2(C\bar{u}+\psi_0)
    \nonumber \\     
    && 
    +4k[(1+ku_1-u_1x)+(-x+k)\bar{u}]
    \nonumber \\
    &&
    \quad \times \tanh(C\bar{u}+\psi_0)]\}\,\,.
\end{eqnarray}

%%%%%%%%%%%%%%%%%%%%%%%%%%%%%%%%%%%%%%%%%%%%%%%%%%%%%%%%%%%%%%%%%%%%%%%%%%%%%%%%%%%%%%%%%%%%%%%%%%%%%%%%%%%%%%%%%%%%%%%%%%
\begin{figure*}[ht]
   \centering
      \includegraphics[scale=0.75]{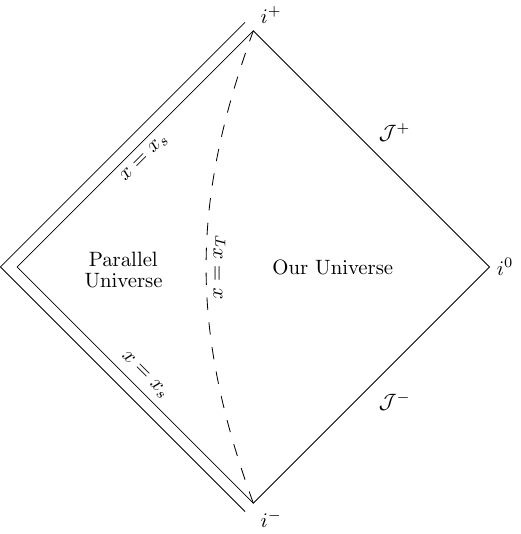}
      \hspace{1cm}
      \includegraphics[scale=0.75]{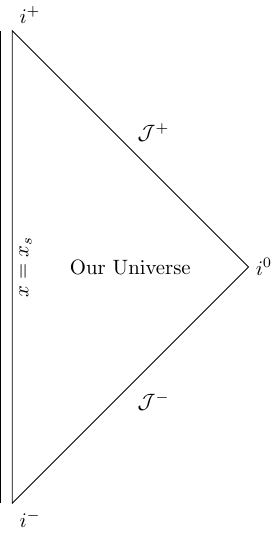}
      \hspace{1cm}
      \includegraphics[scale=0.75]{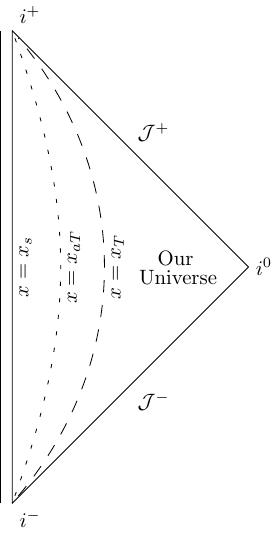}
    \caption{Left plot: Penrose diagram of a naked, light-like singularity ($g_{00}=0$ at $x_s$; diagonal double line), located beyond a throat (long-dashed line), tilted to the left as $x_T$ is closer to $x_s$ than to $x\to\infty$. In this type of diagram, null geodesics can be represented as lines at a 45-degree angle. To the right of the throat lies our universe, with future and past flat spatial infinities (top and bottom diagonal lines, respectively). To the left, there is a parallel universe, with future and past light-like singularities situated at infinity. Middle plot: Penrose diagram of a naked, central, time-like singularity (vertical double line) solution. In this case, the singularity lies in our own universe. Right plot: Penrose diagram of a naked, central, time-like singularity, located beyond a throat and an anti-throat (short-dashed line). To the left of these structures lies a parallel universe.}
    \label{Fig1}
\end{figure*}
%%%%%%%%%%%%%%%%%%%%%%%%%%%%%%%%%%%%%%%%%%%%%%%%%%%%%%%%%%%%%%%%%%%%%%%%%%%%%%%%%%%%%%%%%%%%%%%%%%%%%%%%%%%%%%%%%%%%%%%%%%%%%%%

 By analysing this expression, we find that at $x\to \infty$, or $\bar{u}\to 0^+$, it does not diverge, being null, as well as the remaining terms of $K$, which supports our description of the spatial infinity.
Apart from that, in this term of $K$ we find similar behaviours relative to the previous class. In fact, the first quotient also goes to infinity as $\bar{u} \to \infty$, and so, when $x \to 2k$, while the remaining terms approach $0$ in that limit. The main difference in this case is that the quotient term is always dominant at that point, due to the relations between constants, which means $K_1$ always diverge there. Furthermore, there is neither a second spatial infinity, nor a regular centre before that point.
Nevertheless, that point may not be a singularity if $K_1$, or other terms of $K$ diverge before it. Once again, that is what happens when $u_1<0$, since only in that case the function $\bar{u}+u_1$, that appears in the metric and in $K_1$, has a positive zero at $\bar{u}=-u_1$, which causes $K_1$ to diverge. By analysing the remaining terms of $K$, we end up with the same conclusions. Thus, we are going to split the analysis according to the sign of $u_1$.

Considering first $u_1>0$, we, in fact, find similar behaviour to that in class $[1+]$ with $u_1>0$ and $\abs{C}>h$, but now bearing in mind that $h=0$. In particular, at $x_s=x_\text{min}=2k$, there is a time-like, naked, repulsive central singularity situated beyond a throat and an anti-throat in some cases, or there may be no throats at all. Considering $u_1<0$, we also find similar behaviour to that in class $[1+]$ with $u_1<0$, which means there is a time-like, naked, repulsive central singularity (with no throats), at $u=u_\text{max}=-u_1$. 

Regarding the Penrose diagrams of the spacetimes analysed in this class, they have already been obtained. In the cases in which there are no throats, which occur when $u_1<0$ and in some cases of $u_1>0$, the diagram is the middle plot of Fig. \ref{Fig1}, with ``$u=u_s$" instead of ``$x=x_s$" in the former case, according to our analysis; in the case there is a throat and an anti-throat, when $u_1>0$, the diagram is the right plot of Fig. \ref{Fig1}.

We will now examine the geometry of metrics derived from embedding diagrams within a Euclidean space. The mathematical derivation leading to the equation that relates a two-dimensional curved embedding surface to a general line element is presented in Appendix \ref{ap}. 

At this stage, we aim to visually enhance the Penrose diagrams in Fig. \ref{Fig1}, as well as those presented later in this manuscript, by emphasizing geometric features characteristic of each spacetime—such as local minima and maxima, which we refer to as throats and antithroats. These features are most effectively illustrated through the curved surfaces produced by rotating embedding diagrams around the axial axis.

\begin{figure*}[ht]
   \centering
      \includegraphics[scale=0.47]{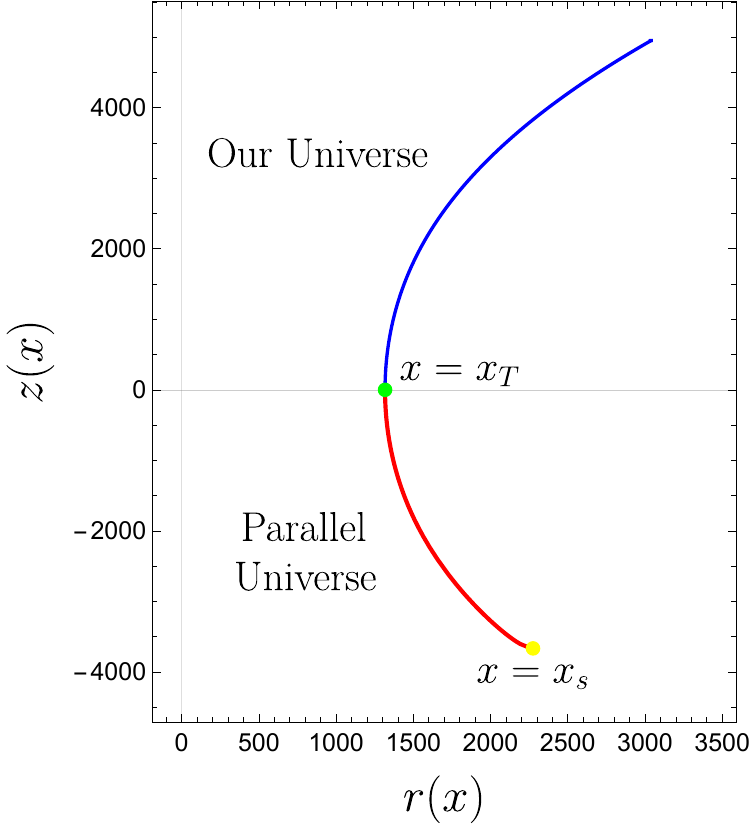}
      \hspace{1.5cm}
     \includegraphics[scale=0.43]{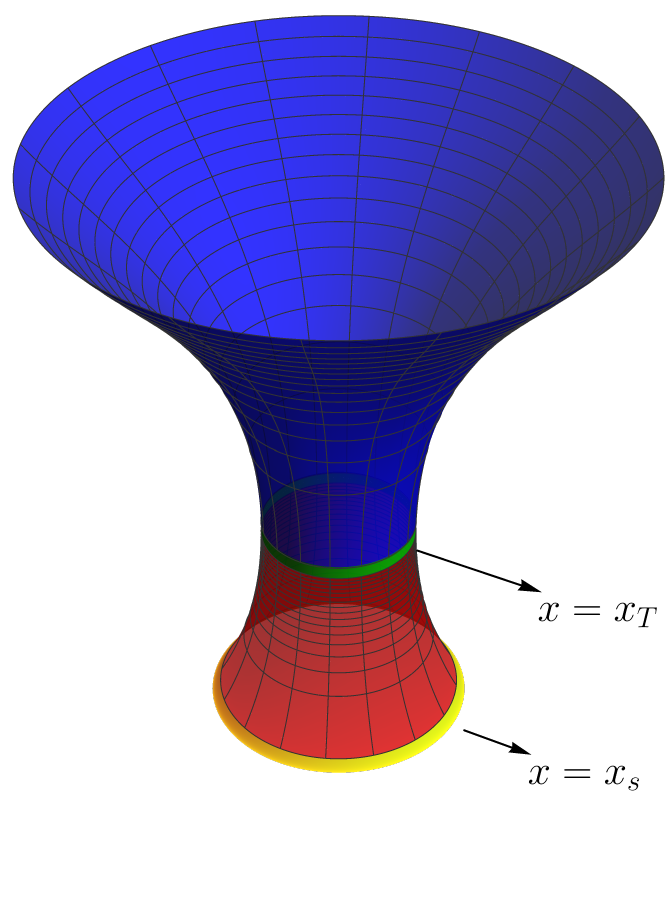}
    \caption{ Left plot: depicts the $z(x)$ \textit{vs} $r(x)$ graph of the two-way traversable wormhole solution. Right plot: 
    	The features in the left plot are most effectively illustrated through the curved surfaces produced by rotating embedding diagrams around the axial axis, which corresponds to the 
    	embedding diagram of the solution.}
    \label{Emb1}
\end{figure*}

Using the line element given in Eq. \eqref{1+}, we can represent all the Penrose diagrams discussed in Fig. \ref{Fig1} through the corresponding embedding diagrams, as shown in Fig. \ref{Emb1}. This is achieved by numerically integrating the solution in Eq. \eqref{z} for our metric, using specific values for the constants involved, and following the procedure outlined in Appendix \ref{ap}.

Figure \ref{Emb1} reproduces the configuration shown on the left side of Fig. \ref{Fig1}, corresponding to the case $u > 0$ with $\mid C\mid < h$, which, as previously discussed, results in an asymmetric two-way traversable wormhole. This setup features a local minimum, interpreted as a throat, as well as a light-like singularity situated at spatial infinity.

In the plot on the left of Fig.~\ref{Emb1}, we present the function $z(x)$ \textit{vs.} $r(x)$ for specific values of the constants, which clearly reveals the presence of a local minimum. The blue curve represents our universe, while the red curve corresponds to a parallel universe. The green dot indicates the position of the throat, and the yellow dot at the end of the red curve marks the location of the singularity.

\begin{figure*}[ht]
   \centering
      \includegraphics[scale=0.39]{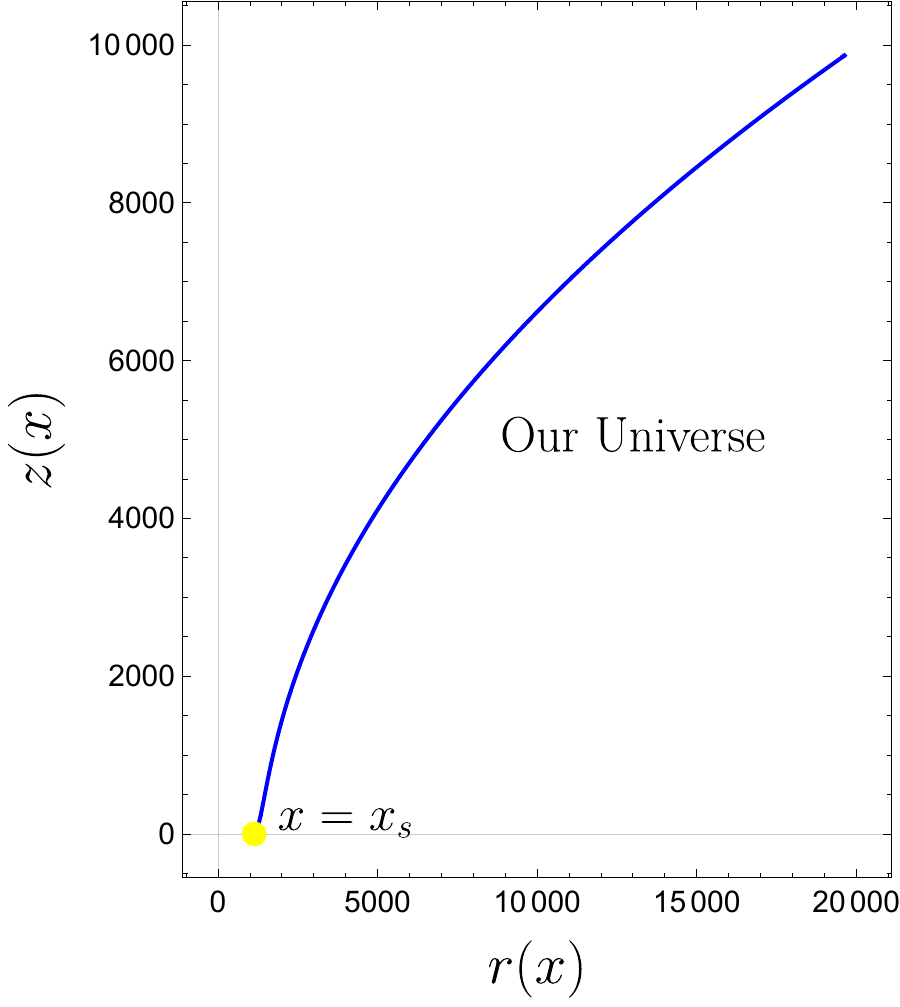}
       \hspace{1cm}
     \raisebox{1.25cm}{ \includegraphics[scale=0.5]{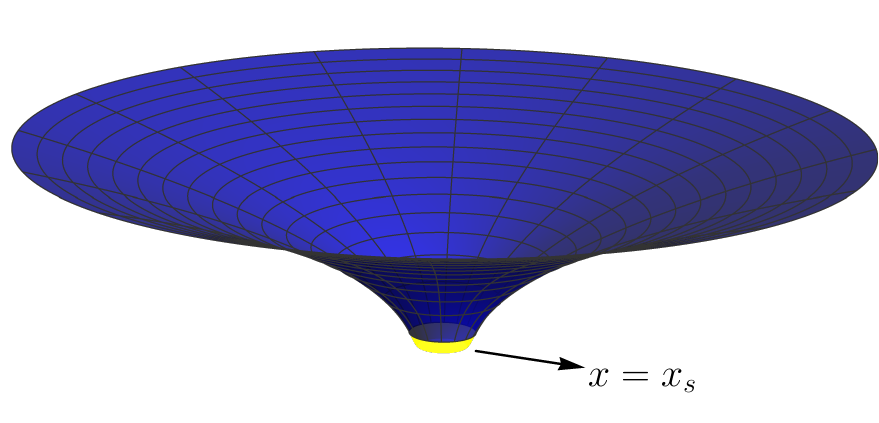}}
    \caption{ Left plot: $z(x)$ \textit{vs} $r(x)$ graph of a naked, central, time-like singularity solution. In this case, the singularity lies in our own universe. Right plot: The embedding diagram obtained by rotating the left plot around the axial axis.}
    \label{Emb2}
\end{figure*}

\begin{figure*}[ht]
   \centering
      \includegraphics[scale=0.555]{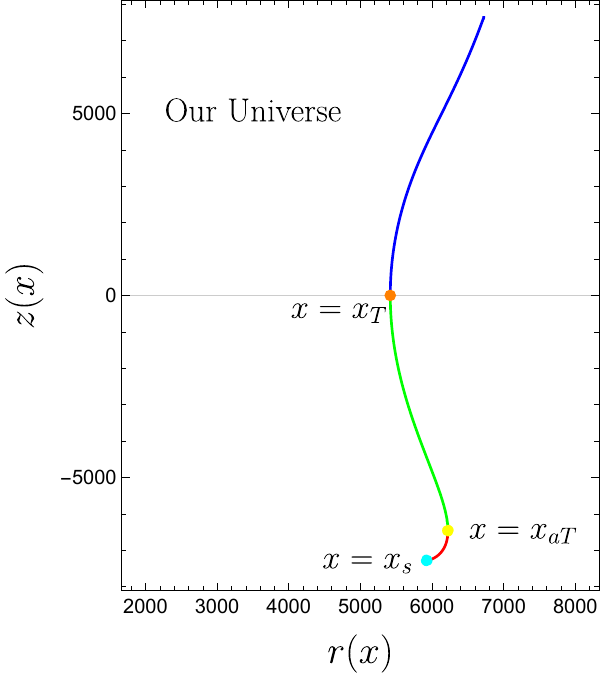}
       \hspace{1cm}
      \raisebox{0.0cm}{\includegraphics[scale=0.555]{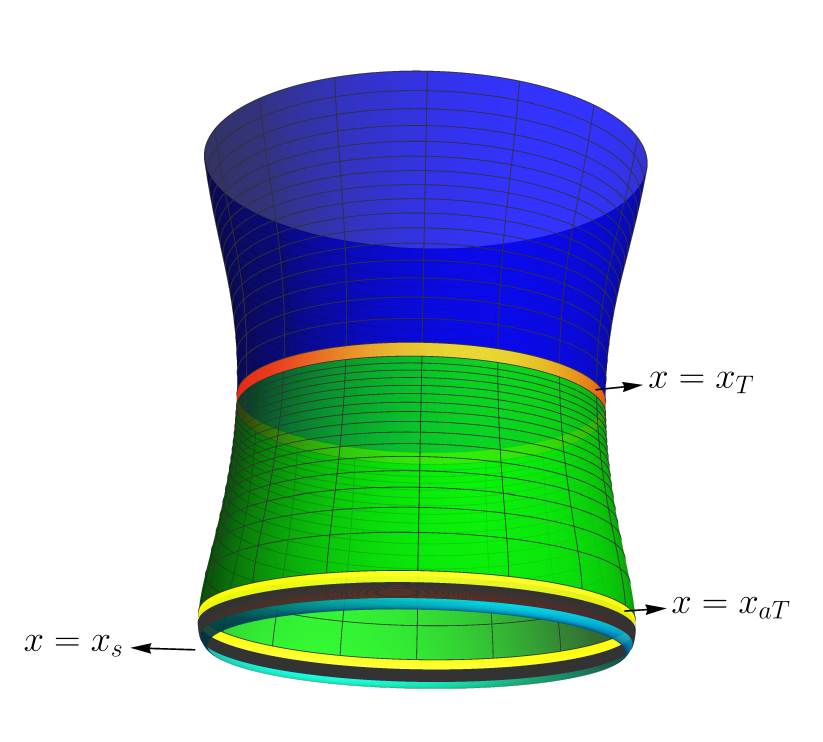}}
    \caption{ Left plot: $z(x)$ \textit{vs} $r(x)$ graph of a naked, central, time-like singularity, located beyond a throat and an anti-throat. Right plot: Embedding diagram.}
    \label{Emb3}
\end{figure*}

By rotating this plot around the $z(x)$-axis, we obtain the corresponding embedding surface diagram, shown on the right of Fig.~\ref{Emb1}. In this representation, the throat $( x = x_T )$ is pictorially highlighted with a green stripe, while the singularity $( x = x_s )$ is marked with a yellow stripe. Using this same procedure, we have also constructed the embedding diagrams corresponding to the middle and right Penrose diagrams in Fig.~\ref{Fig1}, which are displayed in Figs.~\ref{Emb2} and \ref{Emb3}, respectively.

From this point onward, we will refrain from detailing the construction of the subsequent embedding diagrams, as they all follow the same methodological procedure outlined above. A comprehensive analysis can be found in Appendix~\ref{ap}. Additionally, the causal structures are examined in the context of each Penrose diagram. For this reason, we have omitted specific information regarding elements such as curves, colored bands, and the precise locations of throats and singularities.

%%%%%%%%%%%%%%%%%%%%%%%%%%%%%%%%%%%%%%%%%%%%%%%%%%%%%%%%%%%%%%%%%%%%%%%%%%%%%%%%%%%%%%%%%%%%%%%%%%%%%%%%%%%%%%%%%%%%%%%%%%
%%%%%%%%%%%%%%%%%%%%%%%%%%%%%%%%%%%%%%%%%%%%%%%%%%%%%%%%%%%%%%%%%%%%%%%%%%%%%%%%%%%%%%%%%%%%%%%%%%%%%%%%%%%%%%%%%%%%%%%%%%%%%%%

%%%%%%%%%%%%%%%%%%%%%%%%%%%%%%%%%%%%%%%%%%%%%%%%%%%%%%%%%%%%%%%%%%%%%%%%%%%%%%%%%%%%%%%%%%%%%%%%%%%%%%%%%%%%%%%%%%%%%%%%%%
%%%%%%%%%%%%%%%%%%%%%%%%%%%%%%%%%%%%%%%%%%%%%%%%%%%%%%%%%%%%%%%%%%%%%%%%%%%%%%%%%%%%%%%%%%%%%%%%%%%%%%%%%%%%%%%%%%%%%%%%%%%%%%%

%%%%%%%%%%%%%%%%%%%%%%%%%%%%%%%%%%%%%%%%%%%%%%%%%%%%%%%%%%%%%%%%%%%%%%%%%%%%%%%%%%%%%%%%%%%%%%%%%%%%%%%%%%%%%%%%%%%%%%%%%%
%%%%%%%%%%%%%%%%%%%%%%%%%%%%%%%%%%%%%%%%%%%%%%%%%%%%%%%%%%%%%%%%%%%%%%%%%%%%%%%%%%%%%%%%%%%%%%%%%%%%%%%%%%%%%%%%%%%%%%%%%%%%%%%

%%%%%%%%%%%%%%%%%%%%%%%%%%%%%%%%%%%%%%%%%%%%%%%%%%%%%%%%%%%%
\subsection{Class [3+]}
%%%%%%%%%%%%%%%%%%%%%%%%%%%%%%%%%%%%%%%%%%%%%%%%%%%%%%%%%%%%

Regarding this class, we have that it actually encompasses three different signs of $k$. However, by analysing the respective metrics we find that their qualitative behaviour is the same, so we define a unique class to describe all of them. In fact, this class of solutions corresponds to the same spacetime geometry as the previous class, however the analysis of $K_1$ reveals to be unique, and so, it must be done separately.

In the case in which $k>0$, according to Eq. (\ref{25}), we have $k=\sqrt{3C^2-h^2}$, which, in turn, leads to the relation $|h|<\sqrt{3}|C|$. Furthermore, using Eq. (\ref{u1}), we have $u_1=\pm [\arcsin (\frac{h}{q})+c_2\pi]/h$. Here $c_2$ is an arbitrary integer, however, we will always assume $c_2=0$, for simplicity, as its value does not have any effect in the metric. This expression is associated with the sign of $h$ and so it holds for any sign of $k$. In order for this expression to be real-valued, it is required that $\abs{q}\geq\abs{h}$. Once again, given the signs of $n$, $k$ and $h$ and according to the equations (\ref{14}), (\ref{17}) and (\ref{sJ can}), we have the following line element: 
\begin{eqnarray}
    ds_J^2 &=& \cosh^2(Cu+\psi_0)\Bigg\{\frac{h^2dt^2}{q^2 \sin^2[h(u+u_1)]}
		\nonumber \\    
     &&\hspace{-0,5cm}
     -\frac{k^2 q^2 \sin^2[h(u+u_1)]} {h^2 \sinh^2(ku)}\left[\frac{k^2 du^2}{\sinh^2(ku)}+d\Omega^2\right]\Bigg\}.
\end{eqnarray}

Transforming the coordinate $u$ into $x$, using Eq. (\ref{transformation +}), yields the following line element:
\begin{eqnarray}
    ds_J^2 &=& \cosh^2\left(C\bar{u}+\psi_0\right)\Bigg\{\frac{h^2dt^2}{q^2 \sin^2\left[h\left(\bar{u}+u_1\right)\right]}
		\nonumber \\    
    &&-\frac{q^2 \sin^2\left[h\left(\bar{u}+u_1\right)\right]} {h^2}\left[dx^2+x^2\bar{x}\,d\Omega^2\right]\Bigg\}\,,
\end{eqnarray}
where, once again, we have defined $\bar{u}=\frac{\log (\bar{x})}{-2k}$, with $\bar{x} = 1-\frac{2k}{x}$, for notational simplicity.

Now, in the case in which $k=0$, using the same equations as before, we obtain the relation $|h|=\sqrt{3}|C|$ and the line element is given by
\begin{eqnarray}
    ds_J^2 &=& \cosh^2(Cu+\psi_0)\Bigg\{\frac{h^2dt^2}{q^2 \sin^2[h(u+u_1)]}
		\nonumber \\    
    && -\frac{q^2 \sin^2[h(u+u_1)]} {h^2 u^2}\left(\frac{du^2}{u^2}+d\Omega^2\right)\Bigg\}\,.
\end{eqnarray}

As explained before, in this case, we have to use Eq. (\ref{transformation 0}) to transform the coordinate $u$ into $x$, with which we obtain the following line element:
\begin{eqnarray}
    ds_J^2 &=& \cosh^2\left(C/x+\psi_0\right)\Bigg\{\frac{h^2dt^2}{q^2 \sin^2\left[h\left(\frac{1}{x}+u_1\right)\right]}
		\nonumber \\    
    && -\frac{q^2 \sin^2\left[h\left(\frac{1}{x}+u_1\right)\right]} {h^2}\left(dx^2+x^2d\Omega^2\right)\Bigg\}\,.
\end{eqnarray}

Finally, for the case ${k<0}$, we get the relation $k=-\sqrt{h^2-3C^2}$, which yields $|h|>\sqrt{3}|C|$, and the line element is given by 
\begin{eqnarray}
    ds_J^2 &=& \cosh^2(Cu+\psi_0)\Bigg\{\frac{h^2dt^2}{q^2 \sin^2[h(u+u_1)]}
		    	\nonumber \\    
    && \hspace{-0,5cm}
    -\frac{k^2 q^2 \sin^2[h(u+u_1)]} {h^2 \sin^2(ku)}\left[\frac{k^2 du^2}{\sin^2(ku)}+d\Omega^2\right]\Bigg\}\,.
\end{eqnarray}

Now, to transform the coordinate $u$, in this case, we use Eq. (\ref{transformation -}), which yields the line element:
\begin{eqnarray}
    ds_J^2 &=&\cosh^2\left(C \tilde{u} +\psi_0\right)\left\{\frac{h^2dt^2}{q^2 \sin^2[h(\tilde{u}+u_1)]}
    \right.\notag\\
    && \hspace{-0,5cm}
     \left.-\frac{q^2 \sin^2[h(\tilde{u}+u_1)]} {h^2}\left[dx^2+(x^2+k^2)d\Omega^2\right]\right\},
\end{eqnarray}
where we have defined $\tilde{u} = \frac{\arccot(\frac{x}{|k|})+c_1\pi}{|k|}$, for simplicity.

The analysis of these solutions, as aforementioned, is going to be performed together. Analysing the metrics with the coordinate $x$, and using the expression aforementioned for $u_1$, we find, as $x\to\infty$ --- considering $c_1=0$ in the case $k<0$, so that it corresponds to the limit $u\to 0^+$, hence $\tilde{u}\to 0^+$---, that $g_{00}=-g_{11}=\cosh^2\psi_0$, as in the previous classes. Accordingly, these correspond to asymptotically flat spacetimes, in particular Minkowskian if $\psi_0=0$. For the Schwarzschild mass we obtain $m=h \cot(h u_1)\cosh\psi_0 - C\sinh\psi_0$, which, by imposing it to be positive and considering $\psi_0\neq 0$, leads to the constraint $C<h\cot(hu_1)\coth\psi_0$, if $\psi_0>0$, but if $\psi_0<0$, $C$ has to be strictly greater than that. Note that the relations mentioned before, for $h$ and $C$, still need to be considered. Both of these cases allow both signs of $u_1$, being that if it is negative, we know beforehand that $C$ and $\psi_0$ must have opposite signs. On the other hand, if $\psi_0=0$, then $m=h\cot(hu_1)$, which, by requiring $m>0$, imposes $u_1>0$ and also $\abs{q}>\abs{h}$ (the equality leads to $m=0$). As in the previous classes, in the following analysis both signs of $u_1$ are analysed and we always assume combinations of constants that guarantee $m>0$.

Now, to test the regularity of these solutions, we may analyse the Kretschmann scalar. Unlike the analyses carried out before, for this class, and the remaining ones, we are going to compute that scalar using the line elements with the coordinate $u$, since now there are trigonometric functions, whose zeros may correspond to a singularity, for example, and are simpler to analyse with that coordinate. We are going to start by analysing $K_1$, which is given, in each case, respectively, by:
\begin{widetext}
\begin{align}
    K_1=&\frac{1}{k^4 q^2}h^2 \text{sech}^2(C u+\psi_0) \sinh ^4(k u) \csc ^2\left(h [u+u_1]\right) \left\{C^2 \text{sech}^2(C u+\psi_0)+2 k \coth (k u) [C \tanh (C u+\psi_0) \right.\notag\\
    &\left.-h \cot (h [u+u_1])]+h\left[-2 C \tanh (C u+\psi_0) \cot (h [u+u_1])+3 h \csc ^2(h [u+u_1])-2 h\right]\right\}\,, \qquad \text{if } k>0\,\,,\\[10pt]
    K_1=&\frac{1}{q^2}h^2 u^3 \text{sech}^2(C u+\psi_0) \csc ^2(h [u+u_1]) \left\{C^2 u \,\text{sech}^2(C u+\psi_0)-2 C \tanh (C u+\psi_0) [h\,u \cot (h
   [u+u_1])-1] \right.\notag\\
   &\left.+h \left[3 h\,u \csc ^2(h [u+u_1])-2 [\cot (h [u+u_1])+h u]\right]\right\}\,, \qquad \text{if } k=0\,\,,\\
    K_1=&\frac{1}{k^4 q^2}h^2 \text{sech}^2(C u+\psi_0) \sin ^4(k u) \csc ^2(h [u+u_1]) \left\{C^2 \text{sech}^2(C u+\psi_0)+2 C \tanh (C u+\psi_0)
   [k \cot (k u)  \right.\notag\\
   &\left.-h \cot (h [u+u_1])]+h \left[3 h \csc ^2(h [u+u_1])-2 [k \cot (k u) \cot (h [u+u_1])+h]\right]\right\}\,, \qquad  \text{if } k<0\,\,.
\end{align}
\end{widetext}

First of all, analysing these expressions in the limit $u\to 0^+$, we find that all of them are null. Furthermore, when analysing the remaining terms of $K$, we obtain the same results, which sustain our previous description of the spatial infinity.  Apart from that, in the above expressions we have hyperbolic trigonometric functions and also reciprocals, whose behaviours as $u\to\infty$ have already been analysed. However, in this case, we do not need to analyse neither $K$ nor the metrics in that limit, since now there are trigonometric functions whose zeros are very important in our analysis, as aforementioned, and evidently occur before $u\to\infty$. In particular, at the zeros of the $\sin(h [u+u_1])$ function, both $\csc (h [u+u_1])$ and $\cot(h [u+u_1])$ go to infinity, which, in turn, cause $K_1$ to diverge. Accordingly, the first positive one among them has the potential to be a singularity. However, if $K_1$ or other terms of $K$ diverge before it, then it is not a singularity, but that does not happen in any case (we actually draw the same conclusions by analysing any term of $K$). 

Another way for that point not to be a singularity, as already mentioned before, would be if, at a smaller value of $u$, there was a second spatial infinity or a regular centre, but neither is a possibility. The former scenario would be possible in the case of $k<0$, in which each zero of the $\sin(ku)$ function is regular and causes $r$ to approach infinity, if the relations between constants would allow it. However, in particular due to the relation $\abs{h}>\abs{k}$, which is verified in this case, the first positive zero of $\sin(h [u+u_1])$ always occurs first, which means it is, indeed, a singularity. Accordingly, in every case, we define $u_\text{max}=u_s=\frac{\pi}{|h|}-u_1$, if $u_1>0$, and $u_\text{max}=u_s=-u_1$, if $u_1<0$. In this class, unlike the previous ones, the latter sign of $u_1$ does not cause different functions to diverge, however, the difference in the values of $u_\text{max}$ is actually leading to different results in the remaining analysis.

Considering $u_1>0$ and regarding the remaining analysis of the metrics, we actually find similar behaviour to that in class $[2+]$ with $u_1>0$. In fact, at $u=u_\text{max}$ there is a time-like, naked, repulsive central singularity situated beyond a throat, at $u=u_T$, and an anti-throat, at $u=u_{aT}$, with $u_T<u_{aT}<u_s$, or, alternatively, there may be no throats at all.
Now, if we consider $u_1<0$, we find similar behaviour as in classes $[1+]$ and $[2+]$ with $u_1<0$, thus, at $u=u_\text{max}$, there is a time-like, naked, repulsive central singularity (with no throats).

Relatively to the Penrose diagrams of the spacetimes analysed in this class, we have the same as in the previous one. Thus, when there are no throats, which occur when $u_1<0$ and in some cases of $u_1>0$, the diagram is the middle plot of Fig. \ref{Fig1}, with ``$u=u_s$" instead of ``$x=x_s$". When a throat and an anti-throat are present, for $u_1>0$, the diagram is the right plot of Fig. \ref{Fig1}, but also considering ``$u$" instead of ``$x$", according to our analysis.

%%%%%%%%%%%%%%%%%%%%%%%%%%%%%%%%%%%%%%%%%%%%%%%%%%%%%%%%%%%%
\subsection{Class [1-]}
%%%%%%%%%%%%%%%%%%%%%%%%%%%%%%%%%%%%%%%%%%%%%%%%%%%%%%%%%%%%

Now that all the classes of solutions with $n=+1$ have been analysed, for which we only obtained spacetimes with naked singularities, with the possibility of existing a throat or a throat and an anti-throat in some cases, we may start analysing the classes with $n=-1$, which correspond to phantom scalar fields. 

In the upcoming classes, the conformal factor is the main difference relative to the previous ones. While the $\cosh$ function has no zeros, but diverges as its argument goes to infinity, the $\cos$ function never diverges but has uncountably many zeros, each of which may lead to a divergence in the metric. In fact, we may obtain a singularity at the first positive zero, similarly to what was obtained in class [3+], with the $\sin(h [u+u_1])$ function. Nonetheless, this is not the case if spacetime ends (another point of divergence), $r$ goes to infinity, or a regular centre exists, at some other point before that, or even if the remaining functions in the metric ``cancel" that divergence, as has already happened in class [1+], and it corresponds to a regular point instead. This way, a detailed analysis is necessary in each and every case.

That being said, let us start by analysing this class. Similarly to the previous one, this class actually encompasses two different signs of $k$, since the qualitative behaviours of their respective metrics are the same. To clarify this, we will present both line elements, as well as both $K_1$, but they will be analysed together.

In the case of $k>0$, from Eq. (\ref{25}), we get the relation $k=\sqrt{h^2-3C^2}$, which leads to $|h|>\sqrt{3}|C|$. By using Eq. (\ref{u1}), we obtain $u_1=\pm \arcsinh (\frac{h}{q})/h$, which holds for both signs of $k$. Given the signs of $n$, $k$ and $h$ and according to the equations (\ref{14}), (\ref{17}) and (\ref{sJ pha}), we have the following line element in this case:
\begin{eqnarray} \label{1- u}
    ds_J^2 &=& \cos^2(Cu+\psi_0)\Bigg\{\frac{h^2dt^2}{q^2 \sinh^2(h[u+u_1])}
		\nonumber \\    
   && \hspace{-0,85cm} 
   -\frac{k^2 q^2 \sinh^2(h[u+u_1])} {h^2 \sinh^2(ku)}\left[\frac{k^2 du^2}{\sinh^2(ku)}+d\Omega^2\right]\Bigg\}\,.
\end{eqnarray}
Using Eq. (\ref{transformation +}), we are able to transform $u$ into $x$ and obtain the line element:
\begin{eqnarray}
    ds_J^2 &=& \cos^2\left(\bar{u}+\psi_0\right)\Bigg\{\frac{h^2dt^2}{q^2 \sinh^2\left(h\left[\bar{u}+u_1\right]\right)}
		\nonumber \\    
    && \hspace{-0,35cm} 
     -\frac{q^2 \sinh^2\left(h\left[\bar{u}+u_1\right]\right)} {h^2}\left[dx^2+x^2 \bar{x} \, d\Omega^2\right]\Bigg\}\,,\label{metx1-}
\end{eqnarray}
where we have used the definition $\bar{u}=\frac{\log (\bar{x})}{-2k}$, with $\bar{x} = 1-\frac{2k}{x}$, for notational simplicity.

Now, in the case of ${k=0}$, using the same equations as before, we get the relation $|h|=\sqrt{3}|C|$ and the line element:
\begin{eqnarray}
    ds_J^2 &=&\cos^2(Cu+\psi_0)\Bigg\{\frac{h^2dt^2}{q^2 \sinh^2(h[u+u_1])}
		\nonumber \\    
    && -\frac{q^2 \sinh^2(h[u+u_1])} {h^2 u^2}\left(\frac{du^2}{u^2}+d\Omega^2\right)\Bigg\}\,.
\end{eqnarray}
Using Eq. (\ref{transformation 0}) to transform $u$ into $x$, we obtain the line element:
\begin{eqnarray}
    ds_J^2 &=& \cos^2\left(C/x+\psi_0\right)\Bigg\{\frac{h^2dt^2}{q^2 \sinh^2(h[\frac{1}{x}+u_1])}
		\nonumber \\    
     && -\frac{q^2 \sinh^2(h[\frac{1}{x}+u_1])} {h^2}\left(dx^2+x^2d\Omega^2\right)\Bigg\}\,.
\end{eqnarray}

The analysis of these solutions, as aforementioned, will be performed together. Analysing the metrics with the coordinate $x$, as it approaches infinity, in the same manner as in the previous classes, we find $g_{00}=-g_{11}=\cos^2\psi_0$, which is different than what has been obtained until now. In this case, not all values of $\psi_0$ are physically allowed. In particular, if we have $\psi_0=\frac{\pi}{2}+c_1\pi$, where $c_1$ is an integer, then both $g_{00}$ and $g_{11}$ are null and $x\to \infty$ is not even a spacial infinity, because $r$ is finite at that limit (in particular equal to $|C|$). This alone is not sufficient to discard those values of $\psi_0$, since at some $x$ below infinity, corresponding to some $u>0$, or even at some $u<0$ (we may consider both ranges simultaneously, in this case, as explained before), there could be a regular point where $r\to \infty$, at which we would define the spatial infinity instead. The limits $u\to\pm\infty$ would be the only alternatives to $u\to 0$ for representing spatial infinity, since $\sinh(h[u+u_1])$ diverges there and $h>k$ is satisfied, when $k>0$. However, as $u$ approaches those values, the limit of the conformal factor does not exist. Thus, there is no spatial infinity for these values of $\psi_0$. Given our astrophysical knowledge, this is not physically plausible, and so, those cases are going to be discarded and not analysed further. 

Apart from those values of $\psi_0$, $x\to \infty$ is, indeed, a spatial infinity and the metrics are asymptotically flat, being Minkowskian if $\psi_0=c_1 \pi$. The Schwarzschild mass is given by $m=\abs{\cos\psi_0}[h\coth(hu_1)+C\tan\psi_0]$. By imposing it to be positive and considering $\psi_0\neq c_1\pi$, we obtain the constraints $C>-h\cot(\psi_0)\coth(hu_1)$, if $\psi_0\in]c_1\pi,\frac{\pi}{2}+c_1\pi[$ (interval $1$), whereas if $\psi_0\in]\frac{\pi}{2}+c_1\pi,(c_1+1)\pi[$ (interval $2$), $C$ has to be strictly lower. As in the previous classes, both constraints allow both signs of $u_1$. When $\psi_0=c_1 \pi$ the mass is given by $m=h\coth(hu_1)$, which imposes $u_1>0$.

The full range of allowed values of $\psi_0$ changes according to the combination of constants, however, thanks to the previous constraints and the relations mentioned before the line elements, involving $h$ and $C$, we are able to find: for any combination of constants with $u_1>0$ and $C>0$, the range $\psi_0\in]\frac{\pi}{2}+c_1\pi+0.5236,\frac{\pi}{2}+(c_1+1)\pi[$ is always allowed ($m>0$); if $u_1>0$ and $C<0$, that range is given by $\psi_0\in]\frac{\pi}{2}+c_1\pi,\frac{\pi}{2}+(c_1+1)\pi-0.5236[$; for any combination of constants with $u_1<0$ and $C>0$ or $C<0$, we verify that those same ranges exchange their respective sign of $C$ and are now never allowed ($m\leq 0$), being a much more restrictive case. In fact, in this case, we know beforehand that if $\psi_0$ is in interval 1, then $C$ must be positive, whereas if $\psi_0$ is in interval 2, $C$ must be negative. As before, in the following analysis we will consider both signs of $u_1$ and we always assume combinations of constants that guarantee $m>0$.

Now, to test the regularity of the metric, we may analyse the Kretschmann scalars associated with the line elements with $u$. As usual, we are going to start by analysing the term $K_1$, which is given, in each case, respectively, by:
\begin{widetext}
\begin{eqnarray}
%\begin{split}
     K_1 &=&\frac{1}{k^4 q^2}h^2 \sec ^2(C u+\psi_0) \sinh ^4(k u) \text{csch}^2(h [u+u_1]) \left\{-C^2 \sec ^2(C u+\psi_0)-2 k \coth (k u) [C \tan (C u+\psi_0)\right.\\
   &&\left.+h \coth (h [u+u_1])]+h \left[2 (C \tan (C u+\psi_0) \coth (h [u+u_1])+h)+3 h \,\text{csch}^2(h [u+u_1])\right]\right\}\,, \qquad \text{if } k>0\,\,,
   	\nonumber \\
     K_1 &=&\frac{1}{q^2} h^2 u^3 \sec ^2(C u+\psi_0) \text{csch}^2(h [u+u_1]) \{2h^2u\,\coth^2(h[u+u_1])+h^2u\,\csch^2(h[u+u_1])-C[Cu\,\sec^2(Cu+\psi_0)
     \nonumber \\
     &&+2\tan(Cu+\psi_0)]+2h\coth(h[u+u_1])[Cu\,\tan(Cu+\psi_0)-1]\}\,, \qquad \text{if } k=0\,\,.
%\end{split}
\end{eqnarray}
\end{widetext}

By analysing these terms in the limit $u\to 0$, considering $\psi_0\neq \frac{\pi}{2}+c_1\pi$, which is spatial infinity, we find that both of them are null, as well as the remaining terms of $K$, which sustains our previous description of the spatial infinity. Apart from this, and similarly to the previous class, in these equations there are hyperbolic functions that diverge as $u\to \infty$, however, that limit does not need to be considered now, since there are trigonometric functions, whose zeros reveal to be of great importance in our analysis. In fact, the zeros of the conformal factor, $\cos^2(Cu+\psi_0)$, that occur at $u=\frac{\pi/2-\psi_0+c_2\pi}{C}$ ($c_2$ is an integer), cause the $\sec(Cu+\psi_0)$ and $\tan(Cu+\psi_0)$ functions to diverge, in any case (we have already excluded $\psi_0= \frac{\pi}{2}+c_1\pi$), which, in turn, cause $K_1$ to diverge. Furthermore, there is neither a second spatial infinity, as aforementioned, nor a regular centre, before the first positive zero of that function. Accordingly, it is a singularity, unless spacetime ends before it. 

Considering the range $u>0$, as we have done so far, that could only occur when considering $u_1<0$, since, in this case, the $\sinh(h[u+u_1])$ function has a real positive zero at $u=-u_1$, at which $K_1$ also diverges. Nevertheless, the first positive zero of $\cos(Cu+\psi_0)$, whose value depends on $C$ and $\psi_0$ --- the higher is the former and the closer is the latter to $\frac{\pi}{2}+c_1\pi$, the smaller it is ---, always occurs before that point. This is due to the requirement $m>0$, since only certain values of $\psi_0$ are allowed, as analysed before. For instance, higher values of $q$ and $h$ lead to a smaller value of $-u_1$, however, those higher values also cause the allowed values of $\psi_0$ to be closer to $\frac{\pi}{2}+c_1\pi$. If we analyse the remaining terms of $K$, we end up with the same conclusions. Note that if we were to consider the range $u<0$, instead of $u>0$, there would also be another possibility of singularity when considering $u_1>0$, located at the zero of $\sinh(h[u+u_1])$ as well. However, we find the exact same results as when considering $u>0$ and $u_1<0$, since now the expression for the mass is symmetric to $m$, presented before. This way, by requiring that new mass to be positive, we obtain that the first negative zero is always the one of the conformal factor. This is the reason why we may, and will, only consider the range $u>0$.

Accordingly, we find that for both signs of $u_1$ there is a singularity at the first positive zero of $\cos(Cu+\psi_0)$, and so, there we define $u_\text{max}=u_s$. We may define that zero to be located at $u=\frac{\pi/2-\psi_0+f\pi}{C}$, where $f$ is a given integer that causes this to be the first positive zero. Its value is always directly linked to the sign of $C$ and the value of $\psi_0$: considering $\psi_0\in]\frac{\pi}{2}+c_1\pi,\frac{\pi}{2}+(c_1+1)\pi[$, if $C>0$, then $f=c_1+1$, whereas if $C<0$, then $f=c_1$. For example, for $\psi_0\in]-\frac{\pi}{2},\frac{\pi}{2}[$ and $C>0$, we have $f=0$ and the first positive zero is located at $u=\frac{\pi/2-\psi_0}{C}$.

For the remaining analysis of the metric, we are going to consider both signs of $u_1$ separately. Considering first $u_1>0$, we find that $g_{00}$ is null whenever the conformal factor is null, and only there, thus, $g_{00}\neq 0$ at any $u<u_\text{max}$, which means there are no horizons and the metric signature remains unchanged throughout the entire spacetime. However, at the point of singularity, $u=u_\text{max}=u_s$, we have $g_{00}=0$. Furthermore, analysing the radius function, we have that $r\to 0$ as $u\to u_\text{max}$. Apart from that, regarding the existence of throats, there is the possibility of a throat, at $u_T$, and an anti-throat, at $u_{aT}$, with $u_T<u_{aT}<u_s$, or no throats. This depends on the combination of the constants $q$, $h$, $C$ and $\psi_0$, being that there are distinct behaviours according to the values of the latter. In fact, we find that when $C>0$, if $\psi_0\in[c_1\pi,\frac{\pi}{2}+c_1\pi[$, then the derivative of $r$ has no zeros in any case (no throats). For the same sign of $C$, we may split the range $\psi_0\in]\frac{\pi}{2}+c_1\pi,(c_1+1)\pi[$ in two. Accordingly, when $\psi_0\in]\frac{\pi}{2}+c_1\pi,\frac{\pi}{2}+c_1\pi+1.2284[$ the derivative of $r$ has always two zeros, corresponding to a throat and an anti-throat, respectively from lower values of $u$, and when $\psi_0\in[\frac{\pi}{2}+c_1\pi+1.2284,(c_1+1)\pi[$ we verify similar behaviour to that found in previous classes. Thus, for a given combination of $h$, $C$ and $\psi_0$, there is a critical value of $q$, $q_c$, above which there is a throat and an anti-throat and otherwise there are none, as explained in class $[1+]$ with $u_1>0$. Conversely, when $C<0$ those ranges exchange their behaviours, as now if $\psi_0\in]\frac{\pi}{2}+c_1\pi,(c_1+1)\pi]$ there are no throats and the range $\psi_0\in]c_1\pi,\frac{\pi}{2}+c_1\pi[$ is the one that may be split in two, being that when $\psi_0\in]c_1\pi,\frac{\pi}{2}+c_1\pi-1.2284[$ there is a critical value of $q$, in the same way as described before, and when $\psi_0\in[\frac{\pi}{2}+c_1\pi-1.2284,\frac{\pi}{2}+c_1\pi[$ there is always a throat and an anti-throat. In any case, we find that the former is always closer to $u=0$ (spatial infinity), than to $u=u_\text{max}$, while it is possible for the latter to be closer to each of them, or even in the middle, depending on the combination of constants.

In short, considering $u_1>0$, at $u=u_\text{max}$ there is a light-like, naked, attractive central singularity situated beyond a throat and an anti-throat in some cases, or, alternatively, there may be no throats.

Now, considering $u_1<0$, we also find that $g_{00}$ is only null at the point of singularity (first positive zero of the conformal factor, as already discussed). Regarding the radius function, it also approaches $0$ as $u\to u_\text{max}$. Similarly to the previous cases with $u_1<0$, that function has no regular minima. This was already expected from the analysis of $u_1>0$, as the range of $\psi_0$ that is now allowed does not allow any throat in that case. Thus, in this case, at $u=u_\text{max}$ there is a light-like, naked, attractive central singularity (with no throats).

\begin{figure*}[ht]
   \centering
      \includegraphics[scale=0.75]{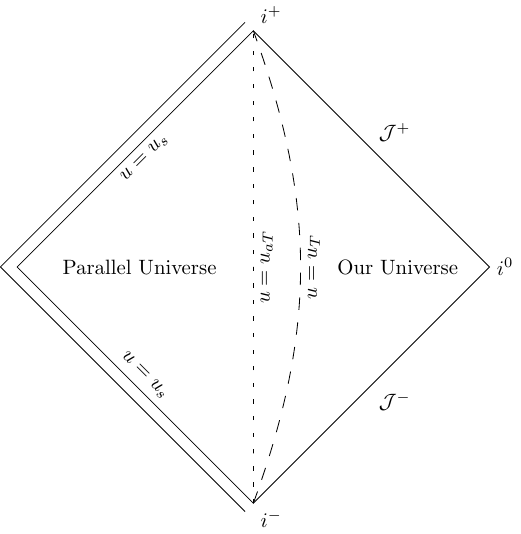}
      \hspace{1cm}
      \includegraphics[scale=0.75]{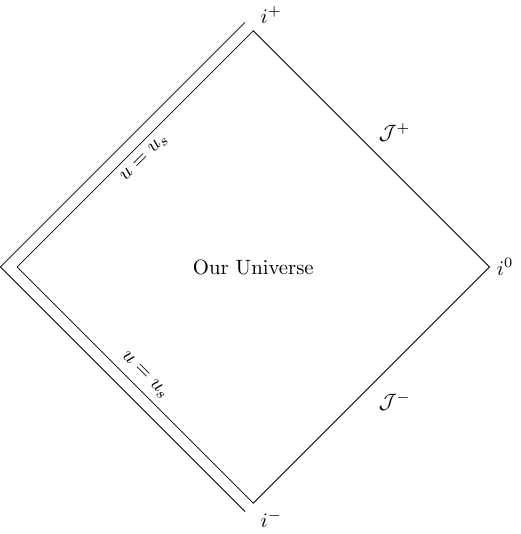}
      \caption{Left plot: Penrose diagram of a naked light-like singularity ($g_{00}=0$ at $u_s$), located beyond a throat, tilted to the right as $u_T$ is closer to $u=0$ than to $u_s$, and an anti-throat, which is in the middle, but could also be tilted to the left or to the right (always to the left of $u_T$). To the right of these structures lies our universe, with future and past flat spatial infinities. To the left, there is a parallel Universe, with future and past central light-like singularities. Right plot: Penrose diagram of a naked, central, light-like singularity solution. In this case, it lies in our own universe.}
    \label{Fig2}
\end{figure*}

At last, relatively to the Penrose diagrams of the spacetimes analysed in this class, we have that they are similar to the left plot of Fig. \ref{Fig1}, since now the naked singularity is light-like as well, however, we need to add an anti-throat and consider the possibility of no throats. Accordingly, for the former case, which may occur when $u_1>0$, we obtain the diagram of the left plot of Fig. \ref{Fig2}. For this case, we only present one of the possibilities regarding the position of the anti-throat, in the middle, relatively to spatial infinity and the singularity, as the other diagrams are obtained by drawing the anti-throat to the left (closer to $u=u_s$) or to the right (closer to $u=0$, but to the left of $u_T$). Now, when $u_1<0$ and in some cases of $u_1>0$, in which there are no throats, the diagram is depicted as the right plot of Fig. \ref{Fig2}.

Following the steps described in Appendix \ref{ap}, we have also reproduced the embedding diagrams corresponding to the Penrose diagrams in Fig. \ref{Fig2}. However, since they have very similar geometrical features to the diagrams in Figs. \ref{Emb2} and \ref{Emb3}, we have decided not to show the corresponding diagrams.

%%%%%%%%%%%%%%%%%%%%%%%%%%%%%%%%%%%%%%%%%%%%%%%%%%%%%%%%%%%%
\subsection{Class [2-]}
%%%%%%%%%%%%%%%%%%%%%%%%%%%%%%%%%%%%%%%%%%%%%%%%%%%%%%%%%%%%

Similarly to the two previous classes, this one now encompasses two different signs of $h$. However, unlike what was seen in those classes regarding $k$, now a different sign of $h$ actually leads to a different expression for the mass. Nonetheless, the remaining analysis is not affected by this, being the qualitative behaviours of the metrics respective to each sign essentially the same, apart from a particular case that is discussed at the end. Thus, we may, and will, analyse them together. Regardless, we present the line elements and corresponding $K_1$ expressions for each case below. 

Starting with the case ${h>0}$, by using the same equations as in the previous class, we obtain $k=-\sqrt{3C^2-h^2}$, which, in turn, yields the relation $|h|<\sqrt{3}|C|$. Additionally, we obtain $u_1=\pm \arcsinh (\frac{h}{q})/h$ and given the signs of $n$, $k$ and $h$ we have the following line element:
\begin{eqnarray}\label{3- u a}
    ds_J^2 &=& \cos^2(Cu+\psi_0)\Bigg\{\frac{h^2dt^2}{q^2 \sinh^2(h[u+u_1])}
		\nonumber \\    
   && \hspace{-0,6cm}
    -\frac{k^2 q^2 \sinh^2(h[u+u_1])} {h^2 \sin^2(ku)}\left[\frac{k^2 du^2}{\sin^2(ku)}+d\Omega^2\right]\Bigg\}\,.
\end{eqnarray}

Using the transformation shown in Eq. (\ref{transformation -}), since in this case $k<0$, we are able to obtain the line element with the coordinate $x$:
\begin{eqnarray}\label{3- x a}
    ds_J^2&=& \cos^2\left(C\tilde{u}+\psi_0\right)\Bigg\{\frac{h^2dt^2}{q^2 \sinh^2[h(\tilde{u}+u_1)]}
    	\nonumber \\
    && \hspace{-0,75cm}
     -\frac{q^2 \sinh^2[h(\tilde{u}+u_1)]} {h^2}\left[dx^2+(x^2+k^2)d\Omega^2\right]\Bigg\}\,,
\end{eqnarray}
where we have defined $\tilde{u} = \frac{\arccot(\frac{x}{|k|})+c_1\pi}{|k|}$, for notational simplicity.

Now, in the case ${h=0}$, we have $k=-\sqrt{3}|C|$ and $u_1=\pm  1/q$. In the coordinate u, the line element is:
\begin{eqnarray}\label{3- u b}
    ds_J^2 &=& \cos^2(Cu+\psi_0)\Bigg\{\frac{dt^2}{q^2 (u+u_1)^2}
		\nonumber \\    
    && -\frac{k^2 q^2 (u+u_1)^2} {\sin^2(ku)}\left[\frac{k^2 du^2}{\sin^2(ku)}+d\Omega^2\right]\Bigg\}\,.
\end{eqnarray}

Using the same transformation as before, we obtain the line element with the coordinate $x$:
\begin{eqnarray}\label{3- x b}
    ds_J^2 &=& \cos^2\left(C\tilde{u} +\psi_0\right)\Bigg\{\frac{dt^2}{q^2 (\tilde{u} +u_1)^2}
		~\nonumber \\    
    && -q^2 (\tilde{u} +u_1)^2\left[dx^2+(x^2+k^2)d\Omega^2\right]\Big\}\,,
\end{eqnarray}
where we have used the definition of $\tilde{u}$ given above.

Analysing the metrics with the coordinate $x$ as in the previous classes, as $x\to \infty$ and now considering also $c_1=0$, which corresponds to $u\to 0^+$, hence to $\tilde{u}\to 0^+$, we find, that $g_{00}=-g_{11}=\cos^2\psi_0$. Similarly to the previous class, when $\psi_0=\frac{\pi}{2}+c_2\pi$, where $c_2$ is an integer, that limit of $x$ does not correspond to a spatial infinity. However, in this class we are not discarding those values, since, unlike before, there are different possibilities of a spatial infinity located at other points of $x$, or $u$, and we have the freedom to choose any of them. For these cases, the analysis of the asymptotic behaviour, as well as the mass, will be carried out when the point of spatial infinity is known. 

If $\psi_0\not=\frac{\pi}{2}+c_2\pi$, then the limit $x\to \infty$ with $c_1=0$ is, indeed, a spatial infinity and both metrics are asymptotically flat, in particular Minkowskian if $\psi_0=c_2\pi$. For these cases we are already able to determine the Schwarzschild mass relative to that limit. If $h>0$ it is given by $m=\abs{\cos\psi_0}[h\coth(hu_1)+C\tan\psi_0]$ and if $h=0$ it is given by $m=\abs{\cos\psi_0}(1/u_1+C\tan\psi_0)$. In the particular case of $\psi_0=c_2\pi$, when $h>0$, the mass is $m=h\cot(hu_1)$, whereas, when $h=0$, it is $m=1/u_1$, and, by imposing $m>0$, both impose $u_1>0$. Using the expression for 
$u_1$ showed before, the latter simplifies to $m=\abs{q}$. By considering $\psi_0\neq c_2\pi$ and imposing $m>0$, when $h>0$, we get the condition $C>-h\cot(\psi_0)\coth(hu_1)$, if $\psi_0\in]c_2\pi,\frac{\pi}{2}+c_2\pi[$ (interval $1$), otherwise, if $\psi_0\in]\frac{\pi}{2}+c_2\pi,(c_2+1)\pi[$ (interval $2$), $C$ has to be strictly lower; when $h=0$, we obtain $C>-\cot(\psi_0)/u_1$, if $\psi_0$ is in interval $1$, whereas $C$ has to be strictly lower than that if $\psi_0$ is in interval $2$. 

Analysing these constraints, we find that both signs of $u_1$ are allowed. Moreover, analysing also all of the relations involving $C$, we find behaviours similar to the previous class, regarding $\psi_0$, but now the ranges to be considered are $\psi_0\in]c_2\pi,\frac{\pi}{2}+c_2\pi[$ and $\psi_0\in]\frac{\pi}{2}+c_2\pi,(c_2+1)\pi[$, instead of $\psi_0\in]\frac{\pi}{2}+c_2\pi+0.5236,\frac{\pi}{2}+(c_2+1)\pi[$ and $\psi_0\in]\frac{\pi}{2}+c_2\pi,\frac{\pi}{2}+(c_2+1)\pi-0.5236[$, respectively (see the analysis carried out in the previous class). In the following analysis, we are going to consider both signs of $u_1$, always assuming combinations of constants that guarantee $m>0$.

As before, we are now interested in analysing the Kretschmann scalars relative to both line elements with $u$. Accordingly, we may start with the term $K_1$, which is given, in each case, respectively, by:
\begin{widetext}
\begin{eqnarray} \label{K1 3-}
%\begin{split}
    K_1 &=&\frac{1}{k^4 q^2}h^2 \sec ^2(C u+\psi_0) \sin ^4(k u) \text{csch}^2(h [u+u_1]) \left\{2 h \coth (h [u+u_1]) [C \tan (C u+\psi_0)-k \cot (k u)]\right.
    \nonumber 	\\
    &&\hspace{-1cm} \left.-C \left[2 k \tan (C u+\psi_0) \cot (k u)+C \sec ^2(C u+\psi_0)\right]+2 h^2 \coth ^2(h[u+u_1])+h^2 \text{csch}^2(h [u+u_1])\right\} , \quad    \text{if } h>0\,\,,
     	\\
    K_1 &=&\frac{1}{k^4 q^2 (u+u_1)^4}\sec ^2(C u+\psi_0) \sin ^4(k u) \left\{(u+u_1) \left(C \left[2 \tan (C u+\psi_0)-C (u+u_1) \sec ^2(C u+\psi_0)\right] \right.\right.
    \nonumber  \\
   & &\left.\left.-2k \cot (k u) [C (u+u_1) \tan (C u+\psi_0)+1]\right)+3\right\}, \quad \text{if } h=0\,\,.
%\end{split}
\end{eqnarray}
\end{widetext}

When analysing these expressions, we find some results similar to those from the analysis of the previous class. In fact, by considering $\psi_0\neq \frac{\pi}{2}+c_2\pi$, in the limit $u\to 0$, which is spatial infinity, we find that both are null, as well as the remaining terms of $K$. Thus, our previous description of that limit, considering those values of $\psi_0$, is supported. Apart from this, the zeros of the conformal factor, $\cos ^2(C u+\psi_0)$, which occur at $u=\frac{\pi/2-\psi_0+c_3\pi}{C}$ ($c_3$ is an integer), also cause $K_1$ to diverge (and $r\to 0$), but now only under a certain condition. This is because, now there is the $\sin(ku)$ function, whose zeros, that occur at $u=\frac{c_4 \pi}{k}$ ($c_4$ is an integer), never cause $K_1$, and the remaining terms of $K$, to diverge and may even cause them to be null and also $r$ to approach infinity, under that same condition (this is why $u=0$ is a spatial infinity, when $\psi_0\neq \frac{\pi}{2}+c_2\pi$). In both cases, that condition refers to the respective function being null on its own, being that if both are null simultaneously, at a given point, then $K_1$, $K$ and $r$ are non-zero finite there (this is why $u=0$ is not a spatial infinity when $\psi_0= \frac{\pi}{2}+c_2\pi$; this type of point may be an double horizon), except in a particular case discussed further below, in which $r$ is zero. Apart from the zeros of the conformal factor, we also find that $K$ diverges at $u=-u_1$, which is the zero of the functions $\sinh(h[u+u_1])$ ($h>0$) and $u+u_1$ ($h=0$). In general, as explained in the previous class, this divergence does not correspond to a singularity, not being relevant, as the divergence due to the conformal factor occurs first, except in a particular case discussed at the end. Apart from these, there are no other divergences in $K$.

\begin{figure}[h!] 
    \centering
    \includegraphics[width=0.95\linewidth]{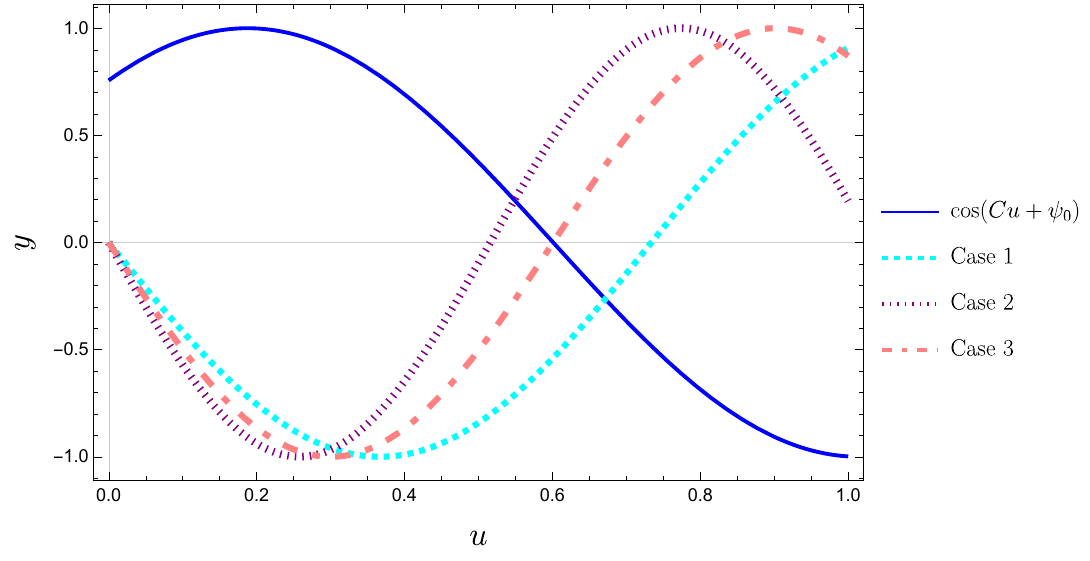}
    \caption{Representation of the three cases of class $[2-]$, in the particular case when $u=0$ is spatial infinity. The curves labelled as Case 1, 2 and 3 correspond to the $\sin(ku)$ function, in different conditions, and their interaction with the curve relative to $\cos(Cu+\psi_0)$ represent each case. 
    In all curves, we have $C=3.8$; in the one of the cosine function we consider $\psi_0$ to be given by Eq. (\ref{psi0=}), with $h=4$ and $f=0$. To distinguish between cases, we have $h=5$ for the curve of Case 1, $h=2.5$ for Case 2 and $h=4$ for Case 3.
    This plot, with appropriate adaptations, may also be used to illustrate the main cases and each subcase of class $[3-]$.}
    \label{Fig5}
\end{figure}

According to which zero occurs first, always relative to spatial infinity, we have that all the possibilities (except for that particular one) are encompassed by three cases: $\cos (C u+\psi_0)=0$ occurs first, being a central singularity (\text{case 1}); $\sin(ku)=0$ occurs first, being a second spatial infinity (\text{case 2}); both functions coincide at their first zero, corresponding to a regular surface with non-zero finite radius, from which the analysis must be continued (\text{case 3}). This phenomenon, in which the Jordan-frame metric continues beyond an endpoint of the Einstein-frame metric, is known as conformal continuation \cite{Bronnikov:2006qj}. In Figure \ref{Fig5}, examples of the interaction between these two functions, each corresponding to one of the three cases, are shown.

%\vspace{-5mm}

Furthermore, we find that the only finite points where the metric function $g_{00}$ vanishes are the zeros of the conformal factor, the other possibility being at $u\to\infty$. This means there are no horizons in case 2 and there is a light-like, naked, attractive singularity in case 1 and an double event horizon in case 3, both located at the respective first zero. Accordingly, the metric signature remains unchanged in all cases, even after that horizon.

Taking into account what was previously discussed, from this point onward we will split the analysis, regarding the value of $\psi_0$, into $\psi_0\neq \frac{\pi}{2}+c_2\pi$ and $\psi_0= \frac{\pi}{2}+c_2\pi$.

%%%%%%%%%%%%%%%%%%%%%%%%%%%%%%%%%%%%%%%%%%%%%%%%%%%%%%%%%%%%
\subsubsection{$\psi_0\neq \frac{\pi}{2}+c_2\pi$}
%%%%%%%%%%%%%%%%%%%%%%%%%%%%%%%%%%%%%%%%%%%%%%%%%%%%%%%%%%%%

As $u=0$ is spatial infinity, we have that all cases are associated with the respective first positive zero, being case 1 associated with the zero $u=\frac{\pi/2-\psi_0+f\pi}{C}<\frac{\pi}{\abs{k}}$ (where $f$ is a given integer that guarantees this expression corresponds to the first positive zero, as explained in class $[1-]$), case 2 with the zero $u=\frac{\pi}{\abs{k}}<\frac{\pi/2-\psi_0+f\pi}{C}$, and in case 3 we have $\frac{\pi/2-\psi_0+f\pi}{C}=\frac{\pi}{\abs{k}}$. This latter scenario is only achieved if:
\vspace{-2mm}
\begin{equation}\label{psi0=}
    \psi_0=-\frac{\pi(2C-\abs{k}-2f\abs{k})}{2\abs{k}}\,\,.
\end{equation}
However, this equality, and hence that correspondence, may not be allowed, since a given combination of constants may lead to $m\leq0$. Nevertheless, even discarding those combinations, it is not always possible, since it depends on $f$, which, in turn, depends on the range that $\psi_0$ is in, as discussed before, thus, its value changes accordingly. In fact, if we choose a sign for $C$ and a range for $\psi_0$, the value of $f$ gets determined (if we fix a value for $f$ instead, it is the range of $\psi_0$ that gets determined). Thus, by using it on the right-hand side of the above equation, we might obtain a value that lies outside its respective range of $\psi_0$. This means that, in such case, requiring $\psi_0$ to be equal to it, causes $f$ to change accordingly, thus altering the value of the expression, making the equality unachievable, as $\psi_0$ is always greater. In fact, it is only achievable for certain combinations of $C$ and $h$. For other combinations, however, it may happen, for example, that the possible correspondence involves the second zero of the cosine function instead of its first one, in which case, that equality would be true if there was a $(f+1)$ instead. 

We also find that, for $C>0$, case $1$ occurs when $\psi_0$ is strictly greater than the expression of Eq. (\ref{psi0=}) and case $2$ when it is strictly lower, whereas for $C<0$ it is the opposite, with only one of the cases being possible in certain combinations of $C$ and $h$, as shown in the previous example. Note that these comparisons have to be carried out inside the ranges $\psi_0\in]\frac{\pi}{2}+(f-1)\pi,\frac{\pi}{2}+f\pi[$, for $C>0$, and $\psi_0\in]\frac{\pi}{2}+f\pi,\frac{\pi}{2}+(f+1)\pi[$, for $C<0$, with $f$ being the same as in Eq. (\ref{psi0=}). 

From now on, we will analyse both signs of $u_1$ separately, as they lead to significantly different results. Starting with $u_1>0$, we have that all three cases aforementioned are possible, depending on the combination of constants. 

Regarding \text{case 1}, we find similar results as in the previous class. The analysis of the zeros of the derivative of $r$ is as described there, however, in this case, we must always consider just the values of $\psi_0$ that lead to the conformal factor being zero first, which may not allow all the ranges, associated with different possibilities, as before, depending on the combination of constants. Apart from this, now the imposition $m>0$ may also not allow that, since for certain combinations the only allowed ranges are $\psi_0\in[c_2\pi,\frac{\pi}{2}+c_2\pi[$, as aforementioned, which are exactly the values for which no throats exist. Nevertheless, for other combinations, the existence of a throat and an anti-throat is allowed. Accordingly, at $u_\text{max}=u_s=\frac{\pi/2-\psi_0+f\pi}{C}$, there is a light-like, naked, attractive central singularity situated beyond a throat and an anti-throat in some cases, which are located at $u_T$ and $u_{aT}$, with $u_T<u_{aT}<u_s$, being also possible that there are no throats. Regarding the location of those structures, relative to $u=0$ and $u=u_s$, we also find the same as in the previous class. The Penrose diagrams, relative to the spacetimes analysed in this case, are the same as those obtained in that class, in Figure \ref{Fig2}, considering also the discussion carried out before them.

Now, in \text{case 2}, we find that $u_\text{max}=\frac{\pi}{\abs{k}}$ corresponds to a second spatial infinity, as mentioned before. Using the line elements (\ref{3- x a}) and (\ref{3- x b}), at the limit $x\to -\infty$ with $c_1=0$ (same as $u$, hence $\tilde{u}$, approaching $u_\text{max}$ from below), we find that both $g_{00}$ and $g_{11}$ are constant, thus, regarding this infinity, spacetime is asymptotically flat as well, however, without the possibility of being Minkowskian. Apart from that, the case $g_{00}=-g_{11}=0$, which, in the analysis of the limit $x\to \infty$ and $c_1=0$, is a problem, now corresponds exactly to case 3, analysed below. The Schwarzschild mass relative to this second spatial infinity is similar to the one of the first spatial infinity for the range $u<0$ --- symmetric to the one shown before ---, but now with additional terms proportional to $u_\text{max}=\frac{\pi}{\abs{k}}$ (see Eq. (\ref{m2mais}), further below, for a similar expression to this case, with $h>0$, but considering the limit $x\to \infty$ with $c_1=2$, which corresponds to $u$ approaching $u=\frac{2\pi}{\abs{k}}$ from above). 

Proceeding with the analysis of the metrics with $u$, we find there are no horizons, as aforementioned, and by analysing the radius function and its derivative, we find there is always a throat, located at $u_T$. Relative to $u=0$ and $u=u_\text{max}$, we find it may be closer, in terms of $u$, to one or the other or even in the middle of them, depending on the combination of constants. Nevertheless, even in this last scenario, in which the radius function is symmetric with relation to $u_T$, spacetime is asymmetric, since $g_{00}$ does not present that symmetry. In short, this is an asymmetric two-way traversable wormhole solution. 

\begin{figure}[ht]
	\centering
	\includegraphics[scale=0.8]{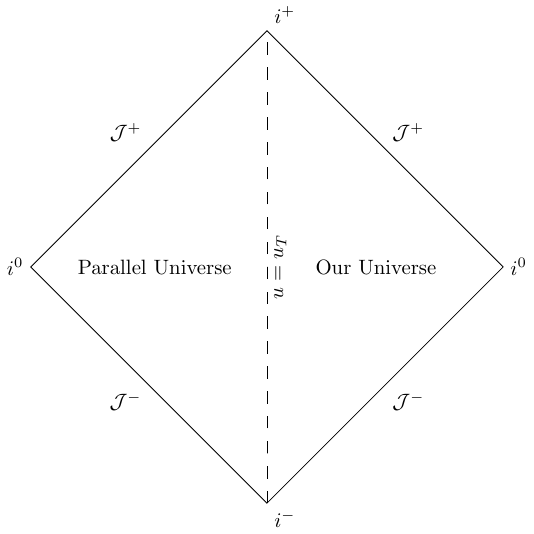}
	\caption{Penrose diagram of a two-way traversable wormhole solution. The throat is in the middle of $u=0$ and $u_\text{max}$, but there is also the possibility of it being tilted to either side. To its right lies our universe, with future and past flat spatial infinities, and a parallel universe, with similar infinities, lies to its left.}
\label{Fig3a}
\end{figure}

At last, for the Penrose diagram of this spacetime we have three possibilities, regarding the position of the throat relative to both infinities, similarly to what we have discussed for the anti-throat in the left plot of Fig. \ref{Fig2}. Nevertheless, we only present the possibility in which the throat is located in the middle, represented in Fig. \ref{Fig3a}, which is similar to the diagram of the well known Morris-Thorne wormhole \cite{Morris:1988cz}.

\begin{figure*}[ht]
	\centering
	\includegraphics[scale=0.5]{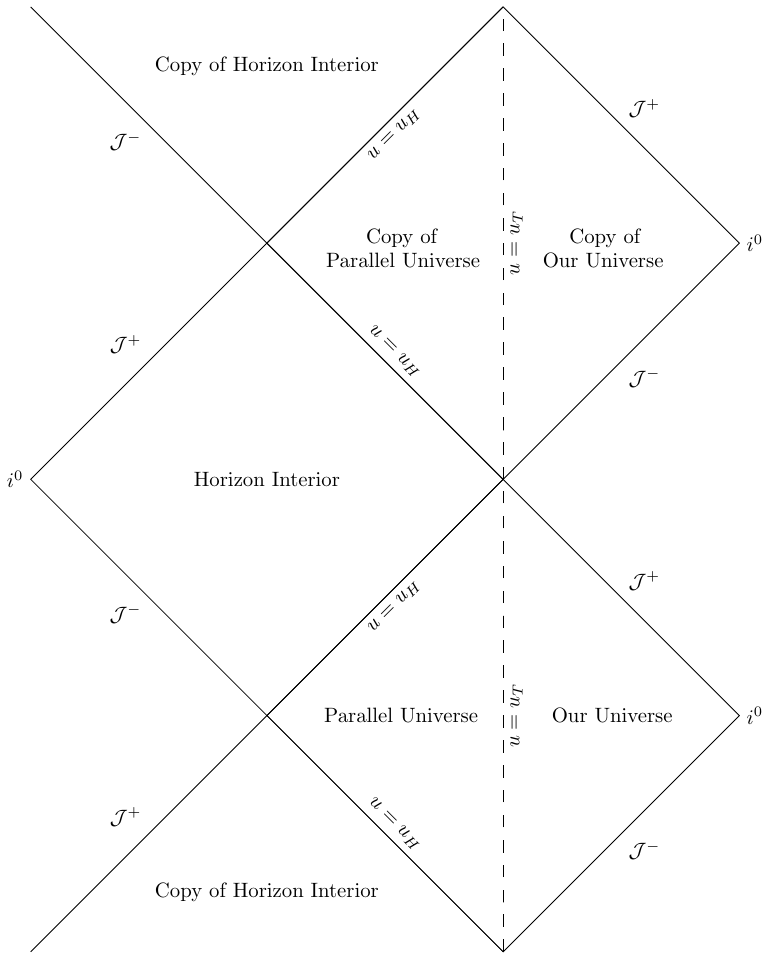}
	\hspace{1.5cm}
	\includegraphics[scale=0.5]{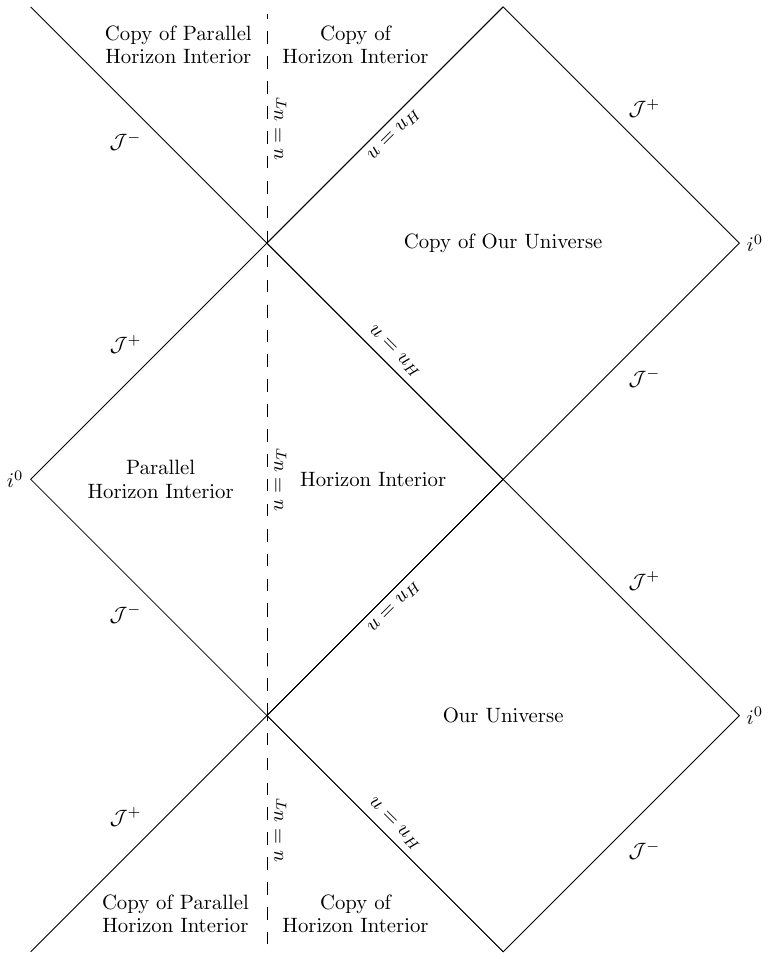}
	\caption{Left plot: Penrose diagram of space-time composed of an double event horizon (diagonal lines at $u=u_H$) inside a throat, which is in the middle of $u=0$ and $u_H$, but could also be tilted to either side. To the left of this throat, there is a parallel universe with future and past double event horizons, containing, in their interior, future and past horizons (to the right) and flat spatial infinities (to the left). There are also copies of both Universes and of the horizon interior, reached by traversing certain horizons. Right plot: Penrose diagram of a black bounce solution, composed of a throat, which is in the middle of $u_H$ and $u_\text{max}$, but could also be tilted to either side, inside of an double event horizon. Outside of it lies our universe. To the right of the throat there are future and past horizons and to its left lies a parallel horizon interior. There are also copies, similarly to the left plot.}
\label{Fig3}
\end{figure*}

Finally, in \text{case 3} we find, as aforementioned, that at $u=\frac{\pi/2-\psi_0+f\pi}{C}=\frac{\pi}{\abs{k}}$ there is an double event horizon and there we define $u=u_H$. Apart from this, analysing in more detail the radius function and its derivative, we find there is always a minimum, hence a throat, at a point $u_T$ between $u=0$ and $u=u_H$. Due to the imposition $m>0$ it is not possible to have $u_T=u_H$ in any case, thus, we always have $0<u_T<u_H$. Now, given the fact that spacetime does not end at $u_H$, we need to carry the analysis further (conformal continuation), to higher values of $u$, and we may start by verifying to which function corresponds the second zero of all. Taking into account that the first positive zero is the intersection of both functions and also that we are still considering $\psi_0\neq \frac{\pi}{2}+c_2\pi$, we find that the second positive zero is always the second zero of $\sin(ku)$, on its own, located at $u=\frac{2\pi}{\abs{k}}$. 
Thus, at that second positive zero, there is a second spatial infinity, and so, we define $u_\text{max}=\frac{2\pi}{\abs{k}}$. Analysing this spatial infinity, considering the limit $x\to -\infty$ with $c_1=1$ when analysing the line elements (\ref{3- x a}) and (\ref{3- x b}), we find results similar to those obtained in the previous case, but now considering a different value for $u_\text{max}$. Thus, considering this spatial infinity, spacetime is also asymptotically flat, and the mass, when $h>0$, is symmetric to Eq. (\ref{m2mais}), shown further below. 

As discussed before, $u_T$ is always closer to $u=0$, whereas $u_H$ is in the middle relatively to both infinities. In short, this solution describes an asymmetric spacetime that presents a second spatial infinity located past a throat and an double event horizon, respectively from the first spatial infinity. Up until now, whenever it was possible for $u=0$ to be the first spatial infinity, we always considered it as such, nevertheless, since both infinities are flat, and the metric signature is always the same, we, in fact, may have the freedom to consider any of them as the first or the second one. However, by doing that, we have to be careful about the mass. Accordingly, considering $u=\frac{2\pi}{\abs{k}}$ to be the first spatial infinity and $u=0$ to be the second one, maintaining the relation $0<u_T<u_H$, we obtain a negative mass, when determined relatively to the new first spatial infinity. Despite that, that solution describes a \text{black bounce}, since now the throat is located past the (extremal) event horizon. 

At last, for the allowed solution, in which there is a throat and then an double event horizon, we obtain the Penrose diagram depicted in the left plot of Fig. \ref{Fig3}, whereas for the black bounce solution, which is interesting to show as well, we obtain the diagram in the right plot of Fig. \ref{Fig3}. Once again, as discussed for Fig. \ref{Fig3a}, there are three possibilities for the position of the throat, but we only show the diagram with it in the middle. Furthermore, bear in mind that the metric signature remains unchanged even after the horizon (this is why there is a throat and not a bounce in the right plot of Fig. \ref{Fig3}).

Considering $u_1<0$, using the value of $\psi_0$ obtained from Eq. (\ref{psi0=}) always leads to $m<0$. Thus, by imposing $m>0$, we always have that $\psi_0$ has to be greater, for $C>0$, and lower, for $C<0$, than that expression. This way, now, only case 1 is possible. Regarding the remaining analysis, we find similar behaviour to that in the previous class, when considering $u_1<0$. Accordingly, in this case, we have $u_\text{max}=u_s=\frac{\pi/2-\psi_0+f\pi}{C}$ and, at that point, there is a light-like, naked, attractive central singularity (no throats). Note that if we were to consider $\psi \equiv 0$, we would obtain this same result, when considering $u_1>0$. At last, for this case the Penrose diagram is the same as the one shown in the right plot of Fig. \ref{Fig2}.

\begin{figure*}[ht]
   \centering
      \includegraphics[scale=0.4]{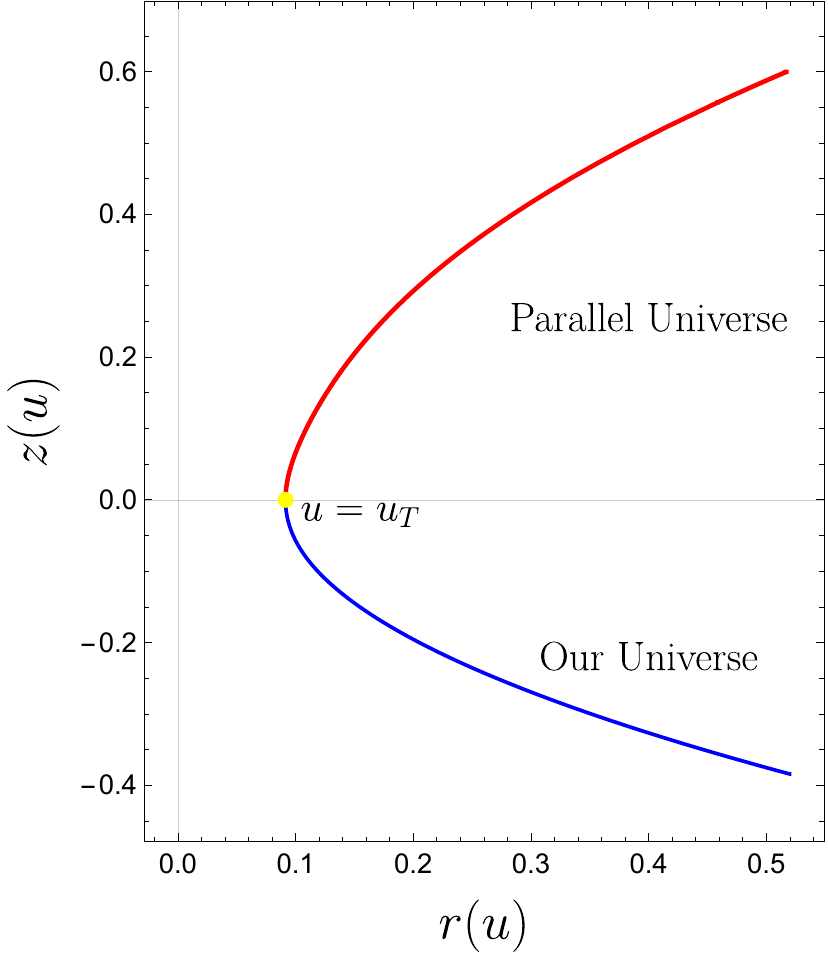}
       \hspace{1cm}
       \raisebox{0.5cm}{
      \includegraphics[scale=0.45]{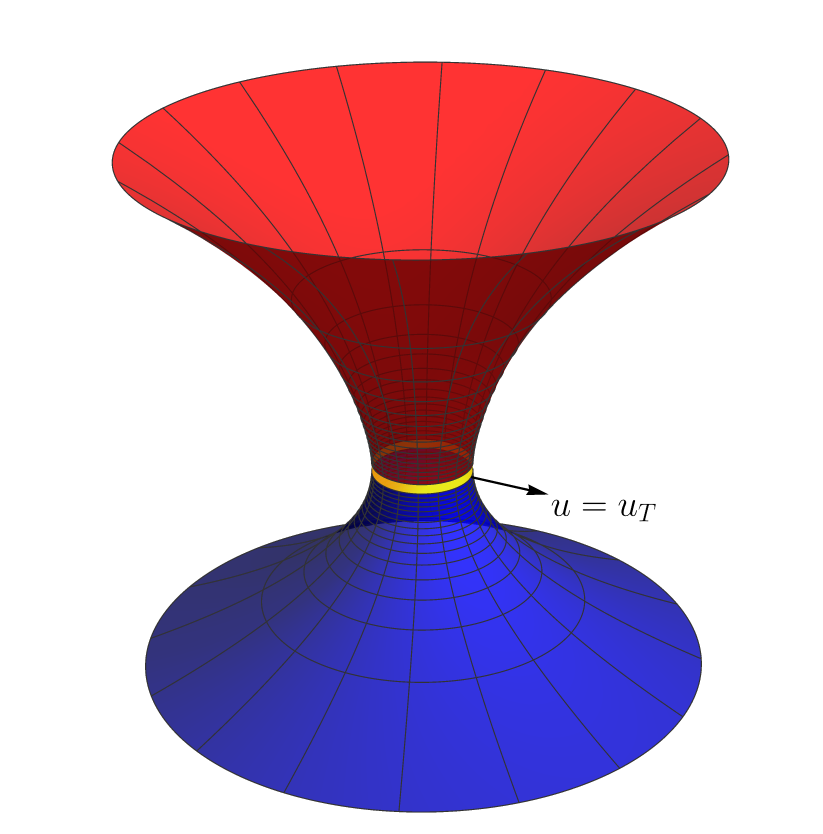}
       }
    \caption{  Left plot: $z(x)$ \textit{vs} $r(x)$ graph of a two-way traversable wormhole solution. Right plot: Embedding diagram.}
    \label{Emb6}
\end{figure*}

\begin{figure*}[ht]
   \centering
      \includegraphics[scale=0.255]{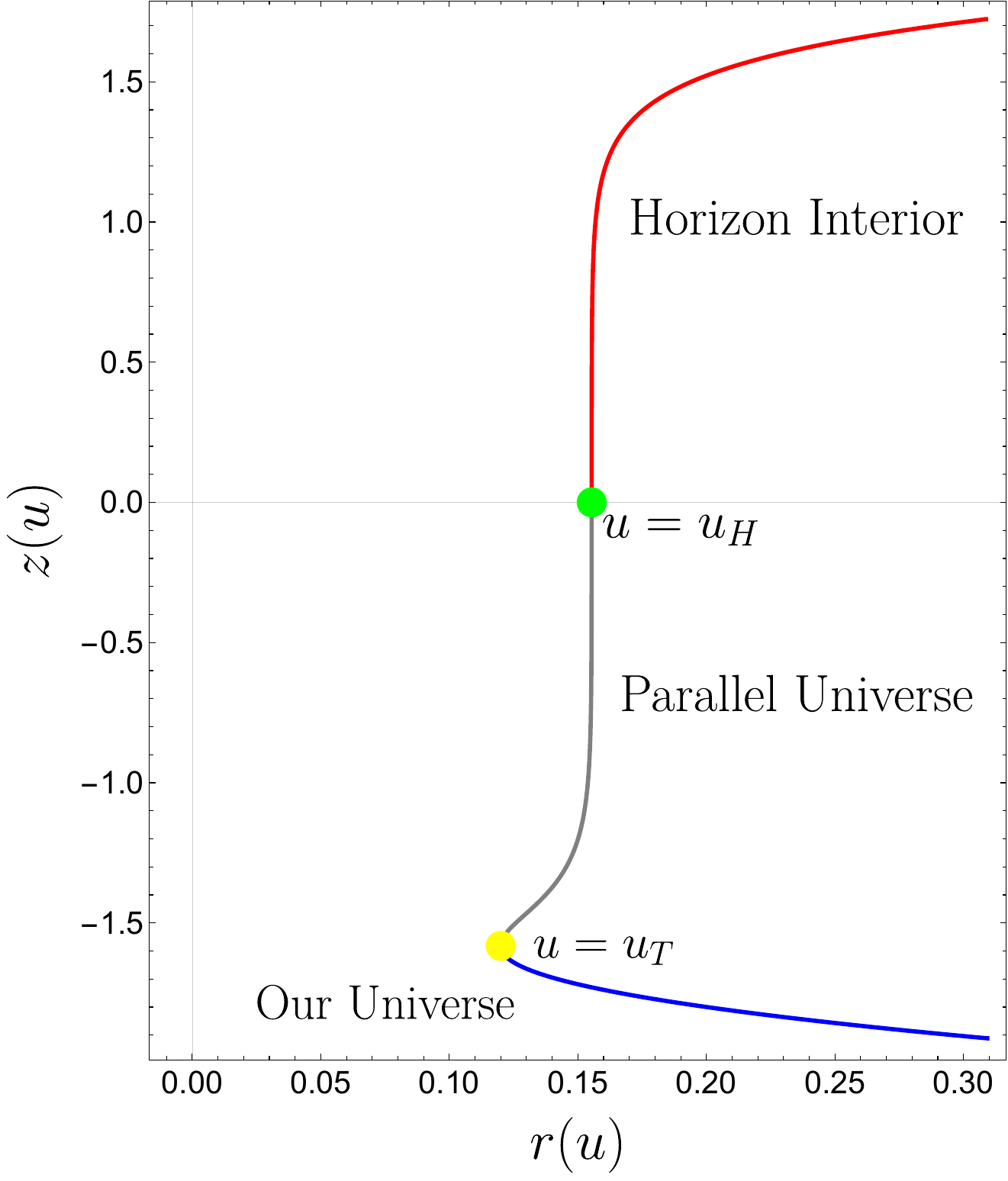}
       %\hspace{0.1cm}
      % \raisebox{-1cm}{
      \includegraphics[scale=0.45]{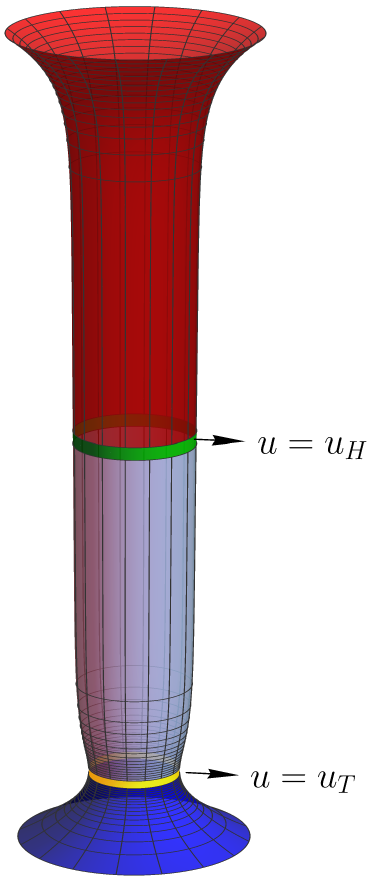}
      \hspace{0.3cm}
      \includegraphics[scale=0.25]{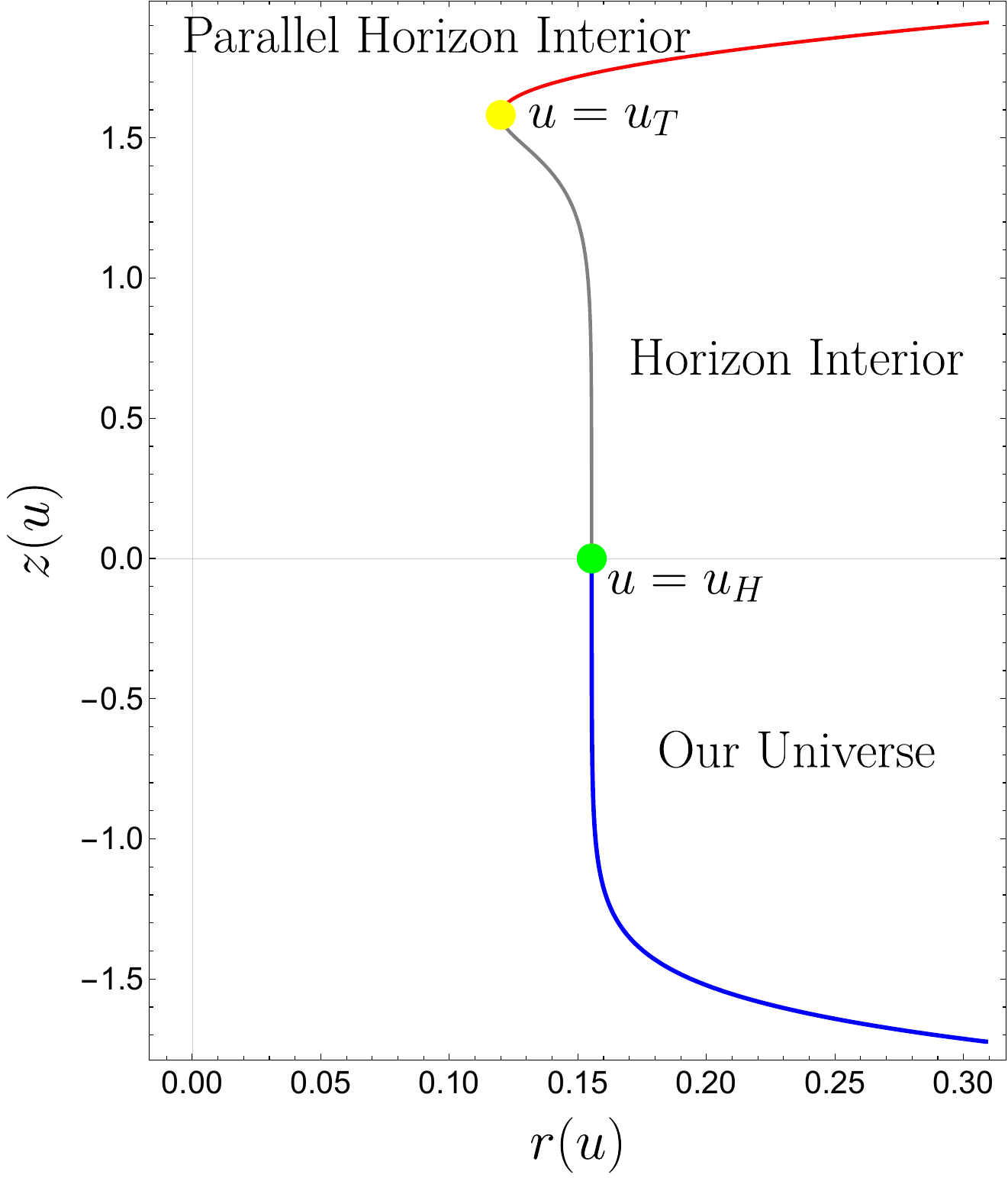}
      % \hspace{0.1cm}
      % \raisebox{-1cm}{
      \includegraphics[scale=0.41]{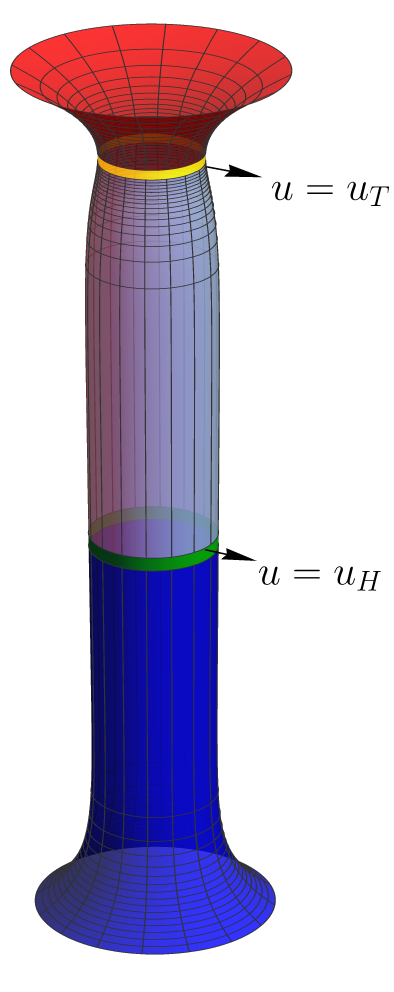}
    \caption{ Embedding diagrams developed from the Penrose diagrams in Fig. \ref{Fig3}.}
    \label{Emb7}
\end{figure*}

We also developed the embedding diagram associated with the Penrose diagram in Fig. \ref{Fig3a}, which represents a two-way traversable wormhole solution. This case is illustrated in Fig. \ref{Emb6}. The embedding diagram for Fig. \ref{Fig3} is shown in Fig. \ref{Emb7}.

%%%%%%%%%%%%%%%%%%%%%%%%%%%%%%%%%%%%%%%%%%%%%%%%%%%%%%%%%%%%%%%%%%%%%%%%%%%%%%%%%%%%%%%%%%%%%%%%%%%%%%%%%%%%%%%%%%%%%%%%%%
%%%%%%%%%%%%%%%%%%%%%%%%%%%%%%%%%%%%%%%%%%%%%%%%%%%%%%%%%%%%%%%%%%%%%%%%%%%%%%%%%%%%%%%%%%%%%%%%%%%%%%%%%%%%%%%%%%%%%%%%%%%%%%%

%%%%%%%%%%%%%%%%%%%%%%%%%%%%%%%%%%%%%%%%%%%%%%%%%%%%%%%%%%%%%%%%%%%%%%%%%%%%%%%%%%%%%%%%%%%%%%%%%%%%%%%%%%%%%%%%%%%%%%%%%%
%%%%%%%%%%%%%%%%%%%%%%%%%%%%%%%%%%%%%%%%%%%%%%%%%%%%%%%%%%%%%%%%%%%%%%%%%%%%%%%%%%%%%%%%%%%%%%%%%%%%%%%%%%%%%%%%%%%%%%%%%%%%%%%

%%%%%%%%%%%%%%%%%%%%%%%%%%%%%%%%%%%%%%%%%%%%%%%%%%%%%%%%%%%%
\subsubsection{$\psi_0= \frac{\pi}{2}+c_2\pi$}
%%%%%%%%%%%%%%%%%%%%%%%%%%%%%%%%%%%%%%%%%%%%%%%%%%%%%%%%%%%%

Now, as aforementioned, we must find a point of spatial infinity, since $u=0$ is not a possibility ($r$ is finite, in particular equal to $|C|$). As was already shown in the previous discussion, any isolated zero of the $\sin(ku)$ function may be assumed as a spatial infinity. We will, in the following analysis, consider $h>0$ and $h=0$ separately, as each require a different approach and lead to slightly different results. 

Taking into account $h>0$ first, we will consider that spatial infinity is located at the second positive zero of that function, $u=\frac{2\pi}{\abs{k}}$, in order to simplify the analysis. This choice allows all possible cases to occur, whereas if we consider, for example, the first zero instead, case 2 is not possible, due to the combination of constants, thus, we would need to choose another point of spatial infinity to account for it. Nevertheless, any zero could have been selected. 
With respect to the spatial infinity that has been chosen, there are two possible ranges of $u$: $u>\frac{2\pi}{\abs{k}}$, similar to the previous $u>0$, as we will see, and $u<\frac{2\pi}{\abs{k}}$, which is similar to the previous $u<0$. By analogy with what we have done so far, we could only analyse the former range, as the latter would lead to the same results. However, that was due to the expression for the mass being symmetric, which we need to verify in this case for the new ranges, and also because of the freedom in the choice of the constants values --- even if always under certain requirements ---, which we do not have in this case, as any value of $\psi_0$ in $\frac{\pi}{2}+c_2\pi$ leads to the same results. Thus, in this case, we must analyse both new ranges of $u$. Despite that, note that considering spatial infinity to be located at some $u<0$, such as $u=-\frac{2\pi}{\abs{k}}$ (symmetric to the one we will consider), may require certain adjustments in the analysis, however, we verify it always leads to the same final results as the respective $u>0$, %. This is, at least in part, because both the conformal factor and $\sin^2(ku)$, as is in the metric, are always symmetric with respect to $u=0$. This means that 
even if there is a spacetime located at $u>0$ and $u<0$ simultaneously. %, we may only consider the range $u>0$ when choosing the point of infinity. 
This being said, in the following analysis we are considering all the possible spacetime solutions.

Regarding the analysis of the spacetime asymptotic behaviour and its mass, it was already addressed, in part, in case 3 above with respect to its second spatial infinity, located at the same point of $u$. Accordingly, in this case, when analysing the line elements (\ref{3- x a}) and (\ref{3- x b}), we have to consider the limit $x\to \infty$ with $c_1=2$, when analysing the range $u>\frac{2\pi}{\abs{k}}$, and $x\to -\infty$ with $c_1=1$, when analysing the range $u<\frac{2\pi}{\abs{k}}$. In both cases, we obtain spacetime is also asymptotically flat. The mass, which is computed considering those same limits, is similar to the one obtained before, for $u=0$, but with additional terms due to the new position of the infinity, as aforementioned. Considering the former limit first, it is given by
\begin{eqnarray}\label{m2mais}
    m &=& \sqrt{\frac{q^2 \cos
   ^2\left(\frac{2 \pi  C}{| k| }+\psi_0 \right) \sinh ^2\left(h \left[\frac{2 \pi }{| k| }+u_1\right]\right)}{h^2}}\times
		\nonumber \\   
   && \hspace{-1cm} \times \left[h \coth \left(h \left[\frac{2 \pi }{| k| }+u_1\right]\right)+C \tan \left(\frac{2 \pi  C}{| k| }+\psi_0 \right)\right].
\end{eqnarray}
Considering the requirement $m>0$, we were able to obtain the following relations, now for $\psi_0$ instead of $C$, due to the complexity of the above expression: when $C>0$, we have $\psi_0>-\arctan\left(h \coth \left[h \left(2 \pi /| k| +u_1\right)\right]/C\right)-2 \pi  C/| k| +2 \pi  c_2$, and $\psi_0$ strictly lower than that, when $C<0$. Both relations must only be used inside a given periodic range of $\psi_0$ --- which must include the considered value of $\psi_0= \frac{\pi}{2}+c_2\pi$ ---, with a length equal to $\pi$, that is dependent on the combination of constants (determined numerically), otherwise they may not be valid. Now, considering the other limit of $x$, we obtain the mass is symmetric to the one above, as well as the relations obtained.

Proceeding with the analysis of the metric, considering first $u>\frac{2\pi}{\abs{k}}$ and $u_1>-\frac{2\pi}{\abs{k}}$, so that the zero of $\sinh(h[u+u_1])$ ($h>0$) or $u+u_1$ ($h=0$) is not in the range of $u$, we find that all cases --- 1, 2 and 3 --- are possible, and we obtain the same final results as before (the Penrose diagrams are also the same). Now, case 3 corresponds to when the second positive zero of the conformal factor and the third one of $\sin(ku)$ coincide, since these correspond to the first zeros relative to spatial infinity, and it occurs when $h=\sqrt{3}\abs{C}/2$. Cases 1 and 2 now occur when $h>\sqrt{3}\abs{C}/2$ and $h<\sqrt{3}\abs{C}/2$. The second spatial infinities of cases 2 and 3 are now located at the third and fourth zeros of $\sin(ku)$ and may be analysed as the remaining ones. 

Now, regarding $u_1<-\frac{2\pi}{\abs{k}}$, we, in fact, find that the imposition $m>0$ does not allow any of those values in any scenario, given the value of $\psi_0$. This is why we need to analyse the range $u<\frac{2\pi}{\abs{k}}$, with $u_1>-\frac{2\pi}{\abs{k}}$. Accordingly, in this range, we find that, due to the imposition on the mass, any value of $u_1$ that is allowed leads to the zero of the conformal factor being always the first to occur (case 1). Thus, in fact, we find similar behaviour to that in the case $u_1<0$ analysed before, for $\psi_0\neq \frac{\pi}{2}+c_2\pi$. Accordingly, the Penrose diagram is the same.

Apart from all of these cases, when $\psi_0= \frac{\pi}{2}+c_2\pi$, there is another interesting possibility. A similar result was obtained in \cite{Bronnikov:2019ugl, Bronnikov:2020vgg}. This is the case in which the conformal factor and the $\sin^2(ku)$ function are exactly the same, being that all of their zeros coincide. This happens whenever $h=\sqrt{2}\abs{C}$, hence, when $k=-\abs{C}$. As discussed before, the intersection of both zeros corresponds to an double horizon, which means, in this case, there is an infinite number of them, each located at $u=\frac{c_3 \pi}{\abs{C}}$. 

\begin{figure*}[ht]
	\centering
	\includegraphics[scale=0.50]{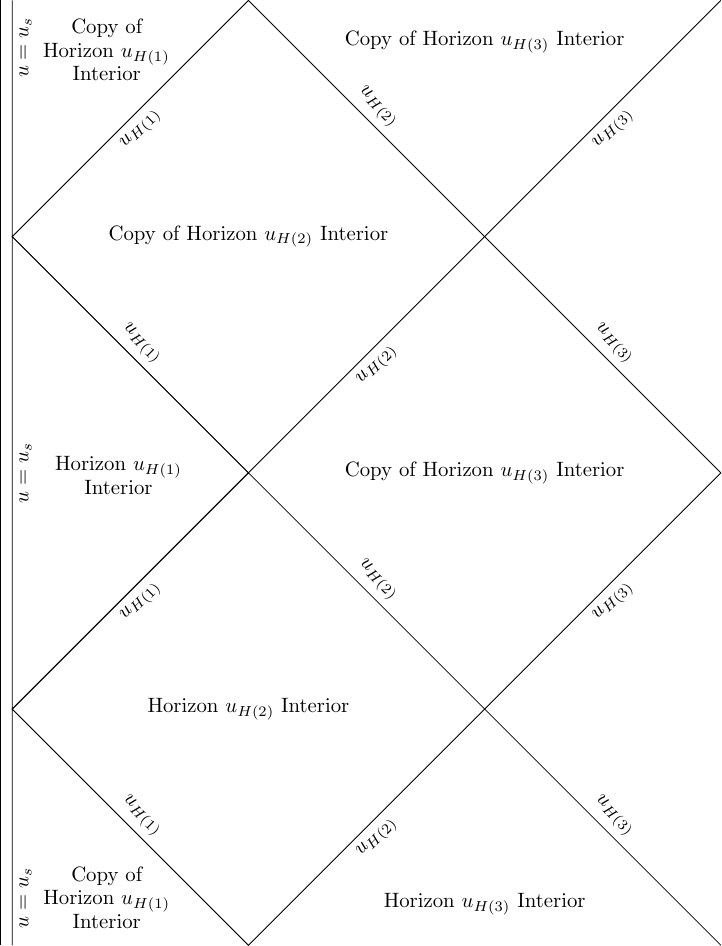}
	\hspace{1.5cm}
	\includegraphics[scale=0.50]{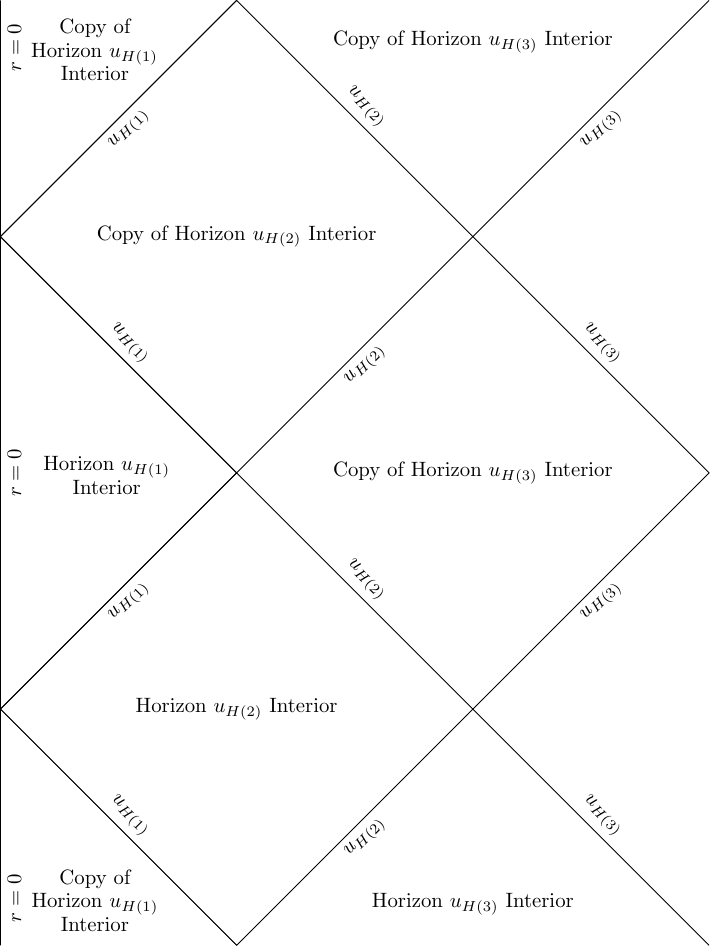}
	\caption{Left plot: Penrose diagram of a space-time composed of infinite double horizons (only three are drawn) with a time-like singularity. Inside the horizon $H(1)$, located at $u_{H(1)}$, lies a central singularity (to the left) and future and past horizons $H(1)$ (to the right), while inside a given horizon $H(i)$ ($i\in\mathbb{N}\setminus\{1\}$) lie future and past horizons located at $u_{H(i-1)}$ (to the left) and $u_{H(i)}$ (to the right). Once inside these horizons, it is also possible to reach copies of their interiors, which emerge in copies of our universe. Right plot: Penrose diagram of a space-time composed of infinite horizons with a regular centre (vertical line).}
	\label{Fig4}
\end{figure*}

We find that even at the limit $u\to -\infty$ (or $\infty$), which corresponds to $r\to \infty$, there is a horizon, as $g_{00}$ and $K$ are both null, which also means this is an asymptotically flat spacetime. Furthermore, in our case, differently from \cite{Bronnikov:2019ugl, Bronnikov:2020vgg} there is a singularity at $u=-u_1$, whether this point is positive or negative and whenever this zero does not coincide with the others. Thus, considering infinity to be at $u\to-\infty$, we may define $u_\text{max}=u_s=-u_1$ (or even $u_\text{min}=u_s=-u_1$, if we considered infinity to be at $u\to\infty$). Apart from that, $g_{00}\to \infty$ when $u\to u_s$, the metric signature remains unchanged in the entire domain and $r$ is null at this point, with no extrema points. This way, this is a time-like, repulsive, central singularity. 
On the other hand, in the case the three zeros coincide, which occurs when $q=\pm \sqrt{2}\abs{C} \csch(\sqrt{2}\pi)$, the point $u_\text{max}=-u_1$ is regular. Furthermore, $r$ is null there, but not a minimum, as its derivative does not vanish. In addition, at the limit $u\to u_\text{max}$, we have that $g_{00}\sim -\frac{C^2}{2} \sinh^2(hu_1)(u-u_\text{max})^2$, and so, it remains non-zero finite there. This way, this point corresponds to a de Sitter-like (due to the asymptotic behaviour of $g_{00}$) regular centre.
At last, the Penrose diagrams for these solutions, shown in the left and right plots of Fig. \ref{Fig4}, respectively, are similar to the one in Figure 3 of \cite{Bronnikov:2020vgg}, but now considering the singularity, which is drawn as we have seen in the middle plot of Fig. \ref{Fig1}, for example, and the regular centre, which is similar to that singularity but with a single line, instead of a double one. In our diagrams, we only represent three of the infinite horizons, being the horizon $u_{H(1)}$ associated with the first value of $u=\frac{c_3 \pi}{\abs{C}}$ to the left of $u_\text{max}$ (singularity or regular centre), $u_{H(2)}$ is associated with the second one, and so on for the remaining horizons until infinity.
In Fig. \ref{Fig4}, however, it was not possible to find all the regions of space-time in which the radicand is positive defined in Eq. \eqref{z}, which makes it impossible to construct embedding diagrams that represent the corresponding Penrose diagrams.

Considering now $h=0$, we are not able to obtain all the results taking into consideration a unique point of spatial infinity, thus, we will consider one for each case. Accordingly, for any combination of constants, to account for case 1, for both equivalents of $u_1>0$ and $u_1<0$, as before, it may be located at the first positive zero, $u=\frac{\pi}{\abs{k}}$; to account for case 2, it may be located at the second positive zero, $u=\frac{2\pi}{\abs{k}}$, being the second spatial infinity at $u_\text{max}=\frac{3\pi}{\abs{k}}$; at last, regarding case 3, it may be located at $u=-\frac{\pi}{\abs{k}}$ (the first negative zero), with the intersection of both zeros, thus, the event horizon, located at $u=0$ and the second spatial infinity at $u=\frac{\pi}{\abs{k}}$. In all three cases, spacetime is in values of $u$ higher than those mentioned for (first) spatial infinity and, regarding the coordinate $x$, all of them correspond to $x\to \infty$, but with $c_1=1$, $c_1=2$ and $c_1=-1$, respectively. All of these spacetimes are asymptotically flat, with the expression for the mass being similar to the one obtained for $u=0$, but with additional terms due to the different locations of the infinity, as seen before. Any other zero of $\sin(ku)$ could have been considered, but all the possible results are being taken into account here. Regarding case 1, we have that now both signs of $u_1$ present a similar behaviour, being similar to the cases $u_1<0$ analysed before, as no throats are possible in any of them, due to the restricted value of $\psi_0$ and also because $h$ is now null. The remaining cases lead to similar behaviour as before. The Penrose diagrams for all of them have already been obtained and discussed.

%%%%%%%%%%%%%%%%%%%%%%%%%%%%%%%%%%%%%%%%%%%%%%%%%%%%%%%%%%%%
\subsection{Class [3-]}
%%%%%%%%%%%%%%%%%%%%%%%%%%%%%%%%%%%%%%%%%%%%%%%%%%%%%%%%%%%%

In this last class, using the same equations as before, we obtain the relation $k=-\sqrt{3C^2+h^2}$, which, as in class $[1+]$, does not lead to any imposition on either $C$ or $h$, allowing for a plethora of different possible solutions. We also obtain $u_1=\pm \arcsin(\frac{h}{q})/h$, as in class $[3+]$, and in order for it to be real-valued, it is required that $\abs{q}\geq \abs{h}$. Given the signs of $n$, $k$ and $h$, we have the line element:
\begin{eqnarray}\label{4- u}
    ds_J^2 &=& \cos^2(Cu+\psi_0)\Bigg\{\frac{h^2dt^2}{q^2 \sin^2(h[u+u_1])}
		\nonumber \\    
    && \hspace{-0,4cm}
     -\frac{k^2 q^2 \sin^2(h[u+u_1])} {h^2 \sin^2(ku)}\left[\frac{k^2 du^2}{\sin^2(ku)}+d\Omega^2\right]\Bigg\}\,.
\end{eqnarray}

Using the transformation of Eq. (\ref{transformation -}), we obtain the line element with the coordinate $x$: 
\begin{eqnarray}
    ds_J^2 &=&\cos^2\left(C\tilde{u}+\psi_0\right)\left\{\frac{h^2dt^2}{q^2 \sin^2[h(\tilde{u}+u_1)]}\right.
    \nonumber \\
   && \hspace{-0,6cm}
   \left.-\frac{q^2 \sin^2[h(\tilde{u}+u_1)]} {h^2}\left[dx^2+(x^2+k^2)d\Omega^2\right]\right\},
\end{eqnarray}
where we have defined, once again, $\tilde{u} = \frac{\arccot(\frac{x}{|k|})+c_1\pi}{|k|}$, for notational simplicity.

Analysing this metric as $x\to \infty$, when $c_1=0$, so that it corresponds to the limit $u\to 0^+$, hence to $\tilde{u}\to 0^+$, we find $g_{00}=-g_{11}=\cos^2\psi_0$. As in the previous class, we will consider the cases $\psi_0\neq\frac{\pi}{2}+c_2\pi$ and $\psi_0=\frac{\pi}{2}+c_2\pi$ ($c_2$ is an integer) separately, since in the former the previous limit of $x$, hence the point $u=0$, corresponds to spatial infinity, being spacetime asymptotically flat, in particular Minkowskian if $\psi_0=c_2 \pi$, whereas in the latter that limit does not correspond to a spatial infinity ($r$ is finite, in particular equal to $|C|$). Despite that, it is possible to find other points of $u$, or points of $x\to \infty$ with other values of $c_1$, that may be considered as such, as analysed before, and so we must also not discard this case. In the former case, the Schwarzschild mass, considering that limit, is given by $m=\abs{\cos\psi_0}(h\cot(hu_1)+C\tan\psi_0)$. In the particular case of $\psi_0=c_2\pi$, it is given by $m=h\cot(hu_1)$, which, by imposing $m>0$, imposes $u_1>0$. If $\psi_0\not=c_2\pi$, by imposing $m>0$, we obtain the condition $C>-h\cot(\psi_0)\cot(hu_1)$ if $\psi_0\in]c_2\pi,\frac{\pi}{2}+c_2\pi[$, otherwise, if $\psi_0\in]\frac{\pi}{2}+c_2\pi,(c_2+1)\pi[$, $C$ has to be strictly lower. Analysing these expressions, we find that both signs of $u_1$ are allowed. Furthermore, we find similar behaviour to that in the previous classes, being that the allowed or forbidden ranges of $\psi_0$, depending on the sign of $u_1$, are the same as in the previous class. Nevertheless, their expressions are different, which means we must always take them into account. As always, in the following analysis, we will consider both signs of $u_1$, always assuming combinations of constants that guarantee $m>0$.

Proceeding with the analysis, we are now interested in analysing the Kretschmann scalar relative to the metric with $u$, being that the term $K_1$ is given by
\begin{eqnarray}
    K_1&=&\frac{1}{k^4 q^2}h^2 \sec ^2(C u +\psi_0) \sin ^4(k u) \csc ^2(h [u+u_1]) 
    \nonumber \\
   && \quad
   \times \left\{-C^2 \sec ^2(C u +\psi_0)-2 k \cot (k u)
    \right.
    \nonumber \\
   && \quad \left.
    \times [C \tan (C u +\psi_0)+h \cot (h [u+u_1])]\right.
    \nonumber \\
   &&\left. + h \left[2 C \tan (C u +\psi_0) \cot (h [u+u_1])
	   \right.\right.
    \nonumber \\
   &&\left.\left. 
   +3 h \csc ^2(h [u+u_1])-2h\right]\right\}\,.
\end{eqnarray}

By analysing this expression at spatial infinity, considering $\psi_0\neq\frac{\pi}{2}+c_2\pi$, we find it is null, as well as the remaining terms of $K$, which supports our previous description of it. Furthermore, as in the previous class, we find that the zeros of the conformal factor, $\cos^2(C u +\psi_0)$, located at $u=\frac{\pi/2-\psi_0+c_3\pi}{C}$, where $c_3$ is an integer, may cause $K_1$ to diverge (and $r$ to be null), however, only under a certain condition. Apart from these zeros, in this case, we have that the ones of the $\sin(h[u+u_1])$ function, located at $u=\frac{c_4 \pi}{h}-u_1$, where $c_4$ is an integer, may cause $\csc (h [u+u_1])$ and $\cot (h [u+u_1])$ to diverge, hence leading to a divergence of $K_1$ (and $r$ to be null), under that same condition. This condition, as before, is related to the $\sin(ku)$ function. Its zeros, located at $u=\frac{c_5 \pi}{k}$, where $c_5$ is an integer, never cause $K_1$, as well as the remaining terms of $K$, to diverge, and may even cause them to be null and $r$ to go to infinity, also under that same condition, which refers to the former two functions being zero at distinct points from this one. Accordingly, if all functions are zero on their own or if $\cos(C u +\psi_0)$ and $\sin(h[u+u_1])$ are null at the same point, but not coinciding with the other function, then the above description for each zero holds; if a zero of $\sin(ku)$ and of one of the other functions coincide, then $K_1$, $K$ and $r$ are non-zero finite there; if all of them coincide, we find the same behaviour for $K$, but now $r$ is null at that point. Furthermore, as in class $[3+]$, different signs of $u_1$ do not lead to different functions diverging. When analysing the remaining terms of $K$, we do not find any other points of divergence, apart from the ones discussed.

Now, to organize the analysis, we will divide all possible solutions into three major cases, according to which zero, between the functions $\cos(C u +\psi_0)$ and $\sin(h[u+u_1])$, occurs first, relative to spatial infinity: $\cos (C u+\psi_0)=0$ occurs first (\text{case 1}); $\sin(h[u+u_1])=0$ occurs first (\text{case 2}); both functions coincide at their first zero (\text{case 3}). We can use Figure \ref{Fig5}, with appropriate adaptations (the $\sin(ku)$ function may be seen as $\sin(h[u+u_1])$), to visualize the interactions between these two functions, corresponding to the three main cases.  \\

%\newpage
Apart from this, within each case, we still need to compare the respective first zero with the first one (relative to spatial infinity) of $\sin(ku)$. Accordingly, each of the cases may be subdivided into three subcases: the zero associated with the main case occurs first (\text{A}), being a central singularity; $\sin(ku)=0$ occurs first (\text{B}), being a second spatial infinity; both first zeros coincide, corresponding to all three in case 3 (\text{C}), either corresponding to a regular centre, or surface with non-zero finite radius from which the analysis must be continued (conformal continuation). In this last subcase we will also find different possibilities for the second zero, being even possible to reach a third one in some cases. Nevertheless, we will not further subdivide this subcase for now, but we will do so when we analyse it in more detail, as it is not that straightforward. Once again, we can use Figure \ref{Fig5}, with appropriate adaptations (the $\cos(Cu+\psi_0)$ function may also be seen as $\sin(h[u+u_1])$ and each case in it must be understood as subcase A, B or C, respectively), to visualize the interactions between the functions of each subcase.

In all subcases, even if there is a horizon, which is going to be extremal, the metric signature remains unchanged throughout the entire spacetime. Apart from this, we find that, at zeros of $\cos (C u+\psi_0)$, and only there, $g_{00}$ vanishes, unless these coincide with the ones of $\sin(h[u+u_1])$, in which case $g_{00}$ is non-zero finite, with its derivative being negative when approaching the zero from the left. In particular, this means there are no horizons in any subcase B; when a singularity occurs at those zeros, which happens, for example, in subcases 1A (former type of zero) and 3A (latter type of zero), it is attractive, being also light-like in the former and time-like in the latter; whenever a zero of $\cos (C u+\psi_0)$ only coincides with $\sin(ku)$ (regular point of $u$ with non-zero finite $r$), which happens, for example, at the first zero of case 1C, there is an double horizon.
Furthermore, at zeros of $\sin(h[u+u_1])$ that do not coincide with $\cos (C u+\psi_0)$, and only there, we have that $g_{00}$ goes to infinity. Thus, when a singularity occurs at those points, which happens, for example, in subcase 2A, it is time-like and repulsive; when the first of those points coincide with $\sin(ku)$ (regular point of $u$ with non-zero finite $r$), which occurs, for example, at the first zero of case 2C, in \cite{Bronnikov:1999wh, Bronnikov:2024uyb} the analysis ends there and it is said that a singularity-free hornlike structure emerges \cite{Banks:1992mi}, with the ``end of the horn" repelling test particles, being infinitely remote and having an asymptotically constant radius. Despite that, in our analysis we will not consider the radius to be asymptotically constant, as we will analyse the metric beyond that point, even if it is repulsive and infinitely distant, since it is neither a singularity, nor a second spatial infinity, nor a regular centre. Thus, we will not take that type of structure into account; instead, we will simply consider that intersection of zeros, whenever it occurs, as a regular repulsive surface. Apart from all of these points, we need to analyse each case in more detail, which is done further below. 

We will split the remaining analysis, according to the value of $\psi_0$, into $\psi_0\neq \frac{\pi}{2}+c_2\pi$ and $\psi_0= \frac{\pi}{2}+c_2\pi$. 

\subsubsection{$\psi_0\neq \frac{\pi}{2}+c_2\pi$}

The first zero, relative to spatial infinity, at $u=0$, of the conformal factor is at $u=\frac{\pi/2-\psi_0+f\pi}{C}$ (where $f$ is a given integer that guarantees this expression corresponds to the first positive zero, as explained in class $[1-]$), of $\sin(h[u+u_1])$ is at $u=\frac{\pi}{\abs{h}}-u_1$ if $u_1>0$, and at $u=-u_1$ if $u_1<0$, and of $\sin(ku)$ is at $u=\frac{\pi}{\abs{k}}$. This means that in case 3 we have $u=\frac{\pi/2-\psi_0+f\pi}{C}=\frac{\pi}{\abs{h}}-u_1$ or $u=\frac{\pi/2-\psi_0+f\pi}{C}=-u_1$, depending on the sign of $u_1$. That is achieved, respectively, when:
\begin{align} \label{case3 condition}
    C=\frac{| h|  (2 f\pi -2 \psi_0+\pi )}{2 (\pi -u_1 | h| )}\quad \text{or} \quad C=\frac{-2 f\pi  +2 \psi_0-\pi }{2 u_1} \,\,.
\end{align}
On the contrary to Eq. (\ref{psi0=}), these equalities can always occur, as now we have a condition on $C$ rather than on $\psi_0$. Despite that, depending on the combination of constants, it may happen that these equalities are not allowed, as they lead to $m\leq0$, resulting in only case 1 or 2 being possible. Nevertheless, there are combinations that lead to $m>0$ as well, allowing case 3 to occur. As explained in class $[1-]$, the value of $f$ changes according to the sign of $C$, changing the value of the above expression. This does not pose any issue, as by using $f$ relative to $C>0$ the above expression is also positive and by using $f$ relative to $C<0$ it is negative. Thus, we find that for each sign of $C$, it retains that same sign. Taking that into account, we find that case 1 occurs when $\abs{C}$ is strictly greater than the above expression, while case 2 occurs when it is strictly lower. 

We will, from this point onward, analyse each main case individually. As we did before, we will analyse both signs of $u_1$ separately, as they lead to different results, but this time, we will separate them within each case, since both of them allow for the three main cases. 

\begin{center}
    Case 1: $\frac{\pi/2-\psi_0+f\pi}{C}<\frac{\pi}{\abs{h}}-u_1$\,\,\,\, or\,\,\,\, $\frac{\pi/2-\psi_0+f\pi}{C}<-u_1$
\end{center}

This case has a lot of similarities to the previous class, $[2-]$, for both signs of $u_1$.

Considering $u_1>0$ first, we find that it allows all subcases, as before, however, in 1C, unlike case 3 from $[2-]$, there are more possibilities regarding the second zero. In that subcase, in the first zero, we have $\frac{\pi/2-\psi_0+f\pi}{C}=\frac{\pi}{\abs{k}}$, which is only possible if Eq. (\ref{psi0=}) is verified. Considering that value for $\psi_0$, there are certain combinations of constants that are not allowed, as they lead to $m\leq0$. Apart from this, we now must always ensure that $C$ has the correct values for case 1, taking into account the discussion performed after Eq. (\ref{case3 condition}). Note that this equation works just as a threshold in this case, thus, Eq. (\ref{psi0=}) depending on the value of $C$ does not pose any problem. Now, by discarding combinations that are not allowed, the correspondence of both zeros is always possible, by choosing the right $\psi_0$, unlike what we found in the previous class, due to the new relation between $C$ and $k$. Subcases 1A ($\frac{\pi/2-\psi_0+f\pi}{C}<\frac{\pi}{\abs{k}}$) and 1B ($\frac{\pi/2-\psi_0+f\pi}{C}>\frac{\pi}{\abs{k}}$) exhibit the same conditions for occurrence as cases 1 and 2 from the previous class, respectively, concerning the sign of $C$ and the relation between the value of $\psi_0$ and Eq. (\ref{psi0=}). 

Regarding subcases 1A and 1B, we find similar results (the Penrose diagrams as well), to those found in cases 1 and 2 from the previous class ($u_1>0$), thus, further analysis is not necessary. Accordingly, in 1A, at $u_\text{max}=u_s=\frac{\pi/2-\psi_0+f\pi}{C}$ there is a light-like, naked, attractive central singularity situated beyond a throat and an anti-throat in some cases, located at $u_T$ and $u_{aT}$, respectively, with $u_T<u_{aT}<u_s$ ($u_{aT}$ may be closer to either endpoint, or in the middle); it is also possible that there are no throats. In subcase 1B, $u_\text{max}=\frac{\pi}{\abs{k}}$ corresponds to a second flat spatial infinity, existing always a throat at $u_T$, that may be closer to either infinity, or in the middle of them, and so, this is a two-way traversable wormhole solution (asymmetric as seen before).

Regarding subcase 1C, as aforementioned, we find an double event horizon at the first zero, where we define $u=u_H$. As before, we need to analyse the metric beyond that point. However, unlike in the previous class, where the second zero was always the second of $\sin(ku)$ (1Cb), now it may also be the first one of $\sin(h[u+u_1])$ (1Ca), or even the intersection of both (1Cc). In this latter subcase, we have $\frac{\pi}{\abs{h}}-u_1=\frac{2\pi}{\abs{k}}$, which is only verified if:
\begin{align} \label{3- case1C}
    C=\pm\sqrt{\frac{3 \pi ^2 h^2-h^4 u_1^2-2 \pi  h^3 u_1}{3 h^2 u_1^2+6 \pi  h u_1+3 \pi^2}}\,\,.
\end{align}
Note that the value for $C$ obtained from this expression is always adequate for the current case. Nevertheless, when considering it, there are certain combinations that lead to $m\leq0$, not being allowed. We find that subcase 1Ca occurs when $\abs{C}$ is strictly lower than that, while subcase 1Cb occurs when it is strictly greater. We need to be careful to only use values of $C$ that correspond to case 1, in particular in subcase 1Ca, as lower values of $\abs{C}$ may correspond to case 2 or 3.

\begin{figure*}[ht]
   \centering
      \includegraphics[scale=0.45]{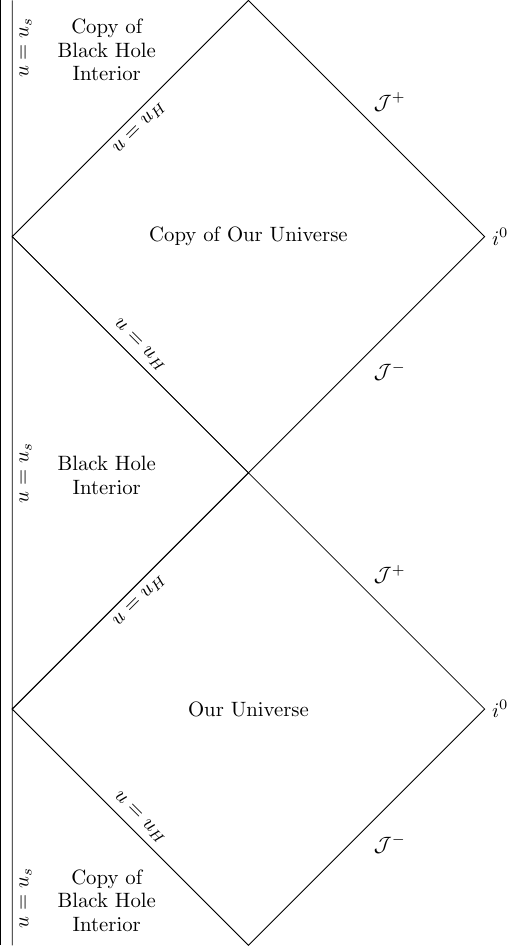}
      \hspace{0.5cm}
      \includegraphics[scale=0.45]{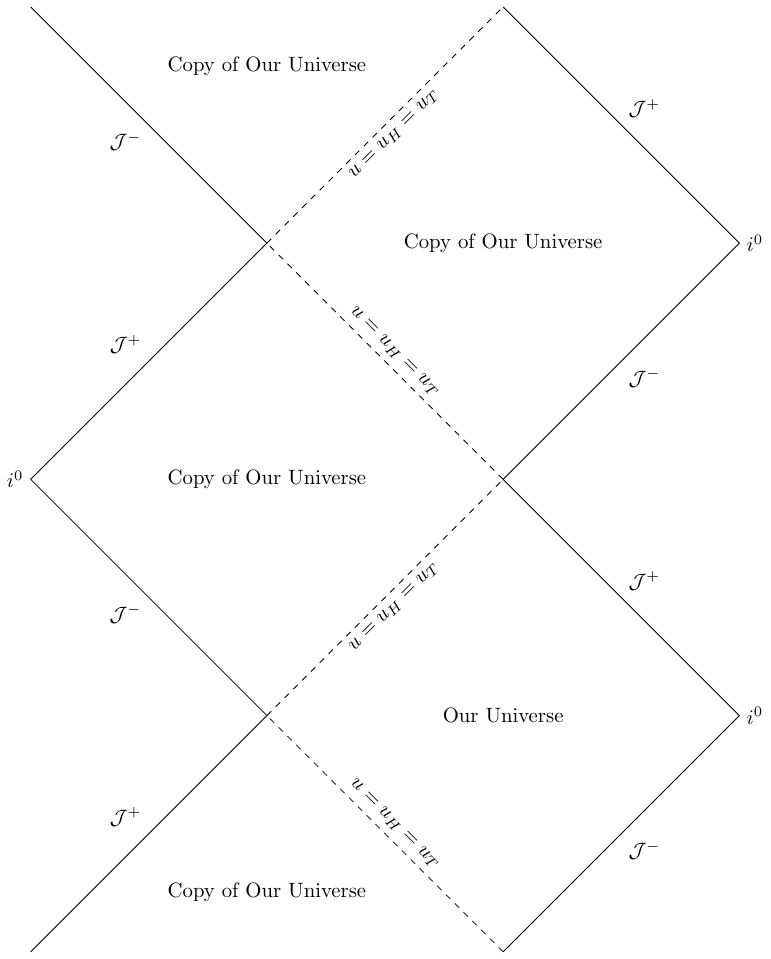}
  	\hspace{0.5cm}
      \includegraphics[scale=0.45]{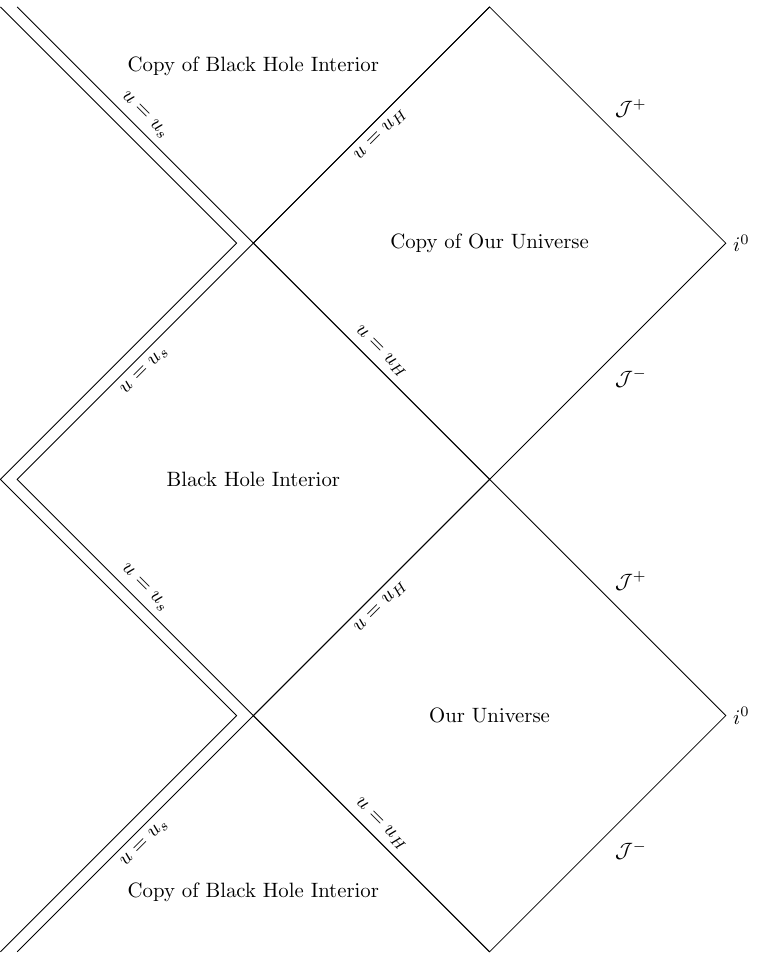}
	  \caption{Left plot: Penrose diagram of a black hole solution with a central, time-like singularity. In our universe are future and past flat spatial infinities (to the right) and double event horizons (to the left), containing, in their interior, future and past horizons (to the right) and a time-like singularity (to the left). There are also copies, reached by traversing certain horizons. Middle plot: Penrose diagram of a one-way traversable wormhole. The throat coincides with the horizon, making it only one-way traversable and being known as extremal-null throat (diagonal dashed lines). Outside of it (to its right) lies our universe, as usual, with future and past throats (to the left), and inside (to its left) is a copy of it (with future and past flat spatial infinities, to the left, and throats, to the right), as spacetime is symmetric with respect to $u_T$. By traversing certain throats, other copies are reached. Right plot: Penrose diagram of a black hole solution with a central, light-like singularity.}
   \label{Fig6}
\end{figure*}

In subcase 1Ca, as discussed before about the zeros of this sine function, we find at $u_\text{max}=u_s=\frac{\pi}{\abs{h}}-u_1$ a time-like and repulsive central singularity. By analysing the radius function and its derivative, we find there are no throats in this subcase. Thus, bearing in mind there is a horizon at $u_H$, this is a black hole solution. Regarding the Penrose diagram of the analysed spacetime, shown in the left plot of Fig. \ref{Fig6}, we find it is the same as that of the Reissner-Nordstr\"{o}m solution, for $q^2=m^2$, which also presents an double event horizon and a time-like singularity \cite{dInverno:1992gxs}.

Subcase 1Cb is the one that most resembles case 3 from class $[2-]$. In fact, as before, there is a second flat spatial infinity at $u_\text{max}=\frac{2\pi}{\abs{k}}$ (see the analysis performed in the previous class about this infinity) and also a throat at $u_T$. The difference is that now, by analysing $r$ and its derivative, we find this throat may be located at any point between $u=0$ and $u_\text{max}$. This way, there are three different possibilities: when $0<u_T<u_H$ we have the solution analysed in that class; when $u_H<u_T<u_\text{max}$ this is a black bounce solution, which, unlike before, is now allowed; when $u_T=u_H$ it is an extremal null throat, thus, this is a symmetric (as in this case all metric functions are symmetric with respect to $u=u_T=u_H$) one-way traversable wormhole solution. For the two first possibilities, the Penrose diagrams have already been obtained, being those in the left and right plots of Fig. \ref{Fig3}, respectively, considering the same discussion about the position of the throat. For the latter possibility it is the one shown in the middle plot of Fig. \ref{Fig6}, with the extremal null throat drawn as a dashed line. In this case, as space-time is symmetric relative to $u_T=u_H$, we have after the throat a copy of our universe, instead of a parallel universe, as we have find until now. A similar spacetime and diagram were obtained in \cite{Simpson:2018tsi} (see Figure 2 there).

At last, in subcase 1Cc, at $u=\frac{\pi}{\abs{h}}-u_1=\frac{2\pi}{\abs{k}}$, a regular repulsive surface emerges. As discussed before, we will analyse the metric beyond it. Accordingly, once again, we have to find which zero comes next. Due to the impositions on $\psi_0$ and $C$ that lead to this subcase, we have that the third zero is always the second one of the conformal factor. By analysing the radius function and its derivative, we find there are no throats in this subcase. This means that, at $u_\text{max}=u_s=\frac{\pi/2-\psi_0+(f+1)\pi}{C}$, there is a light-like and attractive central singularity, situated beyond an double event horizon and that repulsive surface. This type of surface is not represented in Penrose diagrams, thus, the diagram of this solution, shown in the right plot of Fig. \ref{Fig6}, is the same as that of a black hole solution with a light-like singularity and an double event horizon. %(the metric signature remains unchanged even after it).

Now, considering $u_1<0$, we may also use Eq. (\ref{psi0=}) to distinguish between subcases. However, similarly to class $[2-]$, only subcase 1A is possible. This is due to the imposition $m>0$, as before, and also because now we must always ensure that $C$ has the correct values for case 1. Regarding the remaining analysis, we find similar behaviour to that found in the previous class, for the same sign of $u_1$ (same Penrose diagram), which means there are no throats and at $u_\text{max}=u_s=\frac{\pi/2-\psi_0+f\pi}{C}$ there is a light-like, naked, attractive central singularity.

%%%%%%%%%%%%%%%%%%%%%%%%%%%%%%%%%%%%%%%%%%%%%%%%%%%%%%%%%%%%%%%
\begin{center}
    Case 2: $\frac{\pi/2-\psi_0+f\pi}{C}>\frac{\pi}{\abs{h}}-u_1$\,\,\,\, or\,\,\,\, $\frac{\pi/2-\psi_0+f\pi}{C}>-u_1$
\end{center}

Considering $u_1>0$ first, all three subcases are allowed. In subcase 2C, at the first zero, there is the intersection $\frac{\pi}{\abs{h}}-u_1=\frac{\pi}{\abs{k}}$, which is only possible if:
\begin{align} \label{3- case2C}
    C=\pm\sqrt{\frac{-h^4 u_1^2-2 \pi  h^3 u_1}{3 h^2 u_1^2+6 \pi  h u_1+3 \pi^2}}\,\,.
\end{align}
When using the value obtained from this expression, we must be careful, as certain combinations of constants lead to $m\leq0$, and we must also ensure that it holds for case 2, regarding the discussion performed after Eq. (\ref{case3 condition}). We have that subcase 2A ($\frac{\pi}{\abs{h}}-u_1<\frac{\pi}{\abs{k}}$) occurs when $\abs{C}$ is strictly lower than the above expression, while subcase 2B ($\frac{\pi}{\abs{h}}-u_1>\frac{\pi}{\abs{k}}$) occurs when it is strictly greater.

Regarding subcase 2A, as discussed before, at $u_\text{max}=u_s=\frac{\pi}{\abs{h}}-u_1$ there is a time-like, naked, repulsive central singularity. By analysing the radius function and its derivative, we find it is possible that there are a throat and an anti-throat or, alternatively, no throats at all, depending on the combination of constants. Unlike what we found in class $[1-]$, now the entire range of $\psi_0$ that allows zeros of the derivative, for each sign of $C$, has an associated critical value (now there are no values of $\psi_0$ for which there are always a throat and an anti-throat), such as $q_c$, as explained in that class. Regarding the Penrose diagrams, we have the same as in the right and middle plots of Fig. \ref{Fig1}, for the case with a throat and an anti-throat and for the case without throats, respectively.

In subcase 2B, we find similar behaviour to that found in subcase 1B. In particular, at $u_\text{max}=\frac{\pi}{\abs{k}}$ there is a second flat spatial infinity, always beyond a throat at $u_T$, that may be closer to either infinity, or in the middle of them. In this latter case, there are combinations of constants for which not even the radius function is symmetric, however, there are others for which it is, as well as the remaining metric functions, unlike before, hence corresponding to a symmetric spacetime. This way, this subcase describes either an asymmetric or a symmetric two-way traversable wormhole solution. For the Penrose diagram of the asymmetric case we obtain the same as in Fig. \ref{Fig3a}, considering the same discussion about the position of the throat, while in the symmetric case the diagram is similar to that Figure, but now with ``Copy of Our Universe" instead of ``Parallel Universe".

In subcase 2C, as aforementioned and similarly to subcase 1Cc, a regular repulsive surface emerges at $u=\frac{\pi}{\abs{h}}-u_1=\frac{\pi}{\abs{k}}$. As before, we will analyse the metric beyond it, thus, we have to find to which function corresponds the second positive zero of all. In this case, we find it may be the first one of $\cos(Cu+\psi_0)$ (2Ca), the second one of $\sin(ku)$ (2Cb), or the intersection of both (2Cc). This latter subcase, in which $\frac{\pi/2-\psi_0+f\pi}{C}=\frac{2\pi}{\abs{k}}$, is only possible if: 
\begin{align} \label{3- case2Cc}
    \psi_0=-\frac{\pi(4C-\abs{k}-2f\abs{k})}{2\abs{k}}\,\,.
\end{align}
Once again, we must be careful when using this expression, as its value, depending on the combination of constants, may lead to $m\leq0$. Note that here $C$ must always be the one from Eq. (\ref{3- case2C}). Taking that into account, we find that the equality $\abs{q}=\abs{h}$ leads to $\psi_0= \frac{\pi}{2}+f\pi$, which we will only consider later. This way, considering only allowed combinations, the above correspondence is always possible, by choosing the right value of $\psi_0$. We also find that subcases 2Ca and 2Cb follow the same conditions for occurrence as subcases 1A and 1B, respectively. However, once again, we must ensure that $C$ is right for case 2, even being fixed by Eq. (\ref{3- case2C}), as the threshold expression  (\ref{case3 condition}) is affected by the value of $\psi_0$.

Regarding subcase 2Ca, we find at $u_\text{max}=u_s=\frac{\pi/2-\psi_0+f\pi}{C}$ a light-like, naked, attractive central singularity, situated past that repulsive surface. By analysing the radius function and its derivative, we find there are no throats. As that surface is not represented in Penrose diagrams, as aforementioned, in this case we have the same diagram as in the right plot of Fig. \ref{Fig2}.

In subcase 2Cb, at $u_\text{max}=\frac{2\pi}{\abs{k}}$ there is a second flat spatial infinity. By analysing the radius function, and its derivative, we find there is always a throat at $u_T$, that may be closer to either infinity, or in the middle of them, which means it can occur either before, after, or simultaneously with that repulsive surface. When $u_T$ is located in the middle, we find that spacetime is symmetric with relation to it, otherwise it is asymmetric. Regarding the Penrose diagram of this subcase, we find it is similar to the one shown in Fig. \ref{Fig3a}, considering the same discussion about the position of the throat, but now with ``Copy of Our Universe" instead of ``Parallel Universe" when it is located in the middle (symmetric case).

At last, in subcase 2Cc, at $u=u_H=\frac{\pi/2-\psi_0+f\pi}{C}=\frac{2\pi}{\abs{k}}$, there is an double event horizon, as described before. As spacetime is regular here, we must, once again, analyse the metric beyond this point. This way, we have to find to which function corresponds the third zero of all. Due to the impositions on $C$ and $\psi_0$ that lead to this subcase, we have that the third zero is always the second one of $\sin(h[u+u_1])$. In fact, it can also be its correspondence with the third one of $\sin(ku)$ when considering $\abs{q}=\abs{h}$, however, this was already discarded before. By analysing the radius function and its derivative, we find there are no throats in this subcase. Accordingly, at $u_\text{max}=u_s=\frac{2\pi}{\abs{h}}-u_1$, there is a time-like and repulsive central singularity, situated beyond that repulsive surface and an double event horizon. This way, the Penrose diagram is the same as that of a black hole with a time-like singularity, as shown in the left plot of Fig. \ref{Fig6}.

\begin{figure*}[ht]
   \centering
      \includegraphics[scale=0.31]{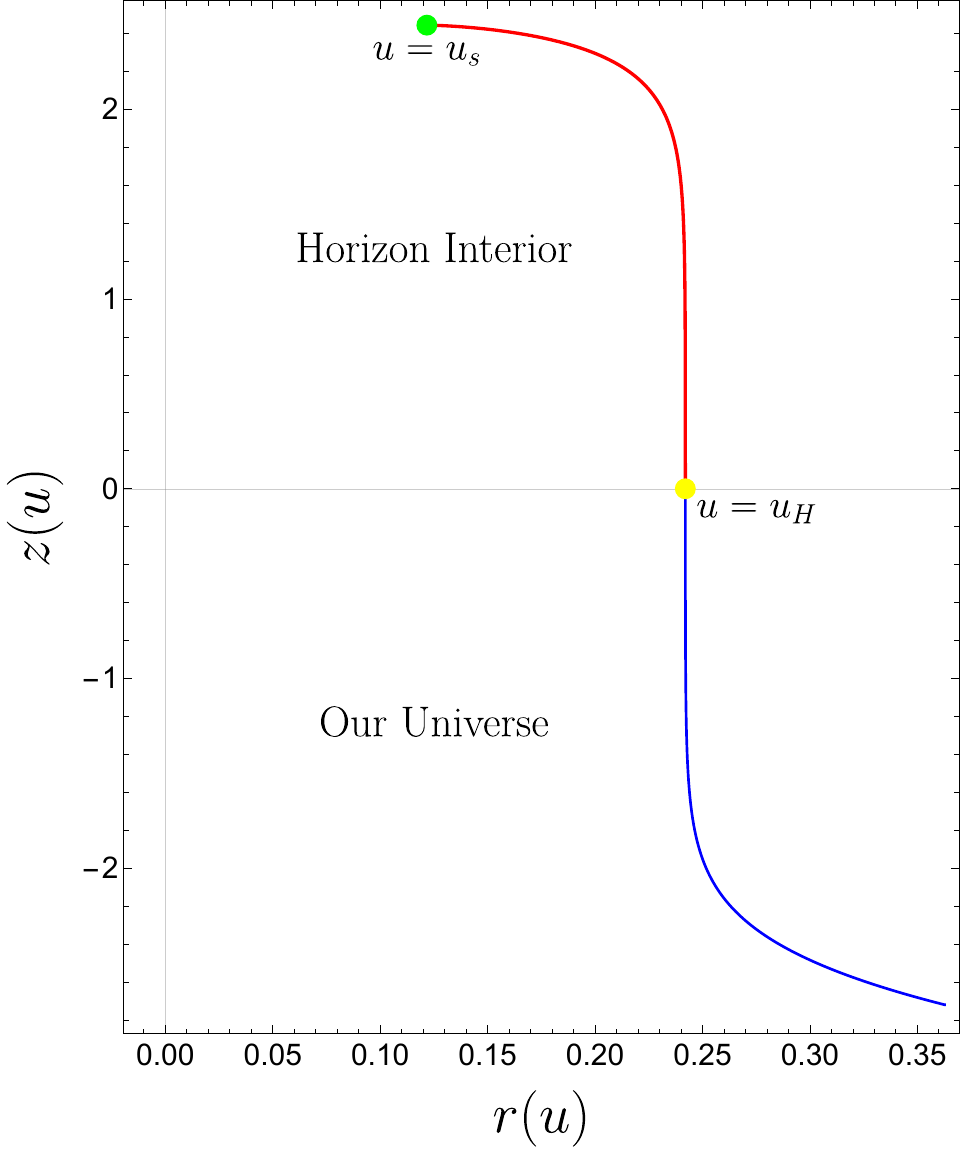}
       \hspace{0.25cm}
      \includegraphics[scale=0.53]{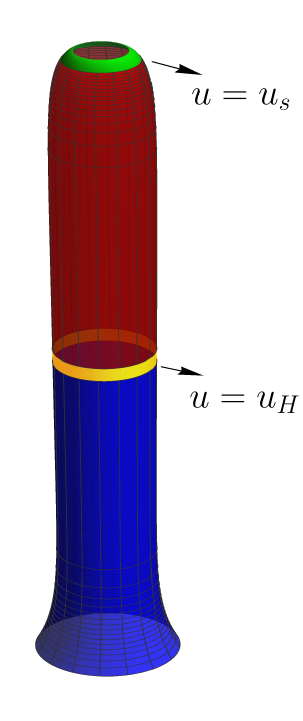}
       \hspace{1.0cm}
      \includegraphics[scale=0.35]{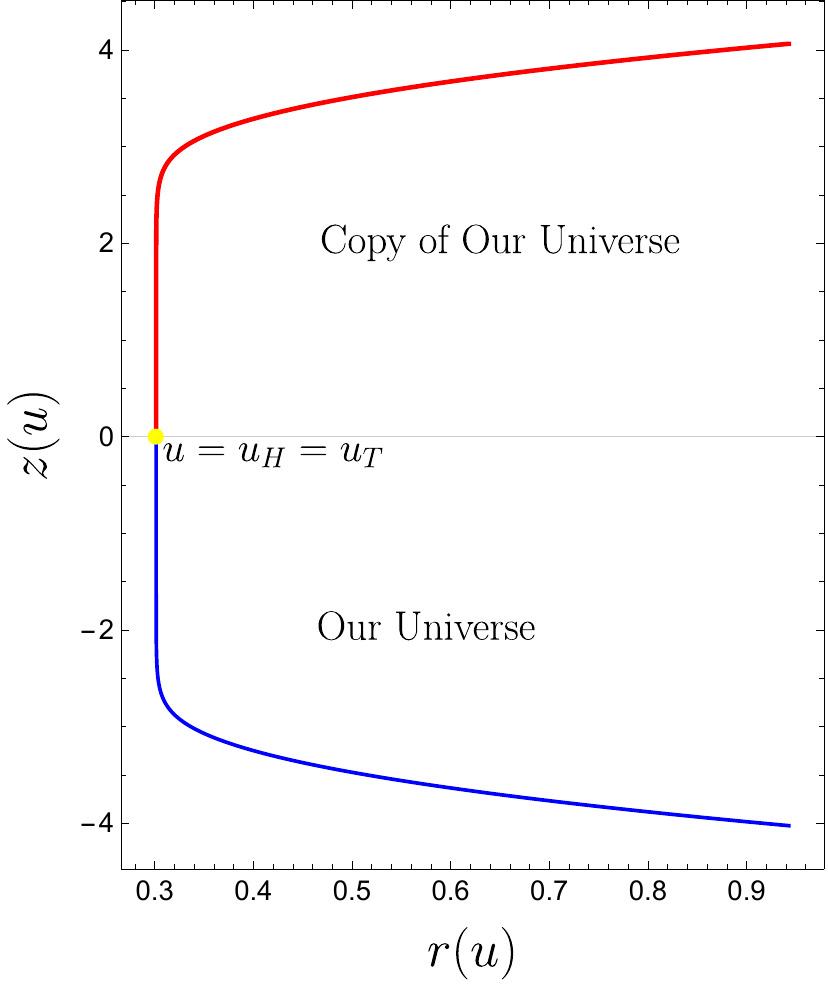}
       \hspace{0.5cm}
      % \raisebox{-1cm}{
      \raisebox{0.3cm}{\includegraphics[scale=0.34]{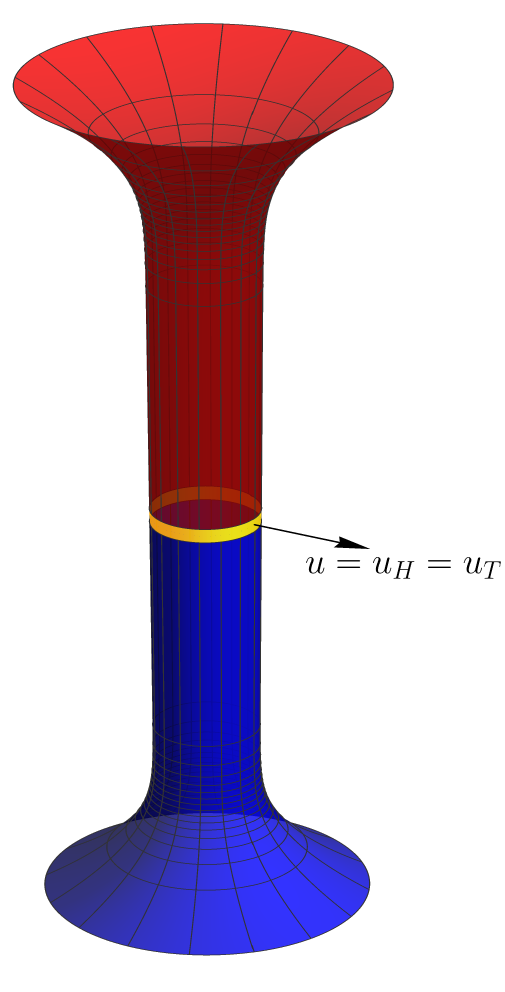}}
    \caption{ Embedding diagrams developed from the Penrose diagrams in Fig. \ref{Fig6}.}
    \label{Emb8}
\end{figure*}

%%%%%%%%%%%%%%%%%%%%%%%%%%%%%%%%%%%%%%%%%%%%%%%%%%%%%%%%%%%%%%%%%
Finally, using the same procedures described in Appendix~\ref{ap}, we have constructed the embedding diagrams corresponding to the Penrose diagrams on the left and center of Fig.~\ref{Fig6}, omitting the rightmost diagram due to its similarity with the left one. The resulting embedding diagrams are shown in Fig.~\ref{Emb8}.

%%%%%%%%%%%%%%%%%%%%%%%%%%%%%%%%%%%%%%%%%%%%%%%%%%%%%%%%%%%%%%%%%%%%%%%%%%%%%%%%%%%%%%
%%%%%%%%%%%%%%%%%%%%%%%%%%%%%%%%%%%%%%%%%%%%%%%%%%%%%%%%%%%%%%%%%%%%%%%%%%%%%%%%%%%%%%%%%%%%%%%%%%%%%%%%%%%%%%%%%%%%%%%%%%%%%%%
%%%%%%%%%%%%%%%%%%%%%%%%%%%%%%%%%%%%%%%%%%%%%%%%%%%%%%%%%%%%%%%%%
Now, when considering $u_1<0$, similarly to the previous case, we find that only subcase 2A is possible. Subcase 2C now would occur when $-u_1=\frac{\pi}{\abs{k}}$, which would only be possible if:
\begin{align} \label{3- case2C u1-}
    C=\pm\sqrt{\frac{\pi ^2-h^2 u_1^2}{3 u_1^2}}\,\,.
\end{align}
This can be used as a threshold for the other subcases, as we have done with Eq. (\ref{3- case2C}). Note that when using this expression for $C$, certain combinations of constants may lead to $m\leq0$. Furthermore, once again, we must always ensure that it is valid for case 2. Accordingly, by imposing $m>0$ and adequate values of $C$ for case 2, we have that only subcase 2A is possible now. Regarding the analysis of $r$, we also find similar behaviour to that found in the previous case with $u_1<0$. This way, at $u_\text{max}=u_s=-u_1$ there is a time-like, naked, repulsive central singularity (no throats). The Penrose diagram is the same as in that case.

\begin{center}
    Case 3: $\frac{\pi/2-\psi_0+f\pi}{C}=\frac{\pi}{\abs{h}}-u_1$\,\,\,\, or\,\,\,\, $\frac{\pi/2-\psi_0+f\pi}{C}=-u_1$
\end{center}

In this case we will always impose $C$ to be equal to Eq. (\ref{case3 condition}), as discussed before.

Starting with case $u_1>0$, we find that the three subcases are possible. Subcase 3C occurs when all first zeros coincide, thus, $\frac{\pi/2-\psi_0+f\pi}{C}=\frac{\pi}{\abs{h}}-u_1=\frac{\pi}{\abs{k}}$, which is only possible if:
\begin{align} \label{3- case3C}
    \psi_0=\frac{1}{2} (2 f\pi  +\pi )\pm \sqrt{\frac{-h^2 u_1^2-2 \pi  h u_1}{3}}\,\,.
\end{align}
Once again, we have to be careful, as certain combinations of constants, using this expression for $\psi_0$, may not lead to $m>0$. In fact, in particular, we find that $\abs{q}=\abs{h}$ causes $m=0$. Since $C$ depends on $\psi_0$, we find that each sign in the above expression changes its value, being its sign the same as that. Bearing in mind the discussions on $f$ (which changes according to the sign of $C$ and the range that $\psi_0$ is in) and the one performed after Eq. (\ref{psi0=}), in which we found that the sign of $C$ changes the behaviour of that threshold, we actually find that subcase 3A occurs when $\psi_0$ lies between the value with the $``-"$ sign and the one with the $``+"$ sign, otherwise, but always inside the ranges of interest of $\psi_0$, according to the value of $f$ as discussed before, it is subcase 3B that occurs.

In subcase 3A, by analysing the radius function and its derivative, we find there are no throats. As discussed before, there is a time-like, naked, attractive central singularity at $u_\text{max}=u_s=\frac{\pi/2-\psi_0+f\pi}{C}=\frac{\pi}{\abs{h}}-u_1$. The Penrose diagram is the same as that shown in the right plot of Fig. \ref{Fig2}.

In subcase 3B we find similar behaviour to that found in 1B.  Accordingly, $u_\text{max}=\frac{\pi}{\abs{k}}$ corresponds to a second flat spatial infinity and there is always a throat at $u_T$, that may be closer to either infinity, or in the middle of them, but not even in this case spacetime is symmetric, and so, this is an asymmetric two-way traversable wormhole solution. The Penrose diagram is the same as that shown in Fig. \ref{Fig3a}, considering also the discussion about the position of the throat that precedes that figure.

Regarding subcase 3C, by analysing the radius function and its derivative, we find there are no extrema points. Accordingly, since we find at the first zero, which is the intersection of the three functions, a regular point, where $r$ is null, as aforementioned, this solution describes a regular centre, without horizons or throats. This way, at this point, we define $u_\text{max}$. Furthermore, in the limit $u\to u_\text{max}$, we have $g_{00}\sim c\, (u-u_\text{max})^2$, where $c$ is a real, positive constant whose value depends on the other constants. Thus, this is an Anti-de Sitter-like regular centre.

\begin{figure}[ht]
	\centering
	\includegraphics[scale=0.80]{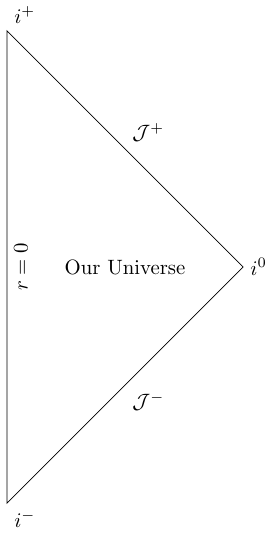}
	\caption{Penrose diagram of a regular centre solution. To the right are the future and past flat spatial infinities of our universe.}
	\label{Fig7}
\end{figure}

Despite being a regular point, this solution does not require further analysis. The Penrose diagram is similar to that of a time-like naked singularity, shown in the middle plot of Fig. \ref{Fig1}, but now without the singularity (double line), as shown in Fig. \ref{Fig7}. 
Although it is also possible to create an embedding diagram for the one illustrated in Fig. \ref{Fig7}, it would not introduce any innovations over the one already shown in Fig. \ref{Emb2} and is therefore omitted for reasons of redundancy.

At last, for $u_1<0$, we find that only subcase 3A is possible. Subcase 3C would occur when  $\frac{\pi/2-\psi_0+f\pi}{C}=-u_1=\frac{\pi}{\abs{k}}$, which would only be possible if:
\begin{align} \label{3- case3C u1-}
    \psi_0=\frac{1}{2} (2 f\pi  +\pi )\pm \sqrt{\frac{-h^2 u_1^2+\pi^2}{3}}\,\,.
\end{align}
The discussion carried out after Eq. (\ref{3- case3C}) is still valid in this case. Accordingly, by imposing $m>0$, as always, we find that only subcase 3A is possible now. By performing the remaining analysis we find similar behaviour to that in 3A with $u_1>0$, which means that, at $u_\text{max}=u_s=\frac{\pi/2-\psi_0+f\pi}{C}=-u_1$, there is a time-like, naked, attractive central singularity (no throats), being the Penrose diagram the same.

\begin{figure}[ht]
	\centering
	\includegraphics[scale=0.475]{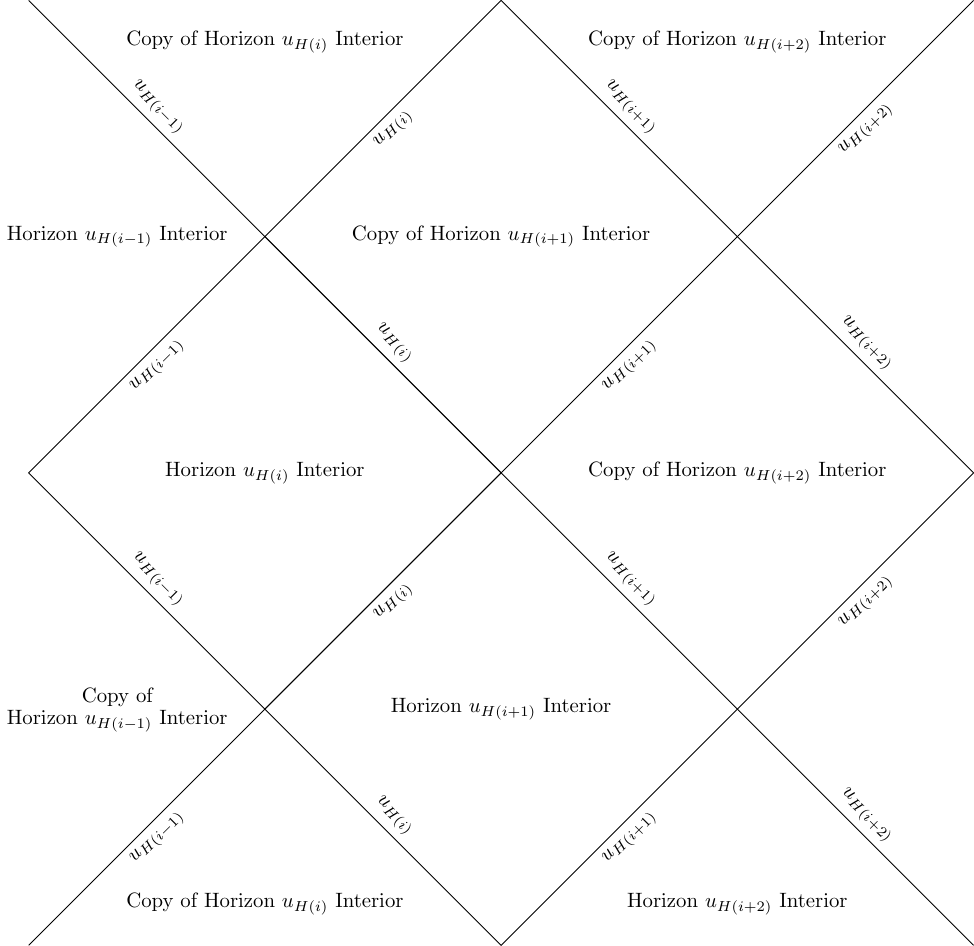}
	\caption{Penrose diagram of a space-time composed of infinite double horizons (only four are drawn), with no singularity or regular centre, unlike before. Inside a given horizon $H(i)$ ($i\in\mathbb{Z}$), located at $u_{H(i)}$, lie future and past horizons $H(i-1)$ (to the left) and $H(i)$ (to the right), being also possible to reach copies of their interiors, which emerge in copies of our universe.}
	\label{Fig7b}
\end{figure}

\subsubsection{$\psi_0= \frac{\pi}{2}+c_2\pi$}

Now, considering these particular values of $\psi_0$ we may perform an analysis similar to that at the end of the previous class. Accordingly, we have to choose another point of first spatial infinity, as $u=0$ is not one in this case. For that, we may choose any of the isolated zeros of the $\sin(ku)$ function, which correspond to flat spatial infinities, as analysed before. 

Similarly to what we have done for the case $h=0$ of the previous class, and unlike the case $h>0$, now, due to the complexity of the metric and the analysis (as now there are three trigonometric functions to consider), in order to account for all the possible solutions, we have to find an adequate point of infinity for each case and subcase, as there is not a unique one that is appropriate for all of them. Following this approach, despite $\psi_0$ being fixed to a unique type of values, in this case and thanks to the relations between all the constants, we, in fact, are able to find the same solutions found for $\psi_0\neq \frac{\pi}{2}+c_2\pi$. 

However, as in the previous class, these values of $\psi_0$ lead to a unique type of solution, not found before, which, however, is non-physical. It is the solution that was mentioned in subcase 2C, when $\abs{q}=\abs{h}$ (see the discussion performed after Eq. (\ref{3- case2Cc})). In this case, we find that any odd zero of $\sin(ku)$ coincide with a zero of $\sin(h[u+u_1])$, while any even zero of the former coincide with a zero of $\cos(Cu+\psi_0)$. Accordingly, at the former zeros regular repulsive surfaces emerge, while at the latter there are double horizons. Apart from this, by analysing the radius function, apart from not finding a spatial infinity, we find it remains constant in the entire domain of $u$, which is clearly non-physical. Despite this, this is a solution with infinite horizons and repulsive surfaces. For the Penrose diagram, as these surfaces are not represented there, we find it is the same as one of a spacetime with infinite horizons, which is shown in Fig. \ref{Fig7b}. Each horizon in that figure represented by $u_{H(i)}$, for example, is associated with a unique value of $u=\frac{c_5 \pi}{k}$.

\begin{table*}[ht]
    \centering
    \caption{Summarized results for all the cases in each class, showing the existence or absence of each space-time feature considered in our analysis. In class $[3-]$, subcase 3C, the absence of all structures indicates the existence of a regular centre solution. Non-physical cases were excluded. In cases in which it is possible for there to be throats and anti-throats, or none of them, only the scenario with both of them is being shown. The solutions of infinite horizons, found in classes $[2-]$ and $[3-]$, are also not represented here.}
    \label{tab:chi}
    \begin{tabular}{|c|c|c|c|c|c|c|c|c|c|c|c|c|}
    \cline{6-13}
        \multicolumn{5}{c|}{} & \makecell{Regular \\ Repulsive Surface} & \makecell{Double \\Horizon} & Throat & \makecell{Anti-throat} & \makecell{Attractive \\ Singularity} & \makecell{Repulsive \\Singularity} & \makecell{Light-like \\Singularity} & \makecell{Time-like \\Singularity} \\\hline
        % 1+ section
        \multicolumn{1}{|c|}{\multirow{3}{*}{$[1+]$}} & \multicolumn{1}{c|}{\multirow{2}{*}{$u_1>0$}} & \multicolumn{3}{c|}{$|C|<h$} &   &  & \checkmark &  & \checkmark  &   &  \checkmark &  \\\cline{3-13}
        \multicolumn{1}{|c|}{} &  & \multicolumn{3}{c|}{$|C|>h$} &  &  & \checkmark & \checkmark &  & \checkmark  &  & \checkmark \\\cline{2-13}
        \multicolumn{1}{|c|}{} & \multicolumn{4}{c|}{$u_1<0$} &  &  &  &  &  & \checkmark & & \checkmark \\\hline
        % 2+ block
        \multicolumn{1}{|c|}{ \multirow{2}{*}{$[2+]$}} & \multicolumn{4}{c|}{ $u_1>0$} &  &  & \checkmark & \checkmark &  & \checkmark &  & \checkmark \\\cline{2-13}
         \multicolumn{1}{|c|}{ }& \multicolumn{4}{c|}{ $u_1<0$} & & & & & & \checkmark &  & \checkmark \\\hline
        % 3+ block
        \multicolumn{1}{|c|}{ \multirow{2}{*}{$[3+]$}} & \multicolumn{4}{c|}{ $u_1>0$} &  &  & \checkmark & \checkmark &  & \checkmark & & \checkmark \\\cline{2-13}
         \multicolumn{1}{|c|}{ }& \multicolumn{4}{c|}{ $u_1<0$} & & & & & & \checkmark &  & \checkmark\\\hline
        % 1- block
        \multicolumn{1}{|c|}{ \multirow{2}{*}{$[1-]$}} & \multicolumn{4}{c|}{ $u_1>0$} &  &  & \checkmark & \checkmark & \checkmark &  & \checkmark & \\\cline{2-13}
         \multicolumn{1}{|c|}{ }& \multicolumn{4}{c|}{ $u_1<0$} & & & & & \checkmark &  & \checkmark & \\\hline
        % 2- block
         \multicolumn{1}{|c|}{\multirow{4}{*}{$[2-]$}} & \multicolumn{1}{c|}{\multirow{3}{*}{$u_1>0$}} & \multicolumn{3}{c|}{Case 1} &  &  & \checkmark & \checkmark  & \checkmark &  & \checkmark  &  \\\cline{3-13}
        \multicolumn{1}{|c|}{} &  & \multicolumn{3}{c|}{Case 2} &  &  & \checkmark &  &  &  &  &  \\\cline{3-13}
        \multicolumn{1}{|c|}{} &  & \multicolumn{3}{c|}{Case 3} &  & \checkmark & \checkmark &  &  &  &  &  \\\cline{2-13}
        \multicolumn{1}{|c|}{} & \multicolumn{4}{c|}{$u_1<0$} &  &  &  &  & \checkmark &  & \checkmark &  \\\hline
        % 3- block 
        \multicolumn{1}{|c|}{\multirow{16}{*}{$[3-]$}} & \multirow{6}{*}{Case 1} & \multirow{5}{*}{$u_1>0$} & \multicolumn{2}{c|}{A} & & & \checkmark & \checkmark & \checkmark & & \checkmark & \\\cline{4-13}
        \multicolumn{1}{|c|}{}&  & & \multicolumn{2}{c|}{B} & &  & \checkmark & & & & &\\\cline{4-13}
        \multicolumn{1}{|c|}{}&  & &  \multirow{3}{*}{C} & a & & \checkmark & & & & \checkmark & & \checkmark \\\cline{5-13}
        \multicolumn{1}{|c|}{}&  & &  & b & & \checkmark & \checkmark & & & & &\\\cline{5-13}
        \multicolumn{1}{|c|}{}&  & &  & c & \checkmark & \checkmark &  & & \checkmark & & \checkmark &\\\cline{3-13}
        \multicolumn{1}{|c|}{} & & \multicolumn{3}{c|}{$u_1<0$} &  &  &  &  & \checkmark &  & \checkmark & \\\cline{2-13}
        \multicolumn{1}{|c|}{} & \multirow{6}{*}{Case 2} & \multirow{5}{*}{$u_1>0$} & \multicolumn{2}{c|}{A} & & & \checkmark & \checkmark & & \checkmark & & \checkmark \\\cline{4-13}
        \multicolumn{1}{|c|}{}&  & & \multicolumn{2}{c|}{B} & & & \checkmark & & & & &\\\cline{4-13}
        \multicolumn{1}{|c|}{}&  & &  \multirow{3}{*}{C} & a & \checkmark & & & & \checkmark & & \checkmark & \\\cline{5-13}
        \multicolumn{1}{|c|}{}&  & &  & b & \checkmark &  & \checkmark & & & & &\\\cline{5-13}
        \multicolumn{1}{|c|}{}&  & &  & c & \checkmark & \checkmark &  & & & \checkmark & & \checkmark\\\cline{3-13}
        \multicolumn{1}{|c|}{} & & \multicolumn{3}{c|}{$u_1<0$} &  &  &  &  &  & \checkmark & & \checkmark\\\cline{2-13}
        \multicolumn{1}{|c|}{} & \multirow{4}{*}{Case 3} & \multirow{3}{*}{$u_1>0$} & \multicolumn{2}{c|}{A} & & & & & \checkmark & & & \checkmark \\\cline{4-13}
        \multicolumn{1}{|c|}{}&  & & \multicolumn{2}{c|}{B} & &  & \checkmark & & & & &\\\cline{4-13}
        \multicolumn{1}{|c|}{}&  & &  \multicolumn{2}{c|}{C} & & & & &  & & &  \\\cline{3-13}
        \multicolumn{1}{|c|}{} & & \multicolumn{3}{c|}{$u_1<0$} &  &  &  &  & \checkmark &  & & \checkmark \\\cline{1-13}
    \end{tabular}
\end{table*}

%%%%%%%%%%%%%%%%%%%%%%%%%%%%%%%%%%%%%%%%%%%%%%%%%%%%%%%%%%%%
\section{Conclusion}\label{Conclusion}
%%%%%%%%%%%%%%%%%%%%%%%%%%%%%%%%%%%%%%%%%%%%%%%%%%%%%%%%%%%%

This paper presented a thorough investigation of exact solutions for electrically charged wormholes, black holes, and black bounces within the framework of hybrid metric-Palatini gravity (HMPG). HMPG merges the metric and Palatini approaches in modified theories of gravity, creating a flexible framework that addresses key challenges in GR, especially in relation to the late-time cosmic acceleration and the dark matter problem. The study explored here considered spherically symmetric solutions under the assumption of a zero scalar field potential, presented in both the Jordan and Einstein conformal frames, with a detailed analysis performed in the former, to encompass a wide range of behaviours.
Our  results reveal a variety of gravitational configurations, such as traversable wormholes, %regular 
black holes with double horizons, and novel ``black universe'' solutions, where regions beyond the event horizon extend into an expanding cosmological structure instead of leading to singularities. 

Each class of solutions is systematically categorized according to the distinct properties exhibited by the scalar field. These classifications are subjected to an in-depth analysis that explores their horizon and throat structures, asymptotic behaviours, and the presence and nature of singularities.
The results of this analysis are comprehensively summarized in Table \ref{tab:chi}, which provides a detailed overview of the existence or absence of the various spacetime features considered. Non-physical cases, which fail to meet the criteria for physical viability, have been excluded from the summary to maintain focus on meaningful solutions.
In scenarios where the possibility arises for both throats and anti-throats to exist, or alternatively for neither structure to be present, only the configuration featuring both structures is included. Furthermore, specific solutions characterized by infinite horizons, which are particular to classes $[2-]$ and $[3-]$, have been omitted from the Table. This exclusion ensures that the representation remains concise and focused on the most illustrative configurations.

This research has revealed new insights into the adaptability of HMPG for modelling a wide variety of astrophysical phenomena and has laid the groundwork for several future directions. Further studies could extend this analysis to incorporate scalar potentials, which would likely yield additional stable black hole configurations and may address issues related to dark energy more directly. 
Additionally, it would be interesting to explore rotating solutions within HMPG, as these are essential for modelling more realistic astrophysical objects such as rotating black holes and accretion disks around compact objects. The stability analysis, particularly in rotating cases, remains a critical area for establishing the physical relevance of these solutions.

Another promising direction lies in applying these HMPG solutions to observational astrophysics. Future work could involve analysing the lensing and gravitational wave signatures of HMPG black holes and wormholes, thus testing the theory's predictions against observational data. Furthermore, exploring cosmological models under HMPG, including potential connections to inflationary dynamics and late-time acceleration, could bridge the gap between theoretical solutions and observable cosmological phenomena. This study demonstrates that HMPG provides a viable alternative framework for addressing limitations in GR, and continued investigation in these areas holds promise for enriching our understanding of gravity on both astrophysical and cosmological scales.

%%%%%%%%%%%%%%%%%%%%%%%%%%%%%%%%%%%%%%%%%%%%%%%%%%%%%%%%%%%%%%%%
\acknowledgments{
%{\bf Acknowledgement}:  
The authors thank Kirill A. Bronnikov for the helpful insights and guidance, which significantly contributed to the development of this work.
FSNL acknowledges support from the Funda\c{c}\~{a}o para a Ci\^{e}ncia e a Tecnologia (FCT) Scientific Employment Stimulus contract with reference CEECINST/00032/2018, and funding through the research grants UIDB/04434/2020, UIDP/04434/2020 and PTDC/FIS-AST/0054/2021.
MER thanks Conselho Nacional de Desenvolvimento Cient\'ifico e Tecnol\'ogico - CNPq, Brazil, for partial financial support. This study was financed in part by the Coordena\c{c}\~{a}o de Aperfei\c{c}oamento de Pessoal de N\'{i}vel Superior - Brasil (CAPES) - Finance Code 001.
}
%%%%%%%%%%%%%%%%%%%%%%%%%%%%%%%%%%%%%%%%%%%%%%%%%%%%%%%%%%%%%%%%

%%%%%%%%%%%%%%%%%% Início do apêndice
\appendix

\section{Embedding diagram}\label{ap}

In this appendix we present in detail the derivation of the equations that allow us to construct the embedding diagrams of the solutions obtained in this work.
At the beginning of this discussion, we start from the following general line element \cite{Bambhaniya:2021ugr}
\begin{equation}
ds^2=g_{00}dt^{2}+g_{11}dx^{2}+g_{22}d\theta^{2}+g_{33} d\phi^{2}.
\end{equation}
For convenience, we restrict the analysis to the equatorial plane, which implies $\theta = \pi/2$, and we consider trajectories with $t = \text{constant}$. Under these conditions, the metric is reduced to a two-dimensional form:
\begin{equation}
ds^{2}=g_{11}dx^{2}+g_{33}d\phi^{2}.\label{ap2}
\end{equation}
Our aim is to integrate this two-dimensional geometry into a three-dimensional Euclidean space. Due to the axial symmetry of the \eqref{2} metric, we assume a cylindrical Euclidean line element:
\begin{equation}
ds^{2}=-dz^{2}-dr^{2}-r^{2}d\varphi^{2},\label{ap3}
\end{equation}
the above metric can be rewritten as
\begin{equation}
ds^{2}=-\left[\frac{dz^{2}}{dx^{2}}+\frac{dr^{2}}{dx^{2}}\right]dx^{2}-r^{2}d\varphi^{2}.\label{ap4}
\end{equation}

By comparing the expressions \eqref{ap2} and \eqref{ap4}, we can identify the correspondences:
\begin{align}
\varphi &= \phi, \\
r(x) &= \sqrt{-g_{33}(x)}, \label{sigmadef} \\
z &= z(x).
\end{align}

From this, we get the following equation for $z(x)$:
\begin{equation}
\frac{dz(x)}{dx} = \pm\sqrt{-g_{11}(x) - \left(\frac{dr(x)}{dx}\right)^2}.
\end{equation}

Substituting Eq. \eqref{sigmadef} directly, we have the final form:
\begin{equation}
\frac{dz(x)}{dx} = \pm \sqrt{-g_{11}(x) - \left( \frac{d\sqrt{-g_{33}(x)}}{dx} \right)^2}
\label{z}
\end{equation}

Equation~\eqref{z} enables the numerical computation of the function $z(x)$, which, together with $r(x)$, defines the embedding surface corresponding to the geometry described by the original metric.

It is clear that the function within the square root in Eq.~\eqref{z} must be non-negative in order to be able to construct the embedding diagrams. However, this condition is not always satisfied in all regions of spacetime, and it is important to observe it to ensure that the solution $z(x)$ is real. If the term is negative, the root becomes imaginary and makes it impossible to construct the embedding surface in that region.

The procedure  for determining the embedding surfaces follows the approach developed in \cite{Paranjape:2003me}. Our spacetime can have various structures, such as spatial infinity, horizons ($x_H/u_H$), throats ($x_T/u_T$), anti-throats ($x_{aT}/u_{aT}$) and singularities ($x_s/u_s$). The presence of these structures requires special care when integrating Eq.~\eqref{z}, as the solution $z(x)$ can diverge if the integration is started on exactly one of these structures.

To avoid this problem, numerical integration should be started at points slightly removed from these structures to ensure the regularity of the function. The different regions can then be connected by suitable initial conditions. In all the cases studied, we have taken one of the characteristic structures as the starting point, for example $x_H$, and imposed the condition $z(x_H) = 0$.

Let us take as an example the Reissner-Nordström (RN) solution, described by $g_{00} = -1/g_{11} = 1 - 2m/x + q^2/x^2$, for a static and spherically symmetric line element. This solution has two structures -- the innermost Cauchy horizon ($x_C$) and the outermost event horizon ($x_H$) -- as well as a singularity in the center ($x_s$). To fully represent this geometry, it is necessary to numerically integrate Eq.~\eqref{z} in three different regions:

1. Outer region: from $x > x_H$ to infinity;

2. Region between the horizons: from $x_C$ to $x_H$, noting the change in the signature of the metric, since $g_{00} < 0$ in this region;

3. Innermost region: from $x_s$ to $x_C$.

Each of these integrations generates an embedding diagram. We emphasize that in order to avoid divergences, the condition $q^2/m^2 > 8/9$ must be guaranteed so that the singularity $x_s$ remains accessible for integration.

\begin{figure*}[!ht]
   \centering
      \includegraphics[scale=0.45]{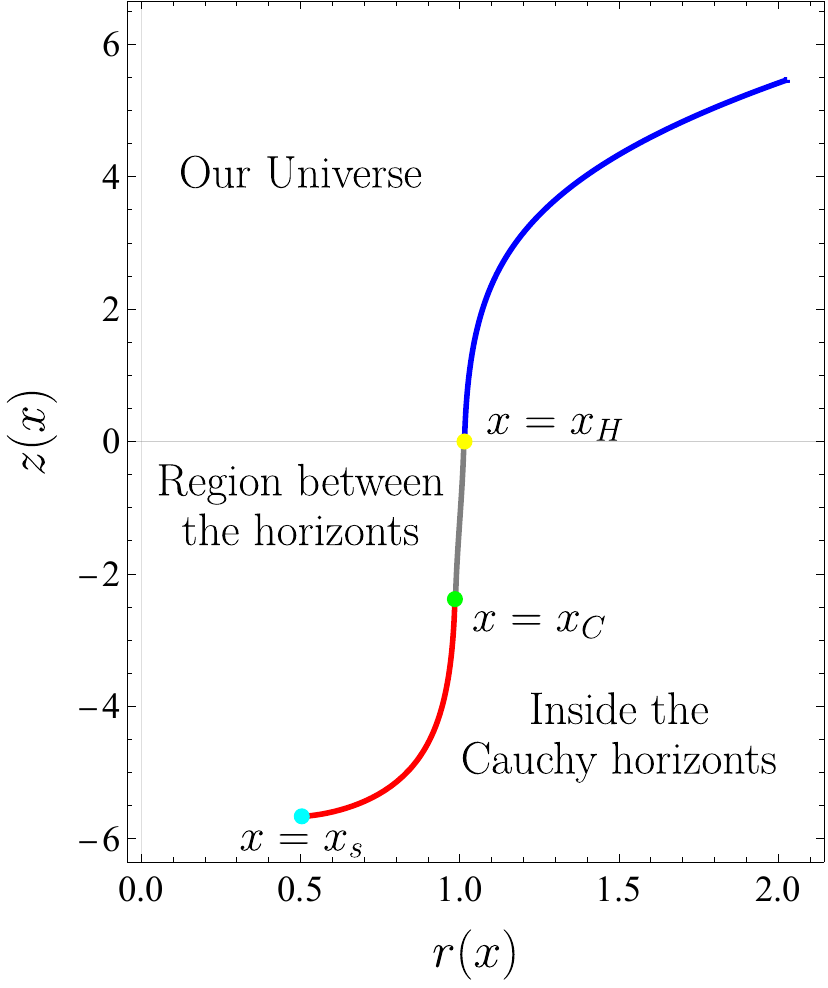}
       \hspace{1.5cm}
      \raisebox{0.15cm}{\includegraphics[scale=0.45]{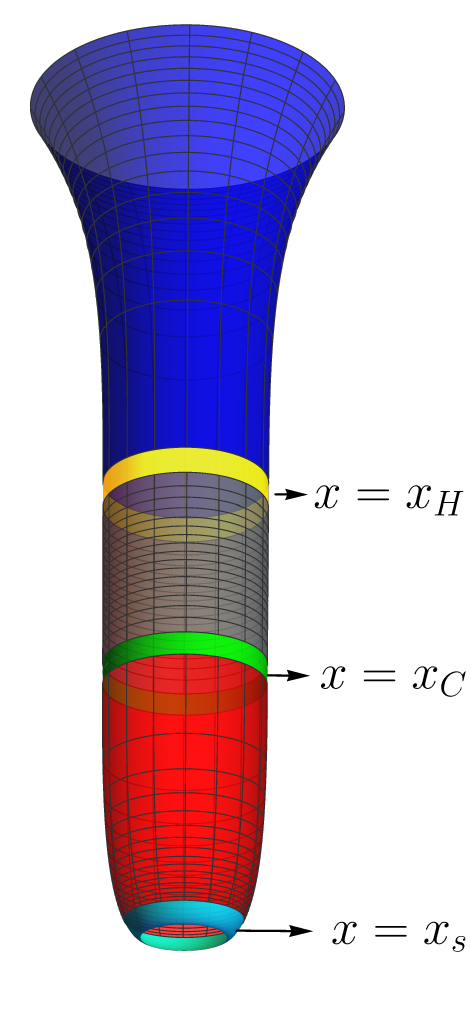}}
    \caption{ Left plot: $z(x)$ \textit{vs} $r(x)$ graph of the Reissner–Nordström solution. Right plot: Embedding diagram}
    \label{EmbRN}
\end{figure*}

The different regions can then be connected with the help of suitable initial conditions. The result is shown in the graph on the left in Fig.~\ref{EmbRN}, which shows the relationship of $z(x)$ \textit{vs} $r(x)$. The blue curve represents the region $x > x_H$, the gray curve corresponds to the range $x_C < x < x_H$, where the yellow dot denotes $x_H$ and the green dot denotes $x_C$. Finally, the red curve shows the region $x_s < x < x_C$ and ends at the cyan-colored point, which indicates the position of the singularity $x_s$.

The corresponding diagram of the surface embedding can be obtained by rotating the graph on the left in Fig.~\ref{EmbRN} around the $z(x)$ axis, as shown in the graph on the right in Fig.~\ref{EmbRN}. The colored tracks -- yellow, green and cyan -- mark the positions of $x_H$, $x_C$ and $x_s$ respectively.

Figure~\ref{EmbRN} illustrates the complete surface diagram for the RN metric. The procedure is the same as for the other solutions in this paper. For example, if a geometry has the structures singularity, throat and spatial infinity, the spacetime must be integrated into three regions: from the singularity to the throat ($x_s < x < x_{aT}$), from the anti-throat to the throat ($x_{aT} < x < x_T$) and beyond the throat ($x > x_T$). The same reasoning is applied to cases with horizons.

We should also point out that in the solutions analyzed, which have more than one horizon, all the cases considered correspond to geometries with extreme horizons. For this reason, it was not necessary to change the signature of the metric ($g_{00} < 0$), as is the case with the RN solution.

To summarize, the embedding diagrams were created in the following steps:

1. Identify the regions: Determine the appropriate intervals for the integration of Eq.~\eqref{z} based on the metric structures.

2. Separate integration: Integrate Eq.~\eqref{z} numerically in each identified region, avoiding points that could cause divergence.

3. Connection between regions: Connect the graphs with appropriate initial conditions by choosing a structure that starts at $z(x_{\text{structure}}) = 0$.

4. Surface generation: Rotate the graph around the $z(x)$ axis to obtain the graph of the surface embedding.

%%%%%%%%%%%%%%%%%%%%%%%%%%%%%%%%%%%%%%%%%%%%%%%%%%%%%%%%%%%%%%%%%%%%%%%%%%%%%%%%%%%%%%%%%%%%%%%%%%%%%%%%%%%%%%%%%%%%%%%%%%
%%%%%%%%%%%%%%%%%%%%%%%%%%%%%%%%%%%%%%%%%%%%%%%%%%%%%%%%%%%%%%%%%%%%%%%%%%%%%%%%%%%%%%%%%%%%%%%%%%%%%%%%%%%%%%%%%%%%%%%%%%%%%%%

%%%%%%%%%%%%%%%%%%%%%%%%%%%%%%%%%%%%%%%%%%%%%%%%%%%%%%%%%%%%

\end{document}